\documentclass[12pt,letterpaper]{article}
\pdfoutput=1
\usepackage[colorlinks=false,
   linkcolor=red, 
   citecolor=blue,
    filecolor=red,
    urlcolor=red,
    linktoc=all, 
    pdfstartview=FitV,
    bookmarksopen=true]{hyperref}
\usepackage[left=2cm,top=1cm,right=3cm,nohead]{geometry}

\usepackage[utf8]{inputenc} 
\usepackage{epsfig,latexsym,amsfonts,amsmath,amsthm,amssymb,amsbsy,multirow,slashed,color,mathrsfs,wasysym,textcomp,subfigure,wrapfig,comment,bbold,array,longtable,multirow}

\usepackage{amsmath}
\usepackage{amssymb}
\usepackage{amsthm}
\usepackage{graphicx}
\usepackage{slashed}
\usepackage{setspace}
\usepackage{mathrsfs}
\usepackage{subfigure}
\usepackage[boxsize=0.5em,aligntableaux=center]{ytableau}
\usepackage[bf]{caption}

\usepackage{tikz}
\usetikzlibrary{positioning}
\usetikzlibrary{intersections} 

\newlength{\PicScale}
\usepackage{breqn}\usepackage[UKenglish]{babel}
\usepackage[toc,page]{appendix}
\usepackage{amsmath}
\usepackage{amssymb}
\usepackage{graphicx}
\usepackage{hhline}
\usepackage{cite}
\usepackage[vcentermath]{youngtab}
\usepackage{geometry}
\usepackage{cite}
\usepackage{datetime,graphicx,textcomp}
\usepackage{bbm} 
 
\newcommand{\mbf}[1]{\mathbf{#1}}
\newcommand{\fund}{\textmd{fund}}
\newcommand{\adj}{\textmd{adj}}
\newcommand{\antisym}{\Lambda^2_{\rm irr}}
\newcommand{\spin}{\textmd{spin}}
\newcommand{\vect}{\textmd{vect}}

\newcommand{\KP}{\bar K_{\mathbb P^1}}

\newcommand{\be}{\begin{equation}}
\newcommand{\ee}{\end{equation}}
\newcommand{\bmat}{\left(\!\!\begin{array}}
\newcommand{\emat}{\end{array}\!\!\right)}
\newcommand{\im}{{\rm Im \,}}
\newcommand{\T}{{\rm T}}
\newcommand{\HH}{{\mathbb{H}}}
\newcommand{\RR}{{\mathbb{R}}}
\newcommand{\CC}{{\mathbb{C}}}
\newcommand{\ZZ}{{\mathbb{Z}}}
\newcommand{\PP}{{\mathbb{P}}}

\newcommand{\kod}[1]{\mathrm{#1}}

\newcolumntype{M}[1]{>{\centering\arraybackslash}m{#1}}
\newcolumntype{N}{@{}m{0pt}@{}}

\numberwithin{equation}{section}

\makeatletter
\def\@cline#1-#2\@nil{
  \omit
  \@multicnt#1
  \advance\@multispan\m@ne
  \ifnum\@multicnt=\@ne\@firstofone{&\omit}\fi
  \@multicnt#2
  \advance\@multicnt-#1
  \advance\@multispan\@ne
  \leaders\hrule\@height\arrayrulewidth\hfill
  \cr
  \noalign{\nobreak\vskip-\arrayrulewidth}}
\makeatother

\graphicspath{{./Images/}}

\begin{document}

\thispagestyle{empty}
 \begin{flushright}
 MPP-2016-61\\
LMU-ASC 15/16
 \end{flushright}
\vspace{1cm}
\begin{center}
{\LARGE {\bf Heterotic T-fects, 6D SCFTs, and F-Theory}}
 \vskip1cm 
Anamar\'ia Font$^{a}$, I\~naki Garc\'ia-Etxebarria$^{b}$, Dieter L\"ust$^{b,c}$,\\ Stefano
Massai$^{c}$, and Christoph Mayrhofer$^{c}$\\
\vskip0.8cm
\emph{$^a$ Departamento de F\'isica, Centro de F\'isica Te\'orica y
  Computacional\\
Facultad de Ciencias, Universidad Central de Venezuela\\
A.P. 20513, Caracas 1020-A, Venezuela}
\vskip0.3cm
\textit{$^b$ Max-Planck-Institut f\"ur Physik\\F\"ohringer Ring 6, 80805
  M\"unchen, Germany}\\
\vskip0.3cm
\textit{$^c$ Arnold Sommerfeld Center for Theoretical Physics,\\
Theresienstra\ss e 37, 80333 M\"unchen, Germany}\\
\vskip0.3cm
\noindent {\small{\texttt{afont@fisica.ciens.ucv.ve, inaki@mpp.mpg.de,
      dieter.luest@lmu.de, stefano.massai@lmu.de, christoph.mayrhofer@lmu.de}}}
\vspace{1cm}
\end{center}

\begin{abstract}

\noindent
We study the $(1,0)$ six-dimensional SCFTs living on defects of
non-geometric heterotic backgrounds (T-fects) preserving a
$E_7\times E_8$ subgroup of $E_8\times E_8$. These configurations can be
dualized explicitly to F-theory on elliptic K3-fibered non-compact
Calabi-Yau threefolds. We find that the majority of the resulting dual
threefolds contain non-resolvable singularities. In those cases in
which we can resolve the singularities we explicitly determine the
SCFTs living on the defect.  We find a form of duality in which
distinct defects are described by the same IR fixed point. For
instance, we find that a subclass of non-geometric defects are
described by the SCFT arising from small heterotic instantons on ADE
singularities.

\end{abstract}

\newpage

\tableofcontents

\section{Introduction}\label{sec:intro}

String theory admits a rich set of supersymmetric compactifications,
giving rise to a vast space of lower dimensional field theories. Most
of the study of these compactifications focuses on regimes where the
background can be understood geometrically by considering a classical
supergravity reduction on the geometry, supplemented with knowledge of
the dynamics on brane stacks. This is far from being the only
possibility, but it is very convenient and very amenable to concrete 
analysis. Nevertheless, it would be interesting to go away from this
geometric class of backgrounds, both to learn more about the
non-classical and non-geometrical properties of string theory, and to gain some insight
about the broader set of possible string vacua.

In this paper we focus on a class of compactifications of the
$E_8\times E_8$ heterotic string which are very non-classical,
involving compactifications on ``spaces'' that cannot be globally
described as geometries, while remaining accessible thanks to duality
with F-theory. We can, in this way, probe many of the properties of the
heterotic string away from the classical regime where it is
conventionally studied.

More concretely, we will focus on cases where the compactification
space for the heterotic string is at a generic point locally
geometric, and described by a $T^2$ fibration. The non-classical
nature of the background arises from the patching between local
descriptions, which we choose to involve non-trivial elements of the
T-duality group acting on the $T^2$ \cite{Hellerman:2002ax}. The
resulting total space is usually referred to as a non-geometric T-fold
\cite{Hull:2004in}. 
In the context of the heterotic string one should note that there is additional gauge bundle data (denoted by
$E_{T^2}$ in the following) which  mixes with the geometric data of the
$T^2$ under generic elements of the T-duality group of the $T^2$.
The patching
will send $(\tau,\rho,E_{T^2})\to (\tau', \rho', E'_{T^2})$, with
$\tau$ the complex structure of the torus, $\rho = \int_{T^2}B+iJ$ its complexified K\"ahler modulus, and $E_{T^2}$ the Wilson line data along the two cycles of the torus.
The primed values arise from the action of the
$O(2,18,\mathbb{Z})$ T-duality on the $T^2$.

Such fibrations will in general have 
defects,
i.e.\ subloci of the compactification space where a local description
in terms of the heterotic string on a smooth $T^2\times\mathbb{R}^n$
with a smooth bundle is no longer possible. For concreteness, we
consider the compactification of the heterotic string to six
dimensions. In this case, we have locally
a $T^2$ fibration over a
complex one-dimensional base. At certain points of the base we have
defects, which will induce a monodromy action on $(\tau,\rho,E_{T^2})$
as we go around them. Our goal in this paper is to describe, for a
particular class of bundles $E_{T^2}$,
the low energy dynamics living on the defect itself.

We will do this by dualizing the configuration to F-theory, where the
dynamics on the defect can be characterized by purely geometric
means. In order to do so in the most explicit way possible, we
restrict the bundle $E_{T^2}$ to have $SU(2)$ structure, so it will
break $E_8\times E_8$ down to $E_8\times E_7$. The bundle data on the
$T^2$ is then described by a single complex number, whose real and
imaginary parts are given by the Wilson line of the $SU(2)$ Cartan
around the one-cycles of the $T^2$. We denote this complexified Wilson
line by $\beta$ in the rest of the paper.

With a single Wilson line turned on, the T-duality group is
$O(2,3,\mathbb{Z})$ and an order four subset of this group can be
identified with $Sp(4,\mathbb{Z})$, which is the action of the mapping
class group of a genus-two curve on the homology. In this paper we
restrict to monodromies in this $Sp(4,\mathbb{Z})$ subgroup, so we
have a formulation in terms of monodromies of genus-two curves. This
correspondence is in fact very deep: as shown recently in
\cite{McOrist:2010jw,Malmendier:2014uka}, there is a very close
connection between the moduli space of genus-two Riemann
surfaces\footnote{The connection between the heterotic moduli space
  with one Wilson line and the associated Siegel modular forms of
  genus-two Riemann surfaces was first noted in \cite{Mayr:1995rx}.}
and the moduli space of elliptically fibered K3 surfaces having an $E_8$
and an $E_7$ point. By duality with F-theory, this is precisely the moduli space of
the heterotic string on $T^2$ with a single Wilson line. Furthermore, the map
has been explicitly worked out in
\cite{Clingher:2146c,Clingher:3503c,Malmendier:2014uka,Gu:2014ova}
(generalizing previous work in the case with unbroken $E_8\times E_8$
symmetry \cite{LopesCardoso:1996hq,McOrist:2010jw}): given a genus-two
Riemann surface, parameterizing the moduli of a heterotic
compactification with unbroken $E_8\times E_7$, there are explicit
expressions --- to be reviewed below --- for the moduli of the dual
K3.

In fact, the existence of the genus-two description for the heterotic
vacua on $T^2$ with a single Wilson line gives us a formal, but
geometric, description of the very non-geometric heterotic
compactifications of interest in this paper. This viewpoint is
particularly fruitful since there exists a classification of the
possible degenerations of genus-two fibers over a complex
one-dimensional base, obtained by Ogg-Namikawa-Ueno 
\cite{ogg66,Namikawa:1973yq}. This is analogous to, but more involved
than, the Kodaira classification of degenerations of genus one
fibrations, which are extensively used in F-theory.

\medskip

We can now summarize the main results of this paper. For each of the
possibilities allowed by the classification of genus-two degenerations
--- or equivalently, for every defect preserving $E_8\times E_7$ and
with monodromy in $Sp(4,\ZZ)$ --- we will apply the heterotic/F-theory
duality map to express the heterotic backgrounds in terms
of 
F-theory compactifications. Generically, the F-theory background dual
to a given 5-brane defect on the heterotic side will be highly
singular. In some cases (the exact criterion is
stated 
in section \ref{sec:classification}) we can resolve the singularity by
performing a finite number of blow-ups in the base of the
fibration. For all the cases where this resolution is possible we
construct the resulting smooth geometry. The blow-ups correspond to
giving vevs to tensor multiplets of the 6d (1,0) theory on the defect,
such that it flows to a Lagrangian description in the IR. In this way,
from the knowledge of the smooth geometry one can understand some
aspects (such as anomaly polynomials
\cite{Ohmori:2014pca,Ohmori:2014kda}) of the strongly coupled CFT
living at the origin of the tensor branch in terms of more ordinary
quantum field theories.
Let us note that as one might have expected, for the cases that we can
resolve we obtain theories that fall into the recent classification
of \cite{Heckman:2013pva,Heckman:2015bfa,DelZotto:2014hpa}.

In order to test our approach we will first consider local genus-two
models that correspond to geometric ADE singularities of a K3 surface,
together with a monodromy $\rho \rightarrow \rho +n$ for the
complexified K\"ahler modulus.
As expected, from the resolution of the dual F-theory models we find a
non-perturbative enhancement of the gauge algebra which agrees with
the theory of pointlike instantons hitting the orbifold singularity
determined in \cite{Aspinwall:1997ye},
 with $n$ related to the number
of instantons at the singular point ($n=0$ corresponds to local
cancellation of the modified Bianchi identity, and thus to having as
many small instantons as the degree of the ADE singularity).
We also determine the matter content from the
dual F-theory geometry, and verify explicitly that it agrees with the
expectation from anomaly cancellation \cite{Intriligator:1997dh}.

We then move on to non-geometric models that involve 
monodromies in the K\"ahler modulus $\rho$ with a non-trivial action
on the torus volume.  We find a form of duality, in that distinct
defects can give rise to the same SCFTs. For instance, we often
encounter the same SCFTs as those describing pointlike instantons on
ADE singularities, even for defects arising from non-geometric
configurations. Understanding the origin of these dualities is an
important open problem. We stress that we also find non-geometric
degenerations which are not dual to pointlike instantons on ADE
singularities, and give SCFTs which are genuinely new in the heterotic
context.

\medskip

This paper is organized as follows. In section \ref{sec:duality} we
review the formulation of heterotic/F-theory duality in terms of a map
between genus-two curves and K3 surfaces, and we discuss how it can be
used to study non-geometric heterotic backgrounds in terms of K3
fibered Calabi-Yau three-folds. In section \ref{sec:geometricmodels}
we apply our formalism to study local heterotic degenerations which
admit a geometric description in some duality frame.  In section
\ref{sec:nongeometricmodels} we discuss truly non-geometric 
models and we show how to construct a global model with such
degenerations. We also explicitly describe various dualities between
different non-geometric and geometric defects.
In section \ref{sec:othermodels} we list the resolutions
of the remaining non-geometric models, considering in particular a
class of models that do not admit a limit with vanishing Wilson line.
Finally in section \ref{sec:classification} we provide the details of
the classification of all possible local heterotic models, both
geometric and non-geometric, admitting F-theory duals that can be
resolved into smooth Calabi-Yau three-folds. We conclude with a
discussion in section \ref{sec:conclusions}. We relegate to appendix
\ref{app:ADE} the resolutions of geometric models that correspond to
pointlike instantons on ADE singularities. In appendix
\ref{app:mapE8E8} we discuss the heterotic/F-theory duality for the
case of vanishing Wilson line.  In appendix
\ref{app:IgusaClebschinvariants} we show the expressions of the
Igusa-Clebsch invariants in terms of coefficients of a sextic that
describe a given genus-two curve. In appendix \ref{app:NUlist} we
reproduce the Namikawa-Ueno classification of singular genus-two
fibers, and for each entry we compute the order of vanishing of the
Igusa-Clebsch invariants. Finally, in appendix
\ref{sec:matter-analysis} we explain how to extract the matter content
from the F-theory resolutions for an explicit example.

\section{Non-geometric heterotic vacua}\label{sec:duality}

In this section we review the formulation of F-theory/heterotic
duality recently discussed in \cite{Malmendier:2014uka,Gu:2014ova}. We
first discuss the duality in eight dimensions and then we show how to
fiber it over a common base to study non-geometric heterotic
compactifications to six-dimensions in terms of F-theory on Calabi-Yau
three-folds. 

\subsection{Heterotic/F-theory duality in 8 dimensions}

It is well known that the $E_8\times E_8$ heterotic string compactified
on $\T^2$ is dual to F-theory compactified on an elliptically fibered K3 surface \cite{Vafa:1996xn}. 
For the heterotic compactification with a Wilson line that breaks the gauge group to $E_7\times E_8$ 
the corresponding K3 is described by a Weierstra\ss{} model of the form
\begin{equation}
y^2=x^3+(a\, u^4 v^4+c\,u^3 v^5)\,x\, w^4+(b\,u^6v^6+d\,u^5v^7+u^7v^5)\,w^6=0\,,
\label{K378}
\end{equation}
where $[u:v]\in\PP^1$ and  $[y:x:w]\in \PP_{3,2,1}$ are the homogeneous coordinates of the 
base and the Weierstra\ss{} equation, respectively. 
For generic values of the
coefficients the fiber has a Kodaira singularity of type $\kod{III^*}$
($E_7$) at $u=0$ and a singularity of type $\kod{II^*}$ ($E_8$) at
$v=0$.
By virtue of the
F-theory/heterotic duality, there must be a map relating the heterotic
moduli\footnote{Recall that the compactification of the  $E_8\times E_8$ heterotic string on $T^2$ comprises eighteen complex moduli: the
sixteen  Wilson line  moduli  $\beta^i$ with $i=1,\ldots, 16$, the complex structure $\tau$ of the torus and   the complexified
K\"ahler modulus $\rho$ of the torus. Since throughout this article we are only interested in compactifications with an unbroken $E_7\times E_8$ non-abelian subgroup, we will drop the superscript of $\beta$. }   $\rho$, $\tau$ and $\beta$ to the K3 coefficients $a$, $b$, $c$ and $d$.

To obtain an understanding for this map, we study certain limits thereof. Consider first the special case $c=0$. One can immediately see
from~\eqref{K378} that this implies that both singularities are of
 type $\kod{II^*}$ ($E_8$). Thus, $c=0$ corresponds to vanishing Wilson line, i.e.\ to
$\beta =0$.  
In this limit, the coefficients $a$, $b$ and $d$ 
are related to the heterotic moduli $\tau$ and $\rho$ in the following way
\begin{align}
j(\tau) j(\rho) &= -1728^2 \frac{a^3}{27d}\, ,\label{map88}\\
(j(\tau) -1728 )( j(\rho)-1728) &= 1728^2 \frac{b^2}{4d}\, ,\nonumber
\end{align}
where $j$ is the $SL(2,\mathbb{Z})$ modular invariant function.
The map for this specific configuration was originally
obtained in \cite{LopesCardoso:1996hq}. 
Note that we can interpret the moduli $\tau$ and $\rho$ as complex
structures of two elliptic curves (one of which is the physical
heterotic torus) which are glued together at one point, i.e.\ a degenerated genus-two curve. 
The map thus can be read as a relation
between $SL(2,\mathbb{Z})$ modular forms and the K3 coefficients, cf.\ appendix \ref{app:mapE8E8}. As
we will now discuss, we can extend this relation to encompass a
non-vanishing Wilson line.

In the general setup, with $c\neq 0$, the map has been recently
established in \cite{Malmendier:2014uka}, using previous findings
about K3 surfaces related to curves of genus two
\cite{Kumar:1669k,Clingher:3335c,Clingher:3503c}. The three heterotic
complex parameters $\rho$, $\tau$, $\beta$ live on the Narain moduli space
\begin{equation}
\label{d23}
{\mathcal M}_{\rm het}={\mathcal D}_{2,3}/O(2,3,\ZZ)\qquad\textmd{with}\qquad  {\mathcal D}_{2,3}:= \frac{O(2,3,\RR)}{O(2,\RR) \times O(3,\RR)} \, ,
\end{equation}
where we used the notation of \cite{Malmendier:2014uka}. We will
consider a subset $O^+(L^{2,3})$ of
the Narain U-duality group $O(2,3,\ZZ)$ which preserves orientations, because we will be ultimately interested in fibering the duality group
holomorphically over a base.

A crucial observation is that there is an isomorphism ${\mathcal D}_{2,3} \cong \HH_2$ between the symmetric space and the genus-two Siegel upper half-plane \cite{Vinberg:2013}:
\begin{equation}
\HH_2 = \left\{\Omega =  \bmat{cc} 
\tau & \beta \\
\beta & \rho
\emat 
\bigg{\vert}  \ \tau, \rho, \beta \in \CC, \ \det\im(\Omega) > 0 , \
\im(\rho) > 0  \right\} \, .
\label{omatrix}
\end{equation}
Since on the same grounds a (two-to-one) relation between $O^+(L^{2,3})$ and $Sp(4,\ZZ)$ can be established, there is a correspondence between the moduli space of the heterotic 
compactification and the quotient of $\HH_2$ by the genus-two modular group $Sp(4,\ZZ)$. The action of $M \in Sp(4,\ZZ)$ on $\Omega$ is given by
\begin{equation}
M(\Omega )= (A \Omega + B)(C \Omega + D)^{-1}\, , \quad
M= \bmat{cc}
A & B\\
C & D
\emat
\label{sp4_action}
\end{equation}
where $A$, $B$, $C$ and $D$ are $2\times 2$ matrices such that $M \in Sp(4,\ZZ)$.
More details about this quotient and the relation to ${\mathcal M}_{\rm het}$ can be found in \cite{Malmendier:2014uka} 
and \cite{Vinberg:2013}. 

A genus-two curve has four linearly independent cycles that can be chosen to span a canonical basis such that the 
intersection form has a symplectic structure (see
e.g. \cite{Blumenhagen:2013fgp}). We indicate the symplectic basis as
$(a_1,a_2,b_1,b_2)$ in figure \ref{Fig:genus2}. The matrix $\Omega$ introduced in eq.~\eqref{omatrix}
can be determined from integrals of the two holomorphic one-forms over the $a_i$, $b_i$ cycles \cite{Namikawa:1973yq}.
 The transformations in $Sp(4,\ZZ)$ are induced by changes of homology basis that preserve the intersection form.

Coming back to the dual F-theory description~\eqref{K378}, it has been
found that the duality map can be expressed in terms of genus-two
Siegel modular forms as
\cite{Kumar:1669k,Clingher:3335c,Clingher:3503c,Malmendier:2014uka}
\begin{equation}\label{map78}
a=-\frac1{48}\psi_4(\Omega) \, , \quad
b=-\frac1{864}\psi_6(\Omega) \, ,\quad
c=-4\chi_{10}(\Omega)\, , \quad
d=\chi_{12}(\Omega)  \, . 
\end{equation}
The definition and properties of the relevant Siegel modular forms can
be found in \cite{Malmendier:2014uka}, see also \cite{Gu:2014ova}.

We also note that in eight dimensions the heterotic/F-theory map we use
naturally geometrizes the extra massless string states appearing at
self-dual points on the moduli space in terms of degenerations of the
dual K3 surface \cite{LopesCardoso:1996hq,Lerche:1998nx}.  A recent discussion on
this from the double field theory point of view appeared in
\cite{Aldazabal:2015yna}.

\subsubsection*{Genus-two curves}

As we have discussed above, the heterotic moduli can be put in correspondence with the moduli
of a hyperelliptic genus-two curve. In turn such curve, denoted $\Sigma$, can be represented by a sextic:
\begin{equation}
\label{eq:hyperelliptic-curve}
y^2=f(x)=\sum_{i=0}^6 c_i\, x^i = c_6 \prod_{i=1}^6 (x - \theta_i) \, .
\end{equation}
In order to connect the $c_i$ coefficients with the $a$, $b$, $c$, $d$ in the dual K3
fibration \eqref{K378}, we need to determine the Siegel modular forms
appearing in the map \eqref{map78} in terms of the $c_i$'s. 
This can be done in a convenient way
by first computing the
Igusa-Clebsch invariants of the sextic \eqref{eq:hyperelliptic-curve} 
and then relating them to the
Siegel modular forms of the corresponding genus-two curve. The Igusa-Clebsch invariants are defined in terms
of the six roots $\theta_i$  of \eqref{eq:hyperelliptic-curve}  as: 
\begin{equation}\label{ICinv}
\begin{split}
I_2 & = c_6^2\sum_{15} (12)^2(23)^2(45)^2 \, ,\\
I_4 &=  c_6^4\sum_{10} (12)^2(23)^2(31)^2(45)^2(56)^2(64)^2 \, , \\
I_6&=  c_6^6\sum_{60}
     (12)^2(23)^2(31)^2(45)^2(56)^2(64)^2(14)^2(25)^2(36)^2 \, ,
      \\
I_{10} &=  c_6^{10}\prod_{i < j} (ij)^2\, , 
\end{split}
\end{equation}
where $(ij):=(\theta_i - \theta_j)$ and the sums are over
permutations \cite{Igusa:1962}. By using a computer algebra program, we find the general expressions for $I_2$, $I_4$, $I_6$,
$I_{10}$ as functions of the coefficients $c_i$ of the sextic
\eqref{eq:hyperelliptic-curve}. These are somewhat involved and are
therefore relegated to appendix \ref{app:IgusaClebschinvariants}. 

In the case of a genus-one curve, the discriminant, as well as the coefficients, of the Weierstra\ss{} cubic, are related to $SL(2,\ZZ)$ modular
forms with argument the modular parameter $\tau$ of the genus-one curve. 
For a curve of genus-two the Igusa-Clebsch invariants are similarly given by Siegel modular forms
as follows \cite{Igusa:1962}:
\begin{equation}
\label{SiegelICmap}
\begin{aligned}
&I_2(c_i) = \frac{\chi_{12}(\Omega)}{\chi_{10}(\Omega)} \,,
&I_4(c_i) = 2^{-4}\cdot 3^{-2} \psi_4(\Omega) \,,\\
&I_6(c_i) = 2^{-6}\cdot 3^{-4} \psi_6(\Omega) + 2^{-4}\cdot 3^{-3}\frac{\psi_4(\Omega) \chi_{12}(\Omega)}{\chi_{10}(\Omega)} \,,
&I_{10}(c_i) = 2^{-1}\cdot 3^{-5} \chi_{10}(\Omega) \,,
\end{aligned}
\end{equation}
with $\Omega$ specified by the three complex moduli of the genus-two curve. From \eqref{map78} we can write the dual K3 coefficients in terms of
the Igusa-Clebsch invariants and thus in terms of polynomials of the
coefficients $c_i$:
\begin{equation}\label{igusa_abcd}
a=-3 I_4\, ,\quad
b= 2(I_4 I_2 - 3 I_6) \, ,\quad
c= - 2^3 3^5 I_{10} \, , \quad
d= 2\, 3^5 I_2 I_{10} \ .
\end{equation}
This  form of the map will be very important for the purpose of studying non-geometric heterotic vacua 
in lower dimensions.

\subsection{From 8 to 6 dimensions: local models and exotic defects}\label{subsec:degenerations}

In the previous section we reviewed the close relation between the moduli space of the heterotic
string on $T^2$ with one complex Wilson line and the moduli
spaces of genus-two curves and elliptically fibered K3 surfaces
developing $E_7\times E_8$ degenerations. This led to a formulation of
F-theory/heterotic duality which is very useful 
once we consider
compactifications to lower dimensions, obtained by allowing the moduli
to vary along a (compact) base variety $\mathcal{B}$. We are interested in the
case were $\mathcal{B}$ is complex one-dimensional, locally parametrized by a complex
coordinate $t\in \CC$. The structure of such a compactification is that of a fibration, with fiber a point in the Narain space or equivalently a genus-two curve which encodes this point and the base given by $\mathcal{B}$. Such fibrations allow for a  varying
$\Omega(t)$ with $Sp(4,\ZZ)$ identifications at chart transitions, or to be more precise along non-contractible loops. Since every consistent genus-two fibration will give us a consistent fibration of $\Omega(t)$, we will restrict our attention in the following to the first one. This has the advantage that we obtain again a geometry with which we can more easily deal.\footnote{This genus-two fibration is however not, at least directly, related to the physical compactification space of the heterotic string.}

To preserve supersymmetry this fibration has to be holomorphic. In the case of a non-trivial fibration this implies that the genus-two curve
$\Sigma(t)$ has to degenerate  at co-dimension one loci on the base. Moreover, following $\Omega(\Sigma(t))$ of the genus-two curve around such a degeneration locus it will experience a monodromy
transformation. Since along a loop we end up with the same fiber that we started with, the monodromy must belong to $Sp(4,\mathbb{Z})$. 
Hence, the moduli fields of the heterotic $T^2$ compactification equal only modulo a duality transformation when we go around such a non-trivial loop.
Since the
duality group includes transformations of the type
$\rho \to -1/\rho$, thereby exchanging small and large volume, the
spaces around such singularities are in general non-geometric
T-folds. Note that the non-geometric structure is here of global kind,
i.e.\ as long as we are probing only the local neighborhood of a
regular point we do not experience any duality transformations.

\begin{figure}[t]
\centering
\subfigure[]{
\includegraphics[scale=0.65]{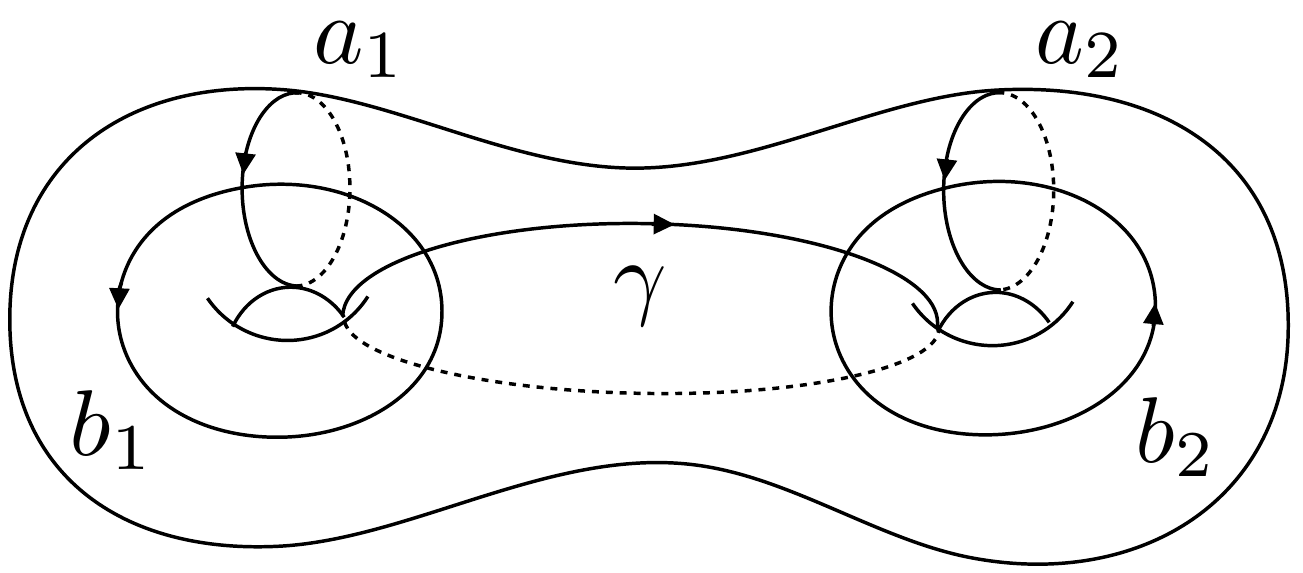}} 
\subfigure[]{
\includegraphics[scale=0.7]{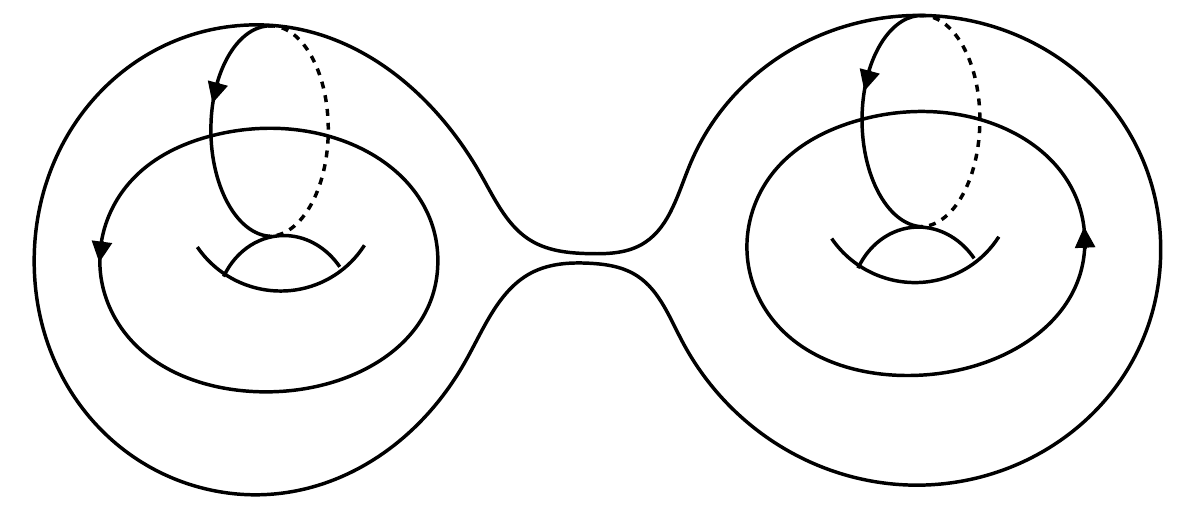}}
\caption{(a) The Humphries generators for a genus-two surface $\Sigma$: any
  element of the mapping class group $\mathcal{M}(\Sigma)$ can be
  written as a product of Dehn twists along the cycles
  $(a_1,b_1,\gamma,a_2,b_2)$. Note that $\gamma = a_1^{-1}a_2$. (b) Switching off the Wilson line
  parameter $\beta$ corresponds to splitting $\Sigma$ into two genus-one
  components. This geometrizes the $SL(2,\mathbb{Z})_{\tau}\times
  SL(2,\mathbb{Z})_{\rho}$ subgroup of the T-duality group $O(2,2,\mathbb{Z})$.}
\label{Fig:genus2}
\end{figure}

In order to get some intuition regarding $\rho$ degenerations, let us
consider the monodromy $\rho \to \rho + 1$. This arises around a point $t_0\in \mathcal B$
at which the cycle $a_2$ of $\Sigma$ shrinks, cf.~figure~\ref{Fig:genus2}. The monodromy corresponds to a Dehn
twist around this vanishing cycle.
In this
case the singularity can be identified with a NS5-brane
\cite{Ooguri:1995wj}. In fact, the corresponding solution $\rho(t) = 1/(2\pi i
)\log(t-t_0)$ coincides with the solution for a NS5-brane on
$\mathbb{C} \times T^2$ if one neglects its position on the
$T^2$. By pinching a different cycle, i.e.~$p \,a_2+q\,b_2$, one gets a more general
$(p,q)$ monodromy in $\rho$, with solution (for $q\neq 0$):
\begin{equation}
\rho (t) = -\frac{2\pi i}{ q^2 \log\left(t-t_0\right)} -
  \frac{p}{q} \, .
\end{equation}
 As an example, we can consider the
monodromy $\rho \rightarrow \rho/(1-\rho)$, corresponding to a $(0,1)$
degeneration. The volume  of the fiber does not come
back to itself after encircling the defect, and hence the solution is 
non-geometric. It coincides with an exotic $5^2_2$-brane
(or Q-brane) \cite{Obers:1998fb,deBoer:2012ma,Hassler:2013wsa}.

More general, as in F-theory, one can have $\rho$ degenerations
described by monodromies in different
conjugacy classes of the duality group. The typical genus-two
degeneration will induce also 
 monodromies for the moduli
$\tau$ and $\beta$. As in\cite{Lust:2015yia}, we refer to such degenerations as T-duality
defects, or T-fects. The aim of this paper is to uncover the
six-dimensional theories that live on such T-fects.

An advantage of mapping the non-geometric fibrations to 
geometric genus-two fibrations is the
existence of a classification of all possible local degenerations of genus-two fibers
due to Ogg and Namikawa-Ueno (NU) \cite{ogg66,Namikawa:1973yq}. This is analogous to the Kodaira classification of genus-one curves \cite{kodaira123} which we
reproduce in table \ref{tab:Kodaira}.
Furthermore, NU give explicit local descriptions of the possible degenerations  in terms of hyperelliptic curves. 
Our strategy will be then to compute the Igusa-Clebsch invariants for each local
genus-two model that realizes a given $Sp(4,\mathbb{Z})$ monodromy,
and use the F-theory/heterotic map \eqref{igusa_abcd} to obtain the
corresponding K3 degeneration in the dual 
F-theory model.

In the following we briefly describe the structure of the NU list for genus-two
degenerations.  For the reader's
convenience, we reproduce this list in appendix \ref{app:NUlist}
(and adopt their notation). For each model we list the order
of vanishing of the Igusa-Clebsch invariants that we compute from the
expressions \eqref{I2coeff}-\eqref{I10coeff}.

The geometric picture is especially useful to understand the different
classes of degenerations and to obtain a decomposition of the
monodromies in terms of a set of generators, in analogy with the ABC
factorization of F-theory \cite{Gaberdiel:1998mv, Blumenhagen:2010at}.
It follows from a theorem of Humphries (see for instance
\cite{mcgprimerbook}) that the mapping class group of $\Sigma$ is
generated by Dehn twists along the set of five cycles
$(a_1,b_1,\gamma,a_2,b_2)$, shown in figure \ref{Fig:genus2}. 
 If we
pick the base $\mathtt{B} = (a_1, a_2, b_1, b_2)$ for
$H_1(\Sigma,\mathbb{Z})$, their symplectic representation is:
\begin{equation}\label{sp4generators}
\begin{split}
A_1 &= \begin{pmatrix} 1 &0 & 1 &0 \\0 &1 & 0 &0 \\0 &0 & 1&0 \\0 &0 & 0
&  1  \end{pmatrix} \, , \quad
B_1 = \begin{pmatrix} 1 &0 & 0 &0 \\0 &1 & 0&0
  \\-1 &0 & 1&0 \\0 &0 & 0 &1  \end{pmatrix} \, , \quad
\Gamma = \begin{pmatrix} 1& 0 & 1 &-1 \\ 0 &1 & -1 &1 \\0 & 0 &1 &0\\0& 0& 0 &1  \end{pmatrix}
\, , \\[10pt]
A_2 &= \begin{pmatrix} 1 &0 & 0 &0 \\0 &1 & 0 &1 \\0 &0 & 1 &0 \\0 &0 & 0
  &1  \end{pmatrix} \, , \quad 
B_2 = \begin{pmatrix} 1 &0& 0 &0 \\0&1 & 0& 0 \\0 &0 & 1 &0 \\0&-1&
  0 &1  \end{pmatrix} \, .   
\end{split}
\end{equation}
The action of these $Sp(4,\mathbb{Z})$ elements on the period matrix
defined by \eqref{sp4_action} gives the following monodromies for the
moduli:
\begin{equation}
\begin{split}
A_1 &:\quad \tau \rightarrow \tau +1 \, ,\quad \rho \rightarrow \rho \, ,
      \quad \beta \rightarrow \beta \, , \\
B_1 & : \quad \tau \rightarrow \frac{\tau}{1-\tau} \, ,\quad \rho \rightarrow \rho+\frac{\beta^2}{1-\tau} \, ,
      \quad \beta \rightarrow \frac{\beta}{1-\tau} \, , \\
\Gamma & : \quad \tau \rightarrow \tau+1 \, ,\quad \rho \rightarrow \rho+1\, ,
      \quad \beta \rightarrow \beta-1 \, , \\
A_2 &:\quad \tau \rightarrow \tau  \, ,\quad \rho \rightarrow \rho +1\, ,
      \quad \beta \rightarrow \beta \, , \\
B_2 & : \quad \tau \rightarrow \tau +\frac{\beta^2}{1-\rho} \, ,\quad \rho \rightarrow \frac{\rho}{1-\rho} \, ,
      \quad \beta \rightarrow \frac{\beta}{1-\rho} \, . 
\end{split}
\end{equation}
Note that when $\beta =0$, $\Sigma$ splits into the two genus-one
components whose mapping class groups are identified with the subgroups
$SL(2,\mathbb{Z})_{\tau}$ and $SL(2,\mathbb{Z})_{\rho}$ of the
T-duality group $O(2,2,\mathbb{Z})$, cf.~see figure \ref{Fig:genus2}. Indeed, in this limit
$(A_1,\,B_1)$ and $(A_2\, , B_2)$ have the expected monodromies that
generate the genus-one modular group. 
A large set of entries in the NU list has monodromies in
$\mathcal{M}(\Sigma)$ that involve only the generators $A_1 ,\, B_1 ,\,
A_2 ,\, B_2$. Thus in the limit $\beta\rightarrow 0$, these models can be thought of as collisions of Kodaira monodromies
for $\tau$ and $\rho$, associated to the two genus-one
components of $\Sigma$. In this case it is simpler to use a different
basis $\tilde{\mathtt{B}} = (a_1,b_1,a_2,b_2)$ for
$H_1(\Sigma,\mathbb{Z})$, in which the symplectic representations for
the $A$ and $B$ twists are block diagonal and each block coincides
with the factorizations listed in table \ref{tab:Kodaira}. 
In the following we will study several NU examples of this
kind, corresponding to the models $\kod{[K_1-K_2-0]}\equiv \kod{[K_1-K_2]}$,
where ${\mathrm K}_i$ is one of the Kodaira type degenerations. The
monodromy of these models is thus of the form
\begin{equation}
M_{[K_1-K_2]} = M_{\mathtt{B} \tilde{\mathtt{B}}} \begin{pmatrix} K_1
  &0\\0 &K_2\end{pmatrix}M_{\mathtt{B} \tilde{\mathtt{B}}} ^{-1} \, ,
\end{equation}
with
\begin{equation}
M_{\mathtt{B} \tilde{\mathtt{B}}} =\begin{pmatrix}
  1&0&0&0\\0&0&1&0\\0&1&0&0\\0&0&0&1 \end{pmatrix} \, .
\end{equation}
We will also discuss a class of models (for
example the elliptic type 1 in the NU notation) whose
monodromies contain the twist $\Gamma$ and mix the $(\tau, \rho,
\beta)$ moduli among themselves.

\begin{table}[t]
\begin{center}
\begin{tabular}{|c|c|c|c|c|c|}
\hline
$\mu(f)$&$\mu(g)$&$\mu(\Delta)$ & Type  &       Singularity     &    Monodromy    \\
\hline \hline
$\geq 0$&$\geq 0$&0 &  $\kod{I_0}$   &$-$       &     ${\small  \begin{pmatrix} 
 1 & 0 \\ 
 0 & 1 \end{pmatrix} } $                    \\ \hline
  0&0&$n$  & $\kod{I}_n$      &     $A_{n-1}$     &  ${\small  A^n=\begin{pmatrix} 
 1 & n \\
 0 & 1 \end{pmatrix} }$                                 \\ \hline
   $\geq 1$ &1&2 & $\kod{II}$ &\text{cusp}              &    ${\small  B A=\begin{pmatrix} 
 1 & 1 \\
 -1 & 0 \end{pmatrix} }$                                  \\ \hline
  1 &$\geq 2$ & 3     & $\kod{III}$ &    $A_{1}$          &    ${\small  BAB=\begin{pmatrix} 
 0 & 1 \\
 -1 & 0 \end{pmatrix} }$                               \\ \hline
$\geq 2$ &2 & 4   & $\kod{IV}$    &    $A_{2}$       &    ${\small  (BA)^2=\begin{pmatrix} 
 0 & 1 \\
 -1 & -1 \end{pmatrix}} $                                \\ \hline
$\geq2 $&$\geq 3$&6     &  $\kod{I_0^{\ast}}$&     $D_4$  &    ${\small  (BA)^3=\begin{pmatrix} 
 -1 & 0 \\
 0 & -1 \end{pmatrix}} $                                \\ \hline
2&3&$n+6$    & $\kod{I}_n^{\ast}$ &  $D_{4+n}$   &  ${\small  (BA)^3A^n =\begin{pmatrix} 
 -1 & -n \\
 0 & -1 \end{pmatrix}} $                 \\ \hline
$\geq 3$ &4&8    & $\kod{IV^{\ast}}$&      $E_6$    &     ${\small  (BA)^4=\begin{pmatrix} 
 -1 & -1 \\
 1 & 0 \end{pmatrix} }$                 \\ \hline
   3  &$\geq 5$ & 9     & $\kod{III^{\ast}}$&   $E_7$  &    ${\small  (BA)^4B=\begin{pmatrix} 
 0 & -1 \\
 1 & 0 \end{pmatrix} }$                \\ \hline
$\geq 4$ &5&10     & $\kod{II^{\ast}}$ &     $E_8$    &    ${\small  (BA)^5=\begin{pmatrix} 
 0 & -1 \\
 1 & 1 \end{pmatrix} }$                 \\ 
\hline 
\end{tabular}
\caption{Kodaira classification of degenerations of
  elliptic fibers. We show the factorization of the monodromy in terms
of Dehn twists $A$, $B$ around the two cycles of the torus, denoted
as $(a_1,\, b_1)$ in figure \ref{Fig:genus2}. Note that $A$
corresponds to the monodromy of a $(1,0)$ 7-brane (the D7 brane) in type IIB,
while $B$ to the monodromy of a  $(0,1)$ 7-brane.}
  \label{tab:Kodaira}\end{center}\end{table}

\subsection{From 8 to 6 dimensions: global models.}\label{sec:global_models}

In the case of a compact base manifold $\mathcal B$ we cannot decouple gravity consistently anymore. This leads to further data defining the genus-two fibration.  To obtain these constrains, we change to the F-theory frame. On the F-theory side we have, as discussed already above, 
an elliptically fibered K3 surface given by \eqref{K378} instead of the K\"ahler, complex structure, and Wilson line moduli of the $T^2$. This K3 is then, similar to the heterotic side, fibered over the same (compact) base --- in the following a $\mathbb P^1$. Since we want to preserve supersymmetry in six dimensions, the total F-theory compactification 
space must be a  Calabi-Yau three-fold.  To this end, we promote the coefficients $a$, $b$,
$c$ and $d$ in  \eqref{K378} to sections of appropriate line bundles over the base
$\mathbb P^1$. Since the monomials $y^2$, $x^3$ and $u^7v^5$ come
without prefactors, they are all sections of the same line bundle with respect to the base. 
This and the Calabi-Yau condition fixes the class of the fibration uniquely as can
be seen from,
\begin{equation}
\begin{aligned}
&[y^2]=[x^3]=[u^7]=[\mathcal L^n]\,,\\
&[y]+[x]+[u]+[\KP]=[y^2]\,,
\end{aligned}
\end{equation}
where the second line is the condition for a vanishing first Chern class of the tangent bundle. If we chose for $n$ the LCM of 2, 3 and 7, we obtain $\mathcal L=\KP$. Furthermore, the coefficients in  \eqref{K378} are sections of the following line bundles:
\begin{equation}\label{eq:scaling-K3-coefficients}
\left[a\right]= 4\,\KP\,,\quad
\left[b\right]= 6\,\KP\,,\quad
\left[c\right]= 10\,\KP\,,\quad
\left[d\right]= 12\,\KP\, .
\end{equation}
In particular  that means that $a$, $b$, $c$, and $d$ are  polynomials of degree
8, 12, 20 and 24, respectively, in the  homogeneous coordinates
$[t_1:t_2] \in \PP^1$ of the base. 

The resulting Calabi-Yau threefold ($\rm{CY}_3$) can be seen as an elliptic fibration over the Hirzebruch surface ${\mathbb F}_{12}$ 
\cite{Malmendier:2014uka}. We recall that F-theory compactified on a $\rm{CY}_3$ realized as an elliptic fibration over 
${\mathbb F}_n$ is dual to a compactification of the $E_8 \times E_8$
heterotic string on K3 with 
instanton numbers $(12+n, 12-n)$ on the $E_8$ factors \cite{Morrison:1996na, Morrison:1996pp}. For $n=12$ there
are 24 instantons embedded in the first $E_8$. Taking the standard embedding \cite{Candelas:1985en, Kachru:1995wm} 
$E_8$  is broken to $E_7$ with 20 half-hypermultiplets in the ${\bf 56}$. 

We now go back to the fibration of the hyperelliptic curve, given by the sextic \eqref{eq:hyperelliptic-curve}, applying the results of the K3 fibration. Since  all the terms in equation \eqref{eq:hyperelliptic-curve}
have to be sections of the same line bundle, we obtain that
\begin{equation}
[c_i]-[c_{i-1}]=[x]\qquad\Rightarrow\qquad [c_i]=[\mathcal P^{-i}\otimes \mathcal M]
\end{equation}
with $ [x]=[\mathcal P]$ and $[\mathcal M]=2\,[y]$ with respect to the base classes. For the scaling of the Igusa-Clebsch invariants we find then
\begin{equation}
[I_k]=k\,[\mathcal P^{-3}\otimes \mathcal M]\qquad \textmd{with}\qquad k=2,4,6,10\, ,
\end{equation}
where we have used the explicit formulas for the invariants given in \eqref{I2coeff}--\eqref{I10coeff}. 
Comparing this with \eqref{eq:scaling-K3-coefficients}, we see that  $c_3$ is a sections of $\KP$. Demanding that the $c_i$'s do not vanish identically, gives the following inequality
\begin{equation}
6\,[\mathcal P]\le[\mathcal M]\qquad\Rightarrow\qquad 3\,[\mathcal P]\le \KP\,.
\end{equation}
Hence, for $\mathbb P^1$ this leads to a trivial bundle for $\mathcal P$ because the inequality can only be fulfilled by a fractional line bundle. All the coefficients $c_i$ are, therefore, sections of the anti-canonical bundle of $\mathbb P^1$, i.e.\ quadratic polynomials in the homogeneous coordinates of $\mathbb P^1$.

The upshot of the preceding discussion is that a \emph{global} non-geometric heterotic compactification can be described by a fibration of the hyperelliptic curve defined by \eqref{eq:hyperelliptic-curve} over $\PP^1$, such that the coefficients $c_i$ are given by
\begin{equation}
c_i(t_1, t_2) = \sum_{j=0}^{2}\gamma_{ij} t_1^j t_2 ^{2-j} \, ,
\label{cicoeffs}
\end{equation}
where the $\gamma_{ij}$ are constant parameters. A natural question is how the hyperelliptic fiber degenerates as 
we move along the base. In this respect, notice that the discriminant of \eqref{eq:hyperelliptic-curve} is a polynomial of degree 20, 
i.e.\ generically the fiber becomes singular over 20 points on the base. These points indicate the position of branes.
The further study and classification of the possible local degenerations will
be the subject of the next sections.
 Note that the derivation presented above assumes that the genus-two
 fiber does not split everywhere into two genus-one components, or equivalently that
 $I_{10}$ does not vanish identically. The analysis for the case with  $I_{10}\equiv 0$ can be
 found in \cite{McOrist:2010jw}. 

It is interesting to point out that the moduli space of genus-two
surfaces  also arises in the so-called G-vacua of
\cite{Martucci:2012jk, Braun:2013yla, Candelas:2014jma, Candelas:2014kma}.
In these solutions the starting point is a type IIB supersymmetric compactification on $\T^4$ with the metric, dilaton, $B$-field and R-R
potentials taking values in $\CC$. In fact, in \cite{Martucci:2012jk}
it was already proposed to construct global models
by fibering a hyperelliptic curve over $\PP^1$. The techniques that we
develop in this paper should also be useful to understand this class
of U-folds.

\section{Geometric models: five-branes on ADE singularities}\label{sec:geometricmodels}

In this section we begin our study of the brane catalog obtained from
the genus-two degenerations in the Ogg-Namikawa-Ueno
classification. We will first consider a subset of heterotic models
that have a trivial monodromy in $\rho$. These are geometric solutions
for which we have a direct understanding on the heterotic side, and
thus are a useful starting point to put the F-theory/heterotic map at
work. More concretely, the models we consider first are of type
$\kod{[I_0-K]}$, where $\mathrm K$ is an ADE singularity in
$\tau$. Note that by fiberwise mirror symmetry, they also describe a
non-geometric model with constant $\tau$ and non-trivial monodromy in
$\rho$. Having a singularity in $\tau$ induces a monodromy in the
$B$-field due to the Bianchi identity:
\begin{equation}\label{hetbianchi}
dH = \frac{\alpha'}{4} \Big[ \text{Tr} (F\wedge F) - \text{Tr}
  (R\wedge R) \Big]\, .
\end{equation}
Having a component of type $\mathrm{I}_0$, meaning trivial monodromy
in $\rho$, and leaving the gauge group unbroken, forces us to have
some small instantons on top of the ADE singularity, with the number
of instantons related to the order of the singularity. We then add
further small instantons, described by the $\kod{[K-I_n]}$ models
with a monodromy $\rho \to \rho + n$. Finally we consider the
$[\kod I_{n-p-q}]$ models, in which the three moduli $(\tau, \rho, \beta)$
shift by an integer.

\subsection{$\kod{[I_0-II^{\ast}]}$ model and $E_8$ singularity}\label{ss:I0E8}

To start, we consider a geometric $E_8$ singularity on the heterotic side,
described by the model $\kod{[I_0-II^{\ast}]}$. We discuss this
example in detail in order to illustrate the main points, while for the remaining models we
summarize the results in appendix \ref{app:ADE}. 
From the NU list we read off the $Sp(4,\mathbb{Z})$ monodromy:
\begin{equation}
M_{[\kod{II^{\ast}}-\kod{I}_0]} = \begin{pmatrix} 0&0&-1&0 \\ 0&1&0&0\\1&0&1&0\\
  0&0&0&1\end{pmatrix} \, .
\end{equation}
This is indeed the action of the product of twists $(B_1A_1)^5$ along
the homology basis of one of the genus-one handles.
Recall that $A_i$ and $B_i$ are twists
around the $a_i$ and $b_i$ cycles shown in figure \ref{Fig:genus2}. 
The action on the heterotic
$(\tau,\rho,\beta)$ moduli can be found from the $Sp(4,\mathbb{Z})$
action on $\Omega$ given in \eqref{sp4_action}:
\begin{equation}
\tau \rightarrow -\frac{1}{1+\tau} \, ,\quad \rho \rightarrow \rho - \frac{\beta^2}{1+\tau}
\, ,\quad \beta \rightarrow
\frac{\beta}{1+\tau}\, .
\end{equation}
Note that when the Wilson line value $\beta$ is turned off, this is
precisely the monodromy of a $\kod{II^{\ast}}$ type fiber of the $\tau$ fibration. 

The genus-two model with this monodromy is given by the following curve:
\begin{equation}
y^2 =\left(t^5+x^3\right) \left(x^2+\alpha  x+1\right) \, ,
\end{equation}
where the local coordinate $t\in \mathcal B$ was chosen such that the
degeneration is at the origin, and $(x,y)$ are coordinates
on the  fiber.
By computing the Igusa-Clebsch invariants from equations
\eqref{I2coeff}-\eqref{I10coeff}, and plugging the result in the
heterotic/F-theory map \eqref{igusa_abcd}, \eqref{K378}, we get the dual K3
fibration:
\begin{equation}
y^3 = x^3 +f(u,v,t) x + g(u,v,t) \, ,
\end{equation}
where:
{\allowdisplaybreaks
\begin{align}
f&=108 (\alpha -2) (\alpha +2) t^5 u^3 v^4 \big[486 t^{25}
         v-972 \alpha ^3 t^{20} v+2916 \alpha  t^{20} v+486 \alpha ^6 t^{15} v-2916 \alpha ^4 t^{15} v\nonumber\\
&\quad+4374 \alpha ^2 t^{15} v+972 t^{15} v-972 \alpha ^3 t^{10}
  v+2916 \alpha  t^{10} v+2 t^5 u+486 t^5 v-\alpha  u\big] \, , \nonumber\\
g &= u^5 v^5 \big[-314928 \alpha ^3 t^{35} v^2+1259712 \alpha
         t^{35} v^2+629856 \alpha ^6 t^{30} v^2-4408992 \alpha ^4
         t^{30} v^2  \nonumber\\
       &\quad +7479540 \alpha ^2 t^{30} v^2+314928 t^{30} v^2-314928 \alpha ^9 t^{25} v^2+3149280 \alpha ^7 t^{25} v^2-10235160 \alpha ^5 t^{25} v^2  \nonumber\\
       &\quad +9605304 \alpha ^3 t^{25} v^2+4408992 \alpha  t^{25} v^2+216 t^{20} u v-78732 \alpha ^8 t^{20} v^2+1417176 \alpha ^6 t^{20} v^2  \nonumber\\
       &\quad-7007148 \alpha ^4 t^{20} v^2+10235160 \alpha ^2 t^{20} v^2+629856 t^{20} v^2-1944 \alpha ^3 t^{15} u v+7452 \alpha  t^{15} u v  \nonumber\\
       &\quad+157464 \alpha ^5 t^{15} v^2-1417176 \alpha ^3 t^{15} v^2+3149280 \alpha  t^{15} v^2+216 \alpha ^6 t^{10} u v-1620 \alpha ^4 t^{10} u v  \nonumber\\
       &\quad+6156 \alpha ^2 t^{10} u v-11880 t^{10} u v-78732 \alpha
         ^2 t^{10} v^2+314928 t^{10} v^2+216 \alpha ^3 t^5 u v-972
         \alpha  t^5 u v+u^2\big]\, .\nonumber 
\end{align}}
We see that at $u=0$ and $v=0$ there are fibers of type $\kod{III^{\ast}}$
and $\kod{II^{\ast}}$ respectively, coming from the perturbative $E_7\times
E_8$ gauge group of the heterotic string. Moreover, close to $u=t=0$ there are additional enhancements, schematically
described by the following leading terms (up to for now unimportant coefficients):
\begin{equation}
y^2 = x^3 + \left[t^{10} u^3+ t^5 u^4 \right] x + t^{10} u^5 +t^5 u^6 +
u^7 \, .
\end{equation}
Clearly, the vanishing orders of $f$, $g$ and $\Delta$ at $u=t=0$
are non-minimal. To resolve the singularity
we need to perform a series of blowups in the base
\cite{Aspinwall:1997ye} as we now explain.

\begin{figure}[h]
\centering
\setlength{\PicScale}{0.7cm}
\begin{tikzpicture}
\draw [name path=line 0] (1.5*0*\PicScale,0) -- node [above right] {$\kod{III^{\ast}}$} ++(-45:3\PicScale);
\draw [name path=line 1] (1.5*1*\PicScale+2.1\PicScale,0) -- node [below right] {$\kod{I_0}$} ++(-135:3\PicScale);
\draw [name path=line 2] (1.5*2*\PicScale,0) -- node [above right] {$\kod{II}$} ++(-45:3\PicScale);
\draw [name path=line 3] (1.5*3*\PicScale+2.1\PicScale,0) -- node
[below right] {$\kod{IV}$} ++(-135:3\PicScale);
\draw [name path=line 4] (1.5*4*\PicScale,0) -- node [above right] {$\kod{I_0^{\ast}}$} ++(-45:3\PicScale);
\draw [name path=line 5] (1.5*5*\PicScale+2.1\PicScale,0) -- node [below right] {$\kod{II}$} ++(-135:3\PicScale);
\draw [name path=line 6] (1.5*6*\PicScale,0) -- node [above right] {$\kod{IV^{\ast}}$} ++(-45:3\PicScale);
\draw [name path=line 7] (1.5*7*\PicScale+2.1\PicScale,0) -- node
[below right] {$\kod{II}$}++(-135:3\PicScale);
\draw [name path=line 8] (1.5*8*\PicScale,0) -- node [above right] {$\kod{I_0^{\ast}}$} ++(-45:3\PicScale);
\draw [name path=line 9] (1.5*9*\PicScale+2.1\PicScale,0) -- node
[below right] {$\kod{IV}$}++(-135:3\PicScale);
\draw [name path=line 10] (1.5*10*\PicScale,0) -- node [above right]
{$\kod{II}$} ++(-45:3\PicScale);
\draw [name path=line 11] (1.5*11*\PicScale+2.1\PicScale,0) -- node
[below right] {$\kod{I_0}$}++(-135:3\PicScale);
\draw [name path=line 12] (1.5*12*\PicScale,0) -- node [above right]
{$\kod{II^{\ast}}$} ++(-45:3\PicScale);
\draw [name path=line 13] (1.5*13*\PicScale+2.1\PicScale,0) -- node
[below right] {$\dots$}++(-135:3\PicScale);
\end{tikzpicture}\caption{Resolution of the dual
  $\kod{[II^{\ast}-I_0]}$ model.}\label{fig:resolutionE8}
\end{figure}
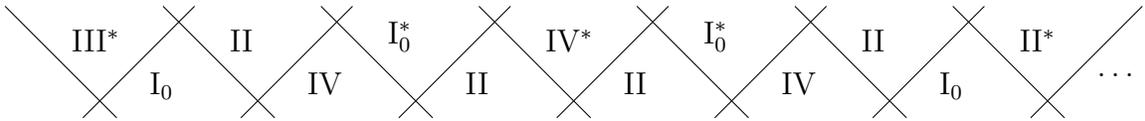

The blowups can be implemented by replacing:
\begin{equation}
\begin{split}
\label{blowup}
x &\rightarrow \left(e_1 e_2^2 \cdots e_{10}^{10} \right)^2x' \, ,\quad y \rightarrow
  \left(e_1 e_2^2 \cdots e_{10}^{10} \right)^3y'\,,
 \\
t &\rightarrow  e_1 e_2 \cdots e_{10} t' \, ,\quad u \rightarrow
    e_1 e_2^2 \cdots e_{10}^{10} u' \, .
\end{split}
\end{equation}
At this stage it is convenient to use the notation of \cite{Heckman:2015bfa} to
identify  each divisor $e_i$ by an integer equal to minus its self-intersection number.
In this notation the above resolution gives a chain 
of the form $1\, 2\,2\,2\,2\,2\,2\,2\,2\,2$.  While this reduces the order
of vanishing of $f$, $g$ and $\Delta$ along each $e_i$ to be
of Kodaira type, at the intersections $e_3\cap e_4$, $e_4\cap e_5$,
$e_5\cap e_6$, $e_6\cap e_7$ the orders of vanishing are still too high and
further blowups are required. We iterate this process until we reach a
smooth model, arriving at the resolution shown in figure
\ref{fig:resolutionE8}.

\noindent We schematically represent the resolution as: 
\begin{equation}\label{10plistE8res}
\mathrm{ [III^{\ast}] - \Box - I_0- II^{\ast}-\Box }\, ,
\end{equation}
where the leftmost factor is the perturbative $E_7$ singularity at $u=0$ and we defined the chain $\Box$ to be:
\begin{equation}\label{smallchainE8}
\Box =  \mathrm{  I_0 - II - IV - I_0^{\ast} -II- IV^{\ast} -II- I_0^{\ast}- IV
- II }\, .
\end{equation}
Dropping the $\mathfrak{e}_7$ factor, the chain in \eqref{10plistE8res} has the self-intersection pattern 
$1\,2\,2\,3\,1\,5\,1\,3\,2\,2\,1\,10\,1$ $2\,2\,3\,1\,5\,1\,3\,2\,2$.

The next step is to figure out the gauge algebras supported on each curve. This amounts
to checking for the presence of monodromies which may reduce the simply laced gauge algebras,
na\"ively expected from the Kodaira classification, to non-simply laced subalgebras thereof \cite{Bershadsky:1996nh}. A detailed description of how this works  in terms of the Weierstra\ss{} model was given in \cite{Grassi:2011hq}. We briefly recall
the procedure for the singularities appearing in our example. Type $\kod{II}$ singularities give no gauge group, while
for type $\kod{IV}$, $\kod{I_0^{\ast}}$, $\kod{IV^{\ast}}$ on a divisor $e_i$ one has to consider
the appropriate monodromy covers, as displayed in table
\ref{tab:monodromycovers}. 
After performing
this analysis on  the chain \eqref{10plistE8res} we finally obtain a
smooth model represented as:
\begin{equation}
\label{10plistE8res2}
\footnotesize{
\begin{tabular}{|cccccccccccccccccccccc|}
\hline
 &  &$\mathfrak{sp}(1)$ &$\mathfrak{g}_2$& &$\mathfrak{f}_4$&
  &$\mathfrak{g}_2$&$\mathfrak{sp}(1)$& & &$\mathfrak{e}_8$& & &$\mathfrak{sp}(1)$ &$\mathfrak{g}_2$& &$\mathfrak{f}_4$&
  &$\mathfrak{g}_2$&$\mathfrak{sp}(1)$& \\
1 & 2&     2
                        &3&1&5&1&3&2&2&1&10&1&2&2&3&1&5&1&3&2&2 \\\hline
\end{tabular}}
\end{equation}

\begin{table}[t]\begin{center}
\renewcommand{\arraystretch}{1.5}
\begin{tabular}{|c|c|c|}
\hline
Type  &     Monodromy  cover & Algebra  \\
\hline \hline
  $\kod{I}_0$, $\kod{I_1}$     &$-$       &     $-$        \\ \hline
$\kod{I_2}$   &$-$       &     $\mathfrak{su}(2)$       \\ \hline
$\kod{I}_n$, $n>2$   &$\psi^2 + \left.\left(9g/2f \right)\right|_{e_i=0}$      &
                                                     $\text{red}:
                                                      \,
                                                      \mathfrak{su}(n)\,
                                                      ,\,\text{irred}:
                                                      \,
                                                      \mathfrak{sp}([n/2])
                                                      $
  \\\hline
$\kod{II}$   &$-$       &     $-$       \\ \hline
$\kod{III}$   &$-$       &     $\mathfrak{su}(2)$       \\ \hline
$\kod{IV}$   &$\psi^2 -\left. \left( g/e_i^2 \right)\right|_{e_i=0}$       &     $\text{red}:
                                                      \,
                                                      \mathfrak{su}(3)\,
                                                      ,\,\text{irred}:
                                                      \,
                                                      \mathfrak{sp}(1)
                                                      $ \\ \hline
$\kod{I_0^{\ast}}$   &$\psi^3 + \psi \left.\left(  f/e_i^2 \right)\right|_{e_i=0}+ \left.
               \left(  g/e_i^3 \right)\right|_{e_i=0}$       &     $ \text{3 comp}: \,
                                              \mathfrak{so}(8) \,
                                              ,\,\text{2 comp}: \,
                                              \mathfrak{so}(7)\,
                                              ,\,\text{irred}: \,
                                              \mathfrak{g}_2  $ \\
  \hline
$\kod{I}_{2n-5}^{\ast}$, $n>2$   &$\psi^2 +\frac14 \left.  \left(\Delta/e_i^{2n+1} \right)\left(2e_if/9g\right)^3\right|_{e_i=0}$       &     $\text{red}:
                                                      \,
                                                      \mathfrak{so}(4n-2)\,
                                                      ,\,\text{irred}:
                                                      \,
                                                      \mathfrak{so}(4n-3)
                                                      $ \\ \hline
$\kod{I}_{2n-4}^{\ast}$, $n>2$   &$\psi^2 + \left.  \left(\Delta/e_i^{2n+2} \right)\left(2e_if/9g\right)^2\right|_{e_i=0}$       &     $\text{red}:
                                                      \,
                                                      \mathfrak{so}(4n)\,
                                                      ,\,\text{irred}:
                                                      \,
                                                      \mathfrak{so}(4n-1)
                                                      $ \\ \hline
$\kod{IV^{\ast}}$   &$\psi^2 - \left.  \left(g/e_i^{4} \right)\right|_{e_i=0}$       &     $\text{red}:
                                                      \,
                                                      \mathfrak{e}_6\,
                                                      ,\,\text{irred}:
                                                      \,
                                                      \mathfrak{f}_4
                                                      $ \\ \hline
$\kod{III^{\ast}}$   &$-$       &     $
                                                      \mathfrak{e}_7
                                                      $ \\ \hline
$\kod{II^{\ast}}$   &$-$       &     $
                                                      \mathfrak{e}_8
                                                      $ \\ \hline
\end{tabular}
\caption{Equations for the monodromy covers of the Kodaira singular
  fibers and the corresponding gauge algebras, adapted from
  \cite{Grassi:2011hq}. For degree 2 covers, we get a bigger algebra
  when the cover is reducible, namely its discriminant has a square
  root. For the type $\kod{I_0^{\ast}}$ the cover has degree 3 and the
    gauge algebra depends on the number of components.  }
  \label{tab:monodromycovers}\end{center}\end{table}

The resulting non-perturbative enhancement
precisely matches the one given by Aspinwall and Morrison in
\cite{Aspinwall:1997ye} for the theory of ten pointlike instantons on
an $E_8$ singularity. This confirms our intuition from the monodromy
of the genus-two model and the heterotic Bianchi identity. The only difference is that we now have a
perturbative algebra $\mathfrak{e}_7$ coming from the broken gauge
group of the heterotic string. Similar chains have been discussed
recently in \cite{DelZotto:2014hpa}. The matter content can be
determined from a closer look at the monodromy covers or by anomaly
cancellation. This was in fact already done in
\cite{Intriligator:1997dh} and we will not repeat the analysis here.

We can now replicate the previous computation for all the models that
have an $\mathrm{I}_0$ component for $\rho$ and an arbitrary elliptic Kodaira
type for $\tau$, for which we obtain the theories of point-like
instantons on ADE singularities derived in \cite{Aspinwall:1997ye}.
Details of this analysis are relegated to appendix \ref{app:ADE}.

\subsubsection{Adding five-branes}
\label{sss:InE8}

We can consider the situation in which
more pointlike instantons sit at the $E_8$ singularity. From the
heterotic perspective this is done by allowing a monodromy
in the $B$-field, in order to satisfy the Bianchi identity
\eqref{hetbianchi}. This corresponds to a parabolic monodromy in
$\rho$, and we thus need to consider the Namikawa-Ueno model
$[\kod{II}^{\ast}-\kod{I}_n]$. The local degeneration can be modeled by the
following curve:
\begin{equation}
y^2 = (t^5+x^3)\left[(x-1)^2 + t^n\right] \, .
\end{equation}
The resolution of the dual F-theory model proceeds in a similar way as
discussed in the previous section. However after performing $10+n$
blowups there is now a chain of $(n+1)$ intersecting $\kod{II^{\ast}}$
fibers. The resolution of these additional intersections are again
similar to the ones in the previous section. We arrive at:
 \begin{equation}
 \label{resoInE8}
[\kod{III}^{\ast}] -\Box-\left(\kod I_0-\kod{II}^{\ast}-\Box
\right)_{(1)}-\dots-\left(\kod I_0- \kod{II}^{\ast}-\Box\right)_{(n+1)} \, ,
\end{equation}
where the chain $\Box$ is defined in \eqref{smallchainE8}. The
non-perturbative gauge algebra is then:
\begin{equation}
 \label{resoInE82}
\mathcal{G}_{np} =\mathfrak{sp}(1) \oplus \mathfrak{g}_2\oplus \mathfrak{f}_4
  \oplus \mathfrak{g}_2\oplus \mathfrak{sp}(1) \oplus
 \left[ \mathfrak{e}_8\oplus \mathfrak{sp}(1)\oplus
  \mathfrak{g}_2\oplus\mathfrak{f}_4\oplus
  \mathfrak{g}_2\oplus\mathfrak{sp}(1)\right]^{\oplus (n+1)} \, .
\end{equation}
This result can again be matched with the theory of $(10+n)$ pointlike
instantons on the $E_8$ singularity given in\cite{Aspinwall:1997ye}.
The pattern of curves and self-intersection numbers is more efficiently determined using the
toric geometry techniques reviewed, and exemplified for this $\kod{[II^{\ast}-I_n]}$ model, 
in section \ref{sec:classification}. In this way we find:
\begin{align}
\label{resoInE83}
&
\footnotesize{
\begin{tabular}{|cccccccccc|}
\hline
& & $\mathfrak{sp}(1)$ &$\mathfrak{g}_2$& & $\mathfrak{f}_4$& &  $\mathfrak{g}_2$& $\mathfrak{sp}(1)$ &  \\
1 & 2& 2& 3& 1& 5 & 1 & 3 & 2 & 2 \\
\hline
\end{tabular}\, 
}
\footnotesize{
\begin{tabular}{|cccccccccccc|}
\hline
 & $\mathfrak{e}_8$& & & $\mathfrak{sp}(1)$ &$\mathfrak{g}_2$& & $\mathfrak{f}_4$& &  $\mathfrak{g}_2$& $\mathfrak{sp}(1)$ &  \\
1 & 11 & 1 & 2& 2& 3& 1& 5 & 1 & 3 & 2 & 2 \\
\hline
\end{tabular}\, 
\times  }   \nonumber  \\ 
& 
\footnotesize{
\begin{tabular}{|cccccccccccc|}
\hline
 & $\mathfrak{e}_8$& & & $\mathfrak{sp}(1)$ &$\mathfrak{g}_2$& & $\mathfrak{f}_4$& &  $\mathfrak{g}_2$& $\mathfrak{sp}(1)$ &  \\
1 & 12 & 1 & 2& 2& 3& 1& 5 & 1 & 3 & 2 & 2 \\
\hline
\end{tabular}^{\, \oplus (n-1)}
\hspace*{-5mm}  \times} \\ \nonumber
& 
\footnotesize{ \times
\begin{tabular}{|cccccccccccc|}
\hline
 & $\mathfrak{e}_8$& & & $\mathfrak{sp}(1)$ &$\mathfrak{g}_2$& & $\mathfrak{f}_4$& &  $\mathfrak{g}_2$& $\mathfrak{sp}(1)$ &  \\
1 & 11 & 1 & 2& 2& 3& 1& 5 & 1 & 3 & 2 & 2 \\
\hline
\end{tabular}\, 
\, . }
\end{align}
We have verified that the matter representations consist
only of  $\frac12(\mathbf{2},\mathbf{1}) \oplus \frac12(\mathbf{2},\mathbf{7})$ for each
$\mathfrak{sp}(1) \oplus \mathfrak{g}_2$, as expected from anomaly cancellation
\cite{Intriligator:1997dh, Heckman:2015bfa}.

\subsection{Five-branes on $\mathbb{C}^2/\mathbb{Z}_k$}
\label{ss:atype}

In the previous section and in appendix \ref{app:ADE} we discuss the
duals of degenerations of elliptic type, which are geometric in some
T-duality frame. 
In order to exhaust all the models that admit a clear geometric interpretation,
we analyze now parabolic models that are associated with A-type singularities.

\subsubsection{$[\kod{I}_{n-p-0}]$ model} \label{sec:I_n-p-0}

We consider a model with a simple parabolic monodromy for  the
moduli, the type $[\kod I_{n-p}]$ in the NU list. In a geometric
frame the monodromy action is just a shift:
\begin{equation}\label{moninp}
\tau \rightarrow \tau + p \, ,\quad \rho \rightarrow \rho +n \, ,
\quad \beta \rightarrow \beta \, .
\end{equation}
From the Bianchi identity \eqref{hetbianchi} we expect this model to
describe $(n+p)$ pointlike instantons on a $\mathbb{C}^2/\mathbb{Z}_p$
singularity. We can verify this explicitly by resolving the dual
F-theory model, as in the previous sections. We start from the local
genus-two fibration given by:
\begin{equation}
y^2 = (x^2 + t^n)(x-1)\left[(x-\alpha)^2+t^p\right] \, .
\end{equation} 
At $t=0$ one homology cycle for each of the two genus-one components shrinks,
giving rise to the monodromy \eqref{moninp}.
The structure of the dual K3 fibration near the intersection $u=t=0$ is described by
the following model:
\begin{equation}
y^2 = x^3 + \left[ t^{n+p} u^3 -3 u^4\right] x + t^{n+p}
u^5+(2+t^p)u^6+u^7 \, ,
\end{equation}
with discriminant
\begin{equation}
\begin{split}
\Delta = -u^9 \big(54 u^3 t^{n+p}&+216 u^2 t^{n+p}+54 u^2 t^{n+2 p}-9
         u t^{2 n+2 p}+4 t^{3 n+3 p}\\
&+27 u^3 t^{2 p}+54 u^4 t^p
         +108 u^3 t^p+27 u^5+108 u^4\big)\, .
\end{split}
\end{equation}
The resolution requires $n+p$ blowups to arrive at a smooth model,
and produces a chain of $(n+p-1)$ curves with self-intersection $(-2)$
supporting singularities of Kodaira type
$\kod{I}_k$, and a $(-1)$ curve at
the end where the chain intersects the $E_7$ singularity. Looking at
the monodromy cover we see that special unitary gauge algebras are realized
and we arrive at the following gauge theory:

\begin{center}\footnotesize{
\begin{tabular}{|cccccccccccccc|}
\hline
&&&&&&&&&&&&&\\[-10pt]
 & &  $\mathfrak{su}(2)$ &&
                                               $\mathfrak{su}(k-1)$&$\widehat{\mathfrak{su}(k)}_1$&$\mathfrak{su}(k)_2$&&$\mathfrak{su}(k)_{m-1}$&$\widehat{\mathfrak{su}(k)}_m$&$\mathfrak{su}(k-1)$&&$\mathfrak{su}(2)$&\\
1 &   2 &2  & $\cdots$&2&     2 &2&$\cdots$&2& 2&     2 &$\cdots$&2&2
  \\\hline 
\end{tabular} } 
\end{center}
where we defined
\begin{equation}
k = \frac{| n+ p |}{2}-\frac{|n - p|}{2}\, , \quad m= |n-p|+1 \, .
\end{equation}
The hat over $\mathfrak{su}(k)_1$ and $\mathfrak{su}(k)_m$ indicates
that these gauge factors do not only have states in the bifundamentals
with their nearest neighbors but also fundamentals coming from the
intersection with the residual discriminant, in accord with anomaly
cancellation. Setting for instance $n>p$ we see that we obtain the
theory on $n+p$ pointlike instantons on a 
$\mathbb{C}^2/\mathbb{Z}_p$ singularity \cite{Aspinwall:1997ye}. 

\begin{figure}[t]
\begin{center}
\includegraphics[scale=0.6]{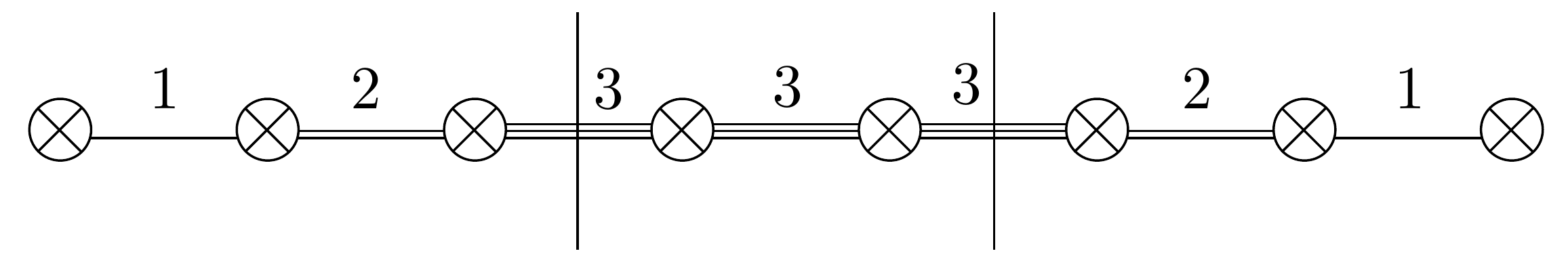}
\caption{The NS5-D6-D8 configuration corresponding to the $\kod{[I_{5-3}]}$
  model: circles with a cross represent NS5 branes, the horizontal lines
are the D6's and the vertical line are D8 branes.}
\label{Fig:i5i3}
\end{center}\end{figure}

It is interesting to note that the same configuration can be
understood from the IIA viewpoint \cite{Hanany:1997gh}, by dualizing
along the circle degenerating on the seven-brane intersections. We now
find a brane system with NS5(12345), D6(123456) and D8(12345789),
shown in figure \ref{Fig:i5i3}. 
The length of the segments wrapped by the D6 branes in the IIA description is determined by the vevs of the scalars in the tensor multiplets.
These in turn are given by the volumes of the base blow-up $\PP^1$'s on the F-theory side.
The D8-branes sit at the boundaries of the ``plateau'' of
$\mathfrak{su}(2)$ factors. The global symmetry from the boundary (not
shown in the figure) can be understood from a non-perturbative
enhancement coming from massless D0 branes (see \cite{Gorbatov:2001pw}
for a review). Due to the effect found in \cite{Hanany:1997sa}, the
brane model is useful to understand the origin of the ``staircase''
behavior of the F-theory chain. Indeed, on the left of the leftmost
D8 brane, and on the right of the rightmost D8 branes we have a unit
of Romans mass and thus we must have one net unit of D6 charge ending
on each NS5. The near horizon geometry of such brane systems has been
discussed recently in \cite{Gaiotto:2014lca,DelZotto:2014hpa}.

\subsubsection{$[\kod{I}_{n-p-q}]$ model}\label{sec:I_n-p-q}

We now discuss a generalization of the previous model that includes a
perturbative monodromy for the Wilson line $\beta$. This is the
$[\kod{I}_{n-p-q}]$ model in the NU list and it is described by the
following fibration:
\begin{equation}
y^2 = (x^2+t^n)\left[ (x-1)^2+t^m\right]\left[(x-2)^2+t^q\right] \, ,
\end{equation}
which has the monodromy:
\begin{equation}
\tau \rightarrow \tau + m+q \, ,\quad \rho\rightarrow \rho+n+q \, ,
\quad \beta\rightarrow \beta -q \, .
\end{equation}
By proceeding as in the previous section, we obtain the following theory
from the resolution of the $u=t=0$ intersection:
\begin{center}\footnotesize{
\begin{tabular}{|ccccccccccccccc|}
\hline
&&&&&&&&&&&&&&\\[-10pt]
 & &  $\mathfrak{su}(2)$ & $\mathfrak{su}(4)$&&$\widehat{\mathfrak{su}(2q)}$&$\mathfrak{su}(2q+1)$&&$\widehat{\mathfrak{su}(k)}$&
 $\mathfrak{su}(k)^{\oplus m}$&$\widehat{\mathfrak{su}(k)}$&$\mathfrak{su}(k-1)$&&$\mathfrak{su}(2)$&\\
1 &   2 &2  & 2&$\cdots$&2&     2 &$\cdots$&2 &$(2)\times m$ & 2&2&$\cdots$&2&2 \\\hline
\end{tabular}}
\end{center}
where we defined
\begin{equation}
k = p+q \, ,\quad m=n-p-1 \, ,
\end{equation}
and we assumed for simplicity that $n>p>q$. In fact, one can check that the
result is completely symmetric under permutations of $(n,p,q)$. The
hat indicates intersection with the residual discriminant and
corresponds to an extra fundamental hypermultiplet.
Note that there are a total of  $(n+p+q)$ nodes in the quiver.

\begin{figure}[t]
\begin{center}
\includegraphics[scale=0.48]{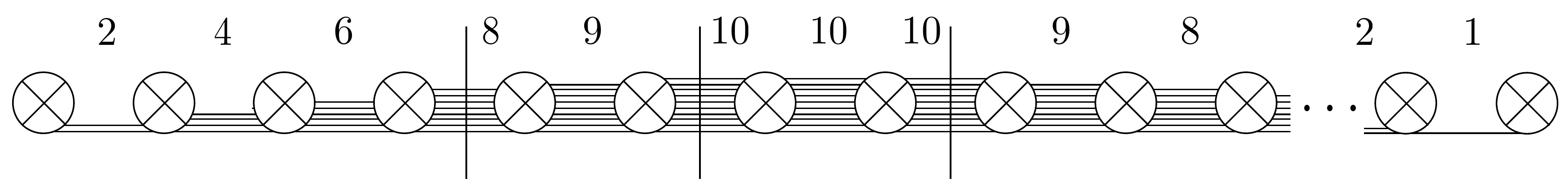}
\caption{The NS5-D6-D8 configuration corresponding to the $[\kod{I}_{8-6-4}]$
  model. We now have an additional D8 on the right hand side, which
  causes additional jumps in the rank of gauge groups. }
\label{Fig:i864}
\end{center}\end{figure}

This type of quiver also appears in IIA brane models with intersecting NS5,
D6 and D8 branes, as we discussed above, and this is again useful to
understand the jumps in the rank of the gauge groups from the presence
of D8 branes. We show in figure \ref{Fig:i864} the brane model giving
the non-perturbative algebra of the model with $n=8, \, p=6\, , q=4$.

We refer to \cite{DelZotto:2014hpa}, section 5.1, for a more detailed
discussion on the relation between the IIA models and the F-theory
geometry. In particular, it is interesting to note that in a IIA model
with multiple D8 branes along a ``staircase'', one can bring all the
D8 branes on one side by Hanany-Witten moves, and there are different
number of D6 branes ending on them. After backreaction this gives
``fuzzy funnels'' (related to the shells of polarized D8 branes in the
solutions of \cite{Gaiotto:2014lca}) which were related in
\cite{DelZotto:2014hpa} to T-branes in the IIB frame. It would be
interesting to understand more directly the relation of this T-brane
data with the $\beta$-monodromies present in the heterotic context.

\section{Non-geometric models and duality web}\label{sec:nongeometricmodels}

We have seen that the explicit formulation of heterotic/F-theory
duality in terms of the map between genus-two and K3 fibrations
reproduces the expectation from the moduli monodromies in a number of
situations where the heterotic side had a clear geometric
interpretation, at least in some duality frame.

In this section we investigate heterotic models with monodromies which
are non-geometric in all T-duality frames. This is the most interesting
situation, since a priori it is not clear if such degenerations are
allowed, and even basic quantities such as the charge of the
corresponding ``exotic'' branes are not obviously available since we
cannot go to a geometric frame, measure the charge,  and dualize back.

One class of non-geometric models is obtained by combining Dehn twists
of the two genus-one components of the genus-two fiber in order to
have monodromies for $\tau$ and $\rho$ that remain non-geometric even
after the exchange of $\tau$ and $\rho$. 
A simple example in the absence of Wilson lines
is a double elliptic T-fold with monodromy $\tau\to-1/\tau \, , \rho
\to- 1/\rho$. We will find that all these
models admit a dual smooth resolution, and moreover the resulting low
energy physics is the same as the one describing the geometric
models studied in the previous section. 
 We believe that this result
can be used as a non-trivial test of any direct description of non-geometric
solutions, for example by using a T-duality covariant formalism such
as double field theory
\cite{Hull:2009mi,Aldazabal:2013sca,Hohm:2013bwa}. We will shortly
analyze in details the $\kod{[III-III]}$ model which is
an interesting example of this class.  

In sections \ref{sec:othermodels} we will 
consider models whose monodromies involve Dehn
twists along the cycle $\gamma$ that links the cycles of the homology
basis (see figure \ref{Fig:genus2}), thus including a non-geometric
mixing of the $\tau$ and $\rho$ moduli. As we will explain in detail, only few 
of these models admit a dual Calabi-Yau
resolution, and for them we again derive the low energy description
from the F-theory side.

\subsection{Double elliptic T-fold: $\kod{[III-III]}$ model}
\label{ss:33}

As an example of a non-geometric degeneration we take the Namikawa-Ueno
$\kod{[III-III]}$ singularity. This model has monodromy:
\begin{equation}
\label{ex33}
\tau \rightarrow \frac{\rho}{\beta^2-\rho\tau} \, ,\quad \rho
\rightarrow \frac{\tau}{\beta^2-\rho\tau} \, ,\quad \beta
\rightarrow -\frac{\beta}{\beta^2-\rho \tau} \, .
\end{equation} 
We see that when $\beta = 0$ we obtain a ``double elliptic''
fibration on an $S^1$ that encircles the heterotic
degeneration, with monodromy $\tau\to-1/\tau \, , \rho
\to- 1/\rho $. Models with such twists have been discussed in the
past from different points of view (see for example \cite{Flournoy:2005xe,Hassler:2014sba,Condeescu:2013yma,Lust:2015yia}).
 The equation for the hyperelliptic curve for such a singularity is:
\begin{equation}\label{eq:hyperell_III-III}
y^2 = x  (x - 1)  (x^{2} + t)  \left[(x-1)^2 + t\right]\, .
\end{equation}
Like for the resolutions in the previous section, we calculate the
Siegel modular forms for this hyperelliptic fiber by using the map
\eqref{SiegelICmap} and the explicit expressions for the Igusa-Clebsch
invariants, and we plug them into equation \eqref{K378}. We find that $f$ and $g$ for this Weierstra\ss{} equation,  look as follows:
\begin{align}
f & = -12 t^2 u^3 v^4 \big(41472 t^{10} v+186624 t^9 v+334368 t^8
    v+300672 t^7 v+139968 t^6 v \nonumber \\ &\quad +31104 t^5 v +16 t^4 u+2592 t^4 v+36 t^3
    u+57 t^2 u+30 t u+9 u\big) \, ,\\
g & = u^5 v^5 \big(-3981312 t^{15} v^2-8957952 t^{14} v^2+11197440
    t^{13} v^2+57542400 t^{12} v^2 +78941952 t^{11} v^2 \nonumber \\
  &\quad +54914112 t^{10} v^2-1024 t^9 u v+21959424 t^9 v^2-3456 t^8 u
    v +5318784 t^8 v^2-288 t^7 u v \nonumber \\ &\quad +746496 t^7
                                                  v^2+9648 t^6 u
                                                  v+46656 t^6 v^2+8640
                                                  t^5 u v+2160 t^4 u
                                                  v+u^2\big) \, .
\end{align}
We are again interested in the enhancements from the intersection of
the residual discriminant with the $E_7$ curve at $t=0$. The terms
relevant for this analysis are
\begin{equation}
\label{K333}
y^2 = x^3 + \left[t^6u^3 + t^2 u^4\right] x + t^4 u^6+  t^6 u^5 +u^7 \, ,
\end{equation}
with a discriminant
\begin{equation}
\Delta = -u^9 \left(4 t^{18} + 12 t^{14} u + 27 t^{12} u + 66 t^{10} u^2 + 58 t^6 u^3 + 
 27 t^8 u^3 + 54 t^4 u^4 + 27 u^5\right)
\end{equation}
We see that at $u=t=0$  vanishing orders of $f$, $g$ and $\Delta$
increase to $(6,7,14)$. To resolve this singularity we proceed as in
\eqref{blowup}, introducing now six divisors $e_i$.

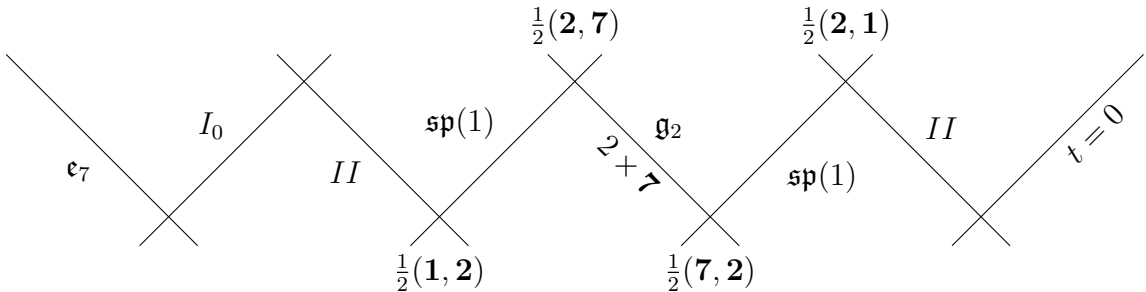
\begin{figure}[h]
\centering
\setlength{\PicScale}{1.2cm}
\begin{tikzpicture}
\draw [name path=line 0] (1.5*0*\PicScale,0) -- node [below left] {$\mathfrak e_7$} ++(-45:3\PicScale);
\draw [name path=line 1] (1.5*1*\PicScale+2.1\PicScale,0) -- node [above left] {$I_0$} ++(-135:3\PicScale);
\draw [name path=line 2] (1.5*2*\PicScale,0) -- node [below left] {$II$} ++(-45:3\PicScale);
\draw [name path=line 3] (1.5*3*\PicScale+2.1\PicScale,0) -- node [above left ] {$\mathfrak{sp}(1)$} ++(-135:3\PicScale);
\draw [name path=line 4] (1.5*4*\PicScale,0) -- node [below,sloped,label=above:{$\mathfrak{g}_2$}] {$2\times \mathbf 7$} ++(-45:3\PicScale);
\draw [name path=line 5] (1.5*5*\PicScale+2.1\PicScale,0) -- node [below right] {$\mathfrak{sp}(1)$} ++(-135:3\PicScale);
\draw [name path=line 6] (1.5*6*\PicScale,0) -- node [above right] {$II$} ++(-45:3\PicScale);
\draw [name path=line 7] (1.5*7*\PicScale+2.1\PicScale,0) -- node [below right, sloped] {$t=0$}++(-135:3\PicScale);
\node [outer sep=.3\PicScale,below,name intersections={of=line 2 and line 3}] at (intersection-1) {$\tfrac12(\mathbf 1,\mathbf 2)$};
\node [outer sep=.3\PicScale,above,name intersections={of=line 3 and line 4}] at (intersection-1) {$\tfrac12(\mathbf 2,\mathbf 7)$};
\node [outer sep=.3\PicScale,below,name intersections={of=line 4 and line 5}] at (intersection-1) {$\tfrac12(\mathbf 7,\mathbf 2)$};
\node [outer sep=.3\PicScale,above,name intersections={of=line 5 and line 6}] at (intersection-1) {$\tfrac12(\mathbf 2,\mathbf 1)$};
\end{tikzpicture}
\setlength{\PicScale}{1cm}
\caption{Pictorial summary of the gauge algebra and matter content
  that arise from the resolution of the dual model of a
  $\kod{[III-III]}$ singularity.}\label{fig:summary_gauge-algebra_matter}
\end{figure}

As a next step, we want to analyze the singularities  that arise at
the new exceptional curves to see what kind of gauge groups and matter
we obtain. From the vanishing orders of $f$, $g$ and $\Delta$ along
this curves we get the chain of $1\,2\,\,2\,2\,2\,2$ curves:
\begin{equation}
\kod{[III^{\ast}]-I_0-II-IV-I_0^{\ast}-IV-II}\,.
\end{equation}
In order to  identify the gauge groups, we look at the conditions in table~\ref{tab:monodromycovers}. The
analysis of the monodromy covers proceeds much as in the previous cases. We see that the $\kod{I_0^*}$ cover does not
factorize, as we expect in the case that the curve is intersected by a
curve with type $\kod{IV}$ singularity \cite{Morrison:2012td}. Hence
we obtain a $\mathfrak g_2$ gauge algebra. For the type $\kod{IV}$ fibers we find
$\mathfrak{sp}(1)$ because both curves have adjacent $\kod{I_0^*}$ and type
$\kod{II}$ singularities \cite{Heckman:2015bfa}. The remaining divisors do not
lead to a contribution to the gauge algebra and we thus find the following
non-perturbative enhancement:
\begin{equation}
\begin{tabular}{|cccccc|}
\hline
 &&  $\mathfrak{sp}(1)$ &$\mathfrak{g}_2$&$\mathfrak{sp}(1)$ &\\
1 & 2&     2 &2&2&2 \\\hline
\end{tabular} \, .
\end{equation}
The matter spectrum for the gauge symmetries can now be deduced from
anomaly cancellation or by studying the monodromy covers in more
detail. From  anomaly cancellation one obtains for $(-2)$-curves that we need four fundamentals for an $\mathfrak{su}(2)$ and four $\mathbf 7$'s for a $\mathfrak g_2$. These states are partitioned into localized and non-local matter. Localized matter arises at the intersections of the curves whereas non-local matter, besides the adjoint, appears in the case of monodromies on the Kodaira fiber.  Therefore, we obtain
\begin{equation}
e_1\cap e_2:\, \tfrac12 (\mathbf 1,  \mathbf 2)\,,\quad e_2\cap e_3:\, \tfrac12 (\mathbf 2,  \mathbf 7)\,,\quad e_3: 2\times \mathbf 7\textmd{ (non-loc.)} \,,\quad e_3\cap e_4:\, \tfrac12 (\mathbf 7,  \mathbf 2)\,,\quad e_4\cap e_5:\, \tfrac12 (\mathbf 2,  \mathbf 1)\,. 
\end{equation}
As a check of the amount of non-local matter, we calculate the genus
of the monodromy cover over $e_3$ (for $e_2$ and $e_4$ checks can be
done in a  similar fashion), which is given by:
\begin{equation}\label{eq:monodromy-cover_expl}
\psi^3 + 2^2 3^3 e_2^2\, e_4^2\,(2^4 3^3\,e_2^2 + e_4^2 - \psi) =0\,,
\end{equation}
where $e_2$ and $e_4$ are the homogeneous coordinates of the rational
line $e_3=0$. From \eqref{eq:monodromy-cover_expl} we see that the
cover is singular at $\psi=e_2=0$ and $\psi=e_4=0$.  Resolving these
two singularities, we find that the genus of the cover is two which
agrees with the two non-local $\mathbf 7$-states that we needed for
anomaly cancellation. We summarise this resolution in figure
\ref{fig:summary_gauge-algebra_matter}.

At this point, we can check that what we obtained is precisely the
same resolution as the one obtained from the NU model
$\kod{[I_0^{\ast}-I_0]}$ in \eqref{eqi0i0star}, giving the theory of
six pointlike instantons on a $D_4$ singularity. At first sight, this seems
very surprising, since we started from two different elements of the
NU list, whose monodromies are not conjugate to each other and there seems to be no
duality that brings a degeneration of type $\kod{[III-III]}$ to a
geometric frame. However, this is in line with the fact that no new
F-theory models are needed to understand the class of non-geometric
heterotic models where
$\tau$ and $\rho$ degenerations do not collide
\cite{McOrist:2010jw}. In the following we will generalize this
observation to obtain a list of all the models described by the same
six-dimensional low energy theory. However, before we detour to describe a global
embedding of the local $\kod{[III-III]}$ model.


\subsection{A global model}


In this section we address the question of global hyperelliptic fibrations. The idea is to first start with the generic situation and continue by tuning the coefficients $c_i$ in  \eqref{eq:hyperelliptic-curve} to obtain different kinds of singularities. In practice we have to choose
the parameters $\gamma_{ij}$ in \eqref{cicoeffs}. We will consider an example that features a 
$\kod{[III-III]}$ singularity, discussed in section \ref{ss:33}, at the origin.

The concrete hyperelliptic curve can be obtained by taking the local equation \eqref{eq:hyperell_III-III} and extending the prefactors of $x^i$ to sections of the anti-canonical bundle of $\mathbb P^1$, cf.\ section \ref{sec:global_models}, which reduce near $t=0$ to the ones from \eqref{eq:hyperell_III-III},
\be
\begin{split}
y^2=
& x^6 \left(\delta_{62} t^2+\delta_{61} t+1\right)+x^5
   \left(\delta_{52} t^2+\delta_{51} t-3\right)+x^4 \left(\delta
  _{4} t^2+2 t+3\right)+\\&+x^3 \left(\delta_{3} t^2-4 t-1\right)
  +\left(t^2+3 t\right) x^2+\left(-t^2-t\right) x \, .
    \end{split}
\label{model33}
\ee
Here $t$ is the affine coordinate on the $\PP^1$ base. When we calculate from  this sextic the vanishing orders of the Siegel modular forms at $t=0$, we find
\begin{equation}
\mu(\psi_4)=2\,,\quad \mu(\psi_6)=3\,,\quad \mu(\chi_{10})=5\,,\quad \mu(\chi_{12})=5\,.
\end{equation}
However, these are the vanishing orders of the $\kod{[II-III]}$ singularity as one can see from Table~\ref{tab:vanishing-orders_ell-type-2}.\footnote{Note, the vanishing orders of $a$, $b$, $c$, $d$ almost uniquely characterize the singularities of the hyperelliptic curve, at least for the ones for which we have an F-theory resolution, cf.\ Section \ref{ss:crit}.}
Therefore, we have to look at the coefficients of the $t^3$, $t^5$, $t^5$ terms in $\psi_6$, $\chi_{10}$, $\chi_{12}$, respectively. All of them are proportional to $\delta_{61}+\delta_{51}$. Hence, we  set $\delta_{51}\equiv-\delta_{61}$ in  \eqref{model33} to obtain indeed a $\kod{[III-III]}$ at $t=0$.

The discriminant of this sextic is found to be
\be
I_{10} =\frac1{2\cdot 3^5}\,\chi_{10}= t^6 (t + 1)^2 P_{12}(t) \, ,
\label{discri33}
\ee
where $P_{12}(t)$ is a polynomial of degree 12 with simple roots, say $r_\ell$, $\ell=1,\ldots, 12$. 
Thus, the fiber degenerates at $t=0$, $t=-1$, and the twelve roots $r_\ell$. There is no singularity at $\infty$. 
To analyze the type of singularities---besides the one at $t=0$ which we know already---we compute the vanishing orders of the Siegel modular forms at the remaining singularities:
\begin{equation}
\begin{split}
t=-1:&\quad \mu(\psi_4)=0\,,\quad \mu(\psi_6)=0\,,\quad \mu(\chi_{10})=2\,,\quad \mu(\chi_{12})=2\,,\\
t=r_\ell:&\quad \mu(\psi_4)=0\,,\quad \mu(\psi_6)=0\,,\quad \mu(\chi_{10})=1\,,\quad \mu(\chi_{12})=1\,.
\end{split}
\end{equation}
From the tables of section \ref{ss:catalog}, we find that the singularity at $t=-1$ is of type $\kod{[I_{2-0-0}]}$ and the singularities  at $t=r_\ell$ are of type $\kod{[I_{1-0-0}]}$.

Let us now examine the global model from the F-theory perspective.
To analyze the singularities on the F-theory side we first determine the discriminant of the elliptic fibration \eqref{K378}
\begin{equation}
\Delta = u^9 \left[4  (a u + c )^3 + 27 u(u^2 + b u  + d)^2\right] =  
u^9  P_{5,60}(u,t) \, ,
\label{discriK378}
\end{equation}
where $P_{5,60}(u,t)$ is the polynomial of degree 5 and 60 in the  affine coordinate $u$ and $t$ of $\mathbb F_{12}$, respectively, read off from the equality. The discriminant clearly exhibits
the $E_7$ and $E_8$ singular fibers at $u=0$ and $u=\infty$, respectively, i.e.\ along the two sections of the Hirzebruch surface. But now we are more interested in locating
additional singular loci. 
Looking at  $P_{5,60}(u,t)$  we find that it does not factorize any further, i.e.\ $P_{5,60}(u,t)=0$ defines an $\kod{I_1}$ locus. This $\kod{I_1}$ curve  intersects the section $u=0$. At the intersection points of $\kod{I_1}$ and $\kod{III^*}$ where also $c$ and $d$ vanish, we obtain singularities of non-minimal type (or non-Kodaira type). Note that these are exactly the loci where also the genus-two curve degenerates. The resolution of these singularities (on the F-theory side) were already analyzed in Section \ref{sec:I_n-p-0} and \ref{ss:33}. 
Besides these points there are no other co-dimension two singularities which render the Calabi-Yau threefold
singular, although there might be other points where the K3, or the elliptic fiber, degenerates.
In particular, we find that the points associated with the enhancement to $SU(2)$ of the heterotic at self-dual points, 
giving $SU(2)$ singularities on the K3 fiber, do not lead to singularities in the total space of the K3 fibration.

\subsection{Dualities}
\label{ss:dual}

We have seen that the resolution of the dual $\kod{[III-III]}$ model gives
the same six-dimensional theory as $\kod{[I_0-I_0^{\ast}]}$, namely
the theory of six pointlike instantons on a $D_4$ singularity. In fact,
this is not an isolated coincidence, as we argue below. 

We first note that the above mentioned duality might be understood
from the monodromy factorization of the two models as we now explain. From
table \ref{tab:Kodaira}, and our discussion in section
\ref{subsec:degenerations}, it follows that the monodromy of the
$\kod{[III-III]}$ model can be written in terms of products of
Dehn twists as $B_1A_1B_1B_2A_2B_2$. Recall that $A_i$ and $B_i$ are
respectively twists
around the $a_i$ and $b_i$ cycles shown in figure \ref{Fig:genus2}. We
can get to the monodromy of the  $\kod{[I_0-I_0^{\ast}]}$ model by
applying the following moves: 
\begin{equation}\begin{split}
\kod{[III-III]}  &= B_1A_1B_1B_2A_2B_2 \\
    &= B_1A_1B_1B_1A_1B_1 \qquad (\rho \rightarrow
                   \tau)  \\
   & =B_1A_1B_1A_1B_1A_1 \qquad (\text{braid})  \\
& = (B_1A_1)^3 = \kod{[I_0-I_0^{\ast}]} \, . 
\end{split}\end{equation}
The last move
follows from braid relations that define the generators of the
mapping class group (see for example \cite{Lust:2015yia}), and it is
the analogous of a collision of two Kodaira fibers of type $\kod{III}$ in
F-theory. The first move replaces locally the $\rho$ fibration with a
fibration in $\tau$. We can check that this move is allowed in the
case of the elliptic models from the direct inspection of the duality
map for the $E_8\times E_8$ case, which is considerably simpler
and it is shown in appendix \ref{app:mapE8E8}. 

 This simple argument also
predicts that the $\kod{[I_0-I_0^{\ast}]} $ model is
equivalent to the $\kod{[IV-II]}$ model, described by the fibration:
\begin{equation}
y^2 = (t+x^3) \left[t^2+(x-1)^3\right] \, ,
\end{equation}
and corresponding to a monodromy $B_1A_1B_1A_1B_2A_2$.
By constructing the dual F-theory model and resolving it, we indeed
find the same six-dimensional theory. 

\begin{table}[t]\begin{center}
\renewcommand{\arraystretch}{1.5}
\begin{tabular}{|c|c|}
\hline
$\mu(I_{10})$ & dual models   \\
\hline \hline
2& $\kod{[I_0-II]}_{0112}$  \\ \hline 
3& $\kod{[I_0-III]}_{0113}$ \\ \hline 
4& $\kod{[I_0-IV]}_{0224}
  \, ,  \,\kod{[II-II]}_{0224}$  \\ \hline 
5&$[\kod{IV}-\kod{I}_1]_{0325} \, , \,\kod{[II-III]}_{0225}$ \\ \hline 
6& $\kod{[I_0-I_0^{\ast}]}_{0226}\, , \,
  \kod{[III-III]}_{0226}\, , \,\kod{[IV-II]}_{0336}$ \\ \hline 
 7& $[\kod{I}_0^{\ast}-\kod{I}_1]_{0227} \, , \,\kod{[IV-III]}_{0337}$ \\ \hline 
8& $\kod{[I_0-IV^{\ast}]}_{0448} \, ,
  \kod{[IV-IV]}_{0448} \, , \,\kod{[I_0^{\ast}-II]}_{0338}$ \\ \hline 
9& $\kod{[I_0-III^{\ast}]}_{0339} \,
  ,  \,\kod{[I_0^{\ast}-III]}_{0339}$ \\ \hline 
10& $\kod{[I_0-II^{\ast}]}_{0\,5\,5\,10} \, ,\,
  \kod{[IV^{\ast}-II]}_{0\,5\,5\,10} \,
  ,\,\kod{[I_0^{\ast}-IV]}_{0\,4\,4\,10}$ \\ \hline 
11& $\kod{[II-III^{\ast}]}_{0\,4\,4\,11} \, ,\,
  \kod{[IV^{\ast}-III]}_{0\,5\,5\,11}$ 
\\ 
\hline 
\end{tabular}
\caption{Dual models: the NU degenerations in the same row give rise
  to the same SCFTs after resolution of the dual F-theory model. We
  indicate as a subscript the vanishing orders of the Igusa-Clebsch
  invariants $I_2\, , I_4 \, , I_6 \, , I_{10}$. }
  \label{tab:dual}\end{center}\end{table}

As a rule, we can find dual models if the sum of the orders of the
discriminant for their two Kodaira components, or equivalently the
order of the Igusa-Clebsch invariant $I_{10}$, is the same. In table \ref{tab:dual} we display
all the models of this type that have the same order. We
indicate as a subscript the order of vanishing of all the
Igusa-Clebsch invariants, listed in appendix \ref{app:NUlist}.
In section \ref{sec:classification} we show that models with higher $\mu(I_{10})$ do not admit dual
smooth Calabi-Yau resolutions. For all the models in table \ref{tab:dual} we
explicitly performed the F-theory resolution  and verified
that for all the degenerations in a row the same theory 
arises. 

We thus see that almost all non-geometric models of type 2 in the NU
list are described by the theory of pointlike instantons on ADE
singularities. 
 It would be interesting to understand better this fact  directly from
the heterotic side, beyond the simple argument given above.
The precise set of dualities that we are finding should also
be an interesting test of T-duality covariant formalisms, such as double
field theory \cite{Hull:2009mi,Aldazabal:2013sca,Hohm:2013bwa}, in
which one might hope to describe 
non-geometric backgrounds. Presumably, these dualities can be clarified
by understanding the non-geometric analog of the Bianchi identity
\ref{hetbianchi}.

We also stress that our findings imply the existence of local
degenerations with monodromies which are non-geometric in all
T-duality frames, thus enlarging the examples of ``exotic'' branes
recently studied for example in \cite{deBoer:2012ma,Hassler:2013wsa},
and provide a dual description of the T-fects constructed in \cite{Lust:2015yia}.

\section{Other models}\label{sec:othermodels}

In the previous sections we explored parabolic models in the NU
classifications that had a clear geometric interpretation, and 
elliptic models of type $\kod{[K_1-K_2]}$, both geometric and
non-geometric, whose dual resolutions can be understood in terms of
the theory of pointlike instantons on ADE singularities.
In this section we consider examples from the remaining NU models, in
particular we explore
models
whose monodromies involve a twist along the cycle $\gamma$ in figure
\ref{Fig:genus2}. As we will show in the next
section, the duals of many models of this kind cannot be resolved,
so we restrict ourselves to the examples that admit a smooth dual  Calabi-Yau
model.

\subsection{Non-geometric degenerations with moduli mixing}
\label{ss:nongeo}

A list of genus-two degenerations with monodromy that mixes the moduli
is provided by the elliptic type 1 models in the NU classification (see
appendix \ref{app:NUlist}). Despite the fact that the corresponding
heterotic models lack a geometric interpretation, the dual F-theory
resolutions are similar to the ones encountered in the previous
sections. For each example we again determine the non-perturbative gauge
algebras. We refer to section \ref{sec:classification} for a detailed
analysis of all the type 1 models, which shows that the models listed
here are the only ones admitting a smooth dual.

\subsubsection*{$\kod{[V]}$ model}

From the NU list we take the local model:
\begin{equation}
y^2 = x^6 + t \, ,
\end{equation}
whose $Sp(4,\mathbb{Z})$ monodromy is
\begin{equation}\label{MV}
M_{\kod{[V]}} = \begin{pmatrix} 0&0&1&0 \\ 0&0&1&1\\-1&1&0&0\\
  0&-1&0&0\end{pmatrix} \, ,
\end{equation}
which acts on the moduli matrix $\eqref{omatrix}$ as
\begin{equation}
\tau \rightarrow  \frac{\rho}{\beta^2-\rho\tau} \, ,\qquad \rho
\rightarrow \frac{\rho+\tau-2\beta}{\beta^2-\rho\tau} \, ,\qquad
\beta\rightarrow  \frac{\rho-\beta}{\beta^2-\rho\tau} \, .
\end{equation}
This is an elliptic monodromy of order six, and up to global
conjugation 
it can be decomposed into the following product
of the $Sp(4,\mathbb{Z})$ generators given in \eqref{sp4generators}:
\begin{equation}
M_{\kod{[V]}} =A_1B_1\Gamma B_2 A_2 \, .
\end{equation}
By computing the Igusa-Clebsch
invariants we find the following dual K3 fibration:
\begin{equation}
y^2= x^3 + 972t^2u^3v^4(-5u+2^73^6  t^3v ) x +
u^5v^5(u^2-2^4 3^6 5 t^3 u v + 2^{11}3^{12}5 t^6 v^2) \, ,
\end{equation}
with discriminant: 
\begin{equation}
\Delta = -27 u^9 v^{10} \left(2^53^6 t^3 v+u\right)^3 \left(2^73^6 t^3
  v-u\right)^2 \, .
\end{equation}
By resolving the intersection $u=t=0$ we get a chain:
\begin{equation}
\mathrm{[III^{\ast}]-I_0 - III - I_0^{\ast} - IV - II } \, .
\end{equation}
The vanishing orders of $f$ and $g$ at each divisor indicate the following gauge algebras:
\begin{equation}
\begin{tabular}{|ccccc|}
\hline
 &  $\mathfrak{su}(2)$ & $\mathfrak{g}_2$&$\mathfrak{sp}(1)$& \\
1 & 2& 2&2 &2\\\hline
\end{tabular} \, .
\end{equation}
The matter content is $\frac12(\mathbf{2},\mathbf{1},\mathbf{1}) \oplus
\frac12(\mathbf{2},\mathbf{7},\mathbf{1}) \oplus 2 (\mathbf{1},\mathbf{7},\mathbf{1})
\oplus \frac12(\mathbf{1},\mathbf{7},\mathbf{2}) \oplus \frac12(\mathbf{1},\mathbf{1},\mathbf{2})$,
in agreement with anomaly cancellation \cite{Intriligator:1997dh, Heckman:2015bfa}.

We find the same result for another model that admits a dual smooth
fibration, the $\kod{[VII]}$ model described by $y^2 = x(x^4+t)$ and
monodromy
\begin{equation}
M_{\kod{[VII]}} =\begin{pmatrix} 0 & 1 & 1 & 0 \\
 1 & -1 & 0 & 1 \\
 -1 & 1 & 1 & 0 \\
 0 & -1 & 0 & 0 \end{pmatrix} \, .
\end{equation}

\subsubsection*{$\kod{[VIII-1]}$ model}

This example is defined by:
\begin{equation}
y^2 = x^5 + t \, ,
\end{equation}
with monodromy matrix
\begin{equation}\label{MV}
M_{\kod{[VIII-I]}} = \begin{pmatrix} 0&1&1&0 \\ 1&0&0&1\\-1&1&1&0\\
  0&-1&0&0\end{pmatrix} \, .
\end{equation}
The action on the moduli is found to be
\begin{equation}
\tau \rightarrow  \frac{\rho}{\beta^2+\rho-\rho\tau} \, ,\qquad \rho
\rightarrow \frac{\tau(\rho+1)-(\beta+1)^2}{\beta^2+\rho-\rho\tau} \, ,\qquad
\beta\rightarrow  \frac{\rho\tau-\beta(\beta+1)}{\beta^2+\rho-\rho\tau} \, .
\end{equation}
In this model only the Igusa-Clebsch invariant $I_{10}$ does not vanish identically. 
The dual F-theory elliptic fibration is then:
\begin{equation}
y^2 =x^3 -2^33^55^5 t^4u^3v^5x + u^7 v^5 \, ,\quad \Delta = -27 u^9
v^{10} \left(u^5-2^{11} 3^{12} 5^{15}t^{12} v^5\right) \, ,
\end{equation}
leading to a simple chain 
\begin{equation}
\kod{[III^{\ast}]-I_0 - III - IV - II}
\end{equation}
with gauge algebras:
\begin{equation}
\begin{tabular}{|cccc|}
\hline
 &  $\mathfrak{su}(2)$ &$\mathfrak{sp}(1)$& \\
1 & 2& 2&2 \\\hline
\end{tabular} \, .
\end{equation}
The matter comprises a bifundamental plus one additional $\mathbf{2}$ for each factor.

\subsubsection*{$\kod{[IX-1]}$ model}

As a last example we consider the NU model:
\begin{equation}
y^2 = x^5 + t^2 \, ,
\end{equation}
with monodromy matrix
\begin{equation}\label{MV}
M_{\kod{[IX-1]}} = \begin{pmatrix} 0&1&1&1 \\ 0&0&1&0\\0&0&0&1\\
  -1&0&0&-1\end{pmatrix} \, ,
\end{equation}
acting on the moduli as:
\begin{equation}
\tau \rightarrow  1+\rho-\frac{(1+\beta)^2}{\tau} \, ,\qquad \rho
\rightarrow -\frac{1}{\tau}\, ,\qquad
\beta\rightarrow  -\frac{\beta+1}{\tau}\, .
\end{equation}
In this case it also happens that $a=b=d=0$ and the dual elliptic fibration is
\begin{equation}
y^2 =x^3 -2^33^55^5 t^8u^3v^5x + u^7 v^5 \, ,\quad \Delta = -27 u^9
v^{10} \left(u^5-2^{11} 3^{12} 5^{15}t^{24} v^5\right) \, .
\end{equation}
Resolving $u=t=0$ now requires a total of 16 blowups and gives the chain:
\begin{equation}
\kod{[III^{\ast}]-I_0 - III - I_0^{\ast} -III-I_0- III^{\ast}-I_0-II-IV-I_0^{\ast}-II-IV^{\ast} -II-I_0^{\ast} - IV -II }\, . 
\end{equation}
The study of monodromy covers, cf.~table~\ref{tab:monodromycovers} leads to the following
non-perturbative gauge algebras:
\begin{equation}
\begin{tabular}{|cccccccccccccccc|}
\hline
 &  $\mathfrak{su}(2)$ &$\mathfrak{so}(7)$& $\mathfrak{su}(2)$&
  &$\mathfrak{e}_7$&&&$\mathfrak{sp}(1)$& $\mathfrak{g}_2$& &$\mathfrak{f}_4$ && $\mathfrak{g}_2$ & $\mathfrak{sp}(1)$&\\
1 & 2&     3 &2&1&8&1&2&2&3&1&5&1&3&2&2 \\\hline
\end{tabular} \, .
\end{equation}
The only matter representations are 
$\frac12(\mathbf{2},\mathbf{8},\mathbf{1}) \oplus
\frac12(\mathbf{1},\mathbf{8},\mathbf{2})$, for $\mathfrak{su}(2) \oplus \mathfrak{so}(7) \oplus \mathfrak{su}(2)$.

As we already mentioned, in the type 1 NU elliptic models
there are no other heterotic degenerations that admit a
dual Calabi-Yau resolution.

\subsection{Parabolic models of type 3}
\label{ss:para3}

In the parabolic type 3 class of the NU list we find additional models that admit
smooth F-theory duals. Below we present the resolution of several  examples. 

\subsubsection*{$[\mathrm{II}_{n-0}]$ model}\label{sec:II_n-0}

The model $[\mathrm{II}_{n-0}]$ is described by the local equation:
\begin{equation}
y^2 = (x^4+\alpha t x^2 + t^2)\left[(x-1)^2+t^{n-1}\right] \, ,
\end{equation}
and has monodromy
\begin{equation}
\tau \rightarrow \tau \, ,\quad \rho \rightarrow \rho +\tau -2\beta + n \, .
\quad \beta \rightarrow \tau -\beta \, .
\end{equation}
The intersection of $\{t=0\}$ with the $E_7$ curve in the dual F-theory
model is described by
\begin{align}
y^2 &= x^3 + \left[t^{5+n}u^3+t^2u^4\right] x + t^{6+n}u^5 +t^3 u^6+
      u^7 \, , \\
\Delta &=-u^9 \left(54 u^3 t^{n+6}+66 u^2 t^{n+9}+39 u t^{2 n+12}+4
         t^{3 n+15}+31 t^6 u^3+54 t^3 u^4+27 u^5\right) \, . \nonumber
\end{align}
The resolution produces a plateau of $\kod{I_0^{\ast}}$
Kodaira fibers, which after further resolution gives the chain:
\begin{equation}
\begin{tabular}{|cccccccccccc|}
\hline
 &   $\mathfrak{su}(2)$ & $\mathfrak{so}(7)$& &
                                                $\mathfrak{so}(8)_1$&&&$\mathfrak{so}(8)_{n-1}$&&$\mathfrak{g}_2$&
 $\mathfrak{sp}(1)$&\\
1 &   2 &3  & 1&4&1&$\cdots$&4&1&3& 2 &2 \\\hline
\end{tabular}
\end{equation}
Chains of type $1\,4\cdots 1\, 4$, with $\mathfrak{so}(8)$ singularities, are described in detail in
\cite{Morrison:2012td}. The resolution is similar to that in the
$[\kod I_0^{\ast}-\kod I_n]$ model (see appendix \ref{app:ADE}), but in this case the
plateau $14\cdots 41$ is connected with the $E_7$ in a different way. There is a chain $123$ instead of $1223$, and
there is a $\mathfrak{su}(2) \times \mathfrak{so}(7)$ with matter 
$\frac12({\mathbf 2}, {\mathbf 8}) \oplus ({\mathbf 1}, {\mathbf 8}) $,
as expected from anomaly cancellation.
It would be interesting to understand better the heterotic interpretation of this model.

\subsubsection*{$[\mathrm{IV}-\mathrm{I}_n^*]$ model}\label{sec:IV-I_n*}

The defining local equation is given by:
\begin{equation}
y^2 = (t+x)(x^2 + t^{n+2})\left[(x-1)^3+t^2\right] \, .
\end{equation}
The monodromy action turns out to be:
\begin{equation}
\tau \rightarrow -\frac{1}{1+\tau} \, ,\quad \rho \rightarrow \rho + n- \frac{\beta^2}{1+\tau}\, ,
\quad \beta \rightarrow  \frac{\beta}{1+\tau}\, .
\end{equation}
Here, and in the remaining examples of this section, we will skip presenting the data of
the dual K3 on the F-theory side. To resolve we proceed as explained before.  
We obtain a resolution precisely equal to that of $[\mathrm{II^*}-\mathrm{I}_n]$,
corresponding to $(10+n)$ pointlike
instantons on an $E_8$ singularity, discussed in sections \ref{ss:I0E8} and \ref{sss:InE8}.
We find that other examples of type $[\mathrm{K}-\mathrm{I}_n^*]$ do match models, analyzed in appendix \ref{app:ADE},
associated to pointlike instantons on $E_7$, $E_6$ and $D_4$ singularities. Indeed, 
the resolutions of $[\mathrm{III}-\mathrm{I}_n^*]$ and $[\mathrm{III^*}-\mathrm{I}_n]$, 
 $[\mathrm{II}-\mathrm{I}_n^*]$ and $[\mathrm{IV^*}-\mathrm{I}_n]$,  as well as
 $[\mathrm{I}_0-\mathrm{I}_n^*]$ and $[\mathrm{I}_n-\mathrm{I}_0^*]$,  do coincide.

\subsubsection*{$[\mathrm{IV^*}-\mathrm{II}_n]$ model}\label{sec:IV*-II_n}

According to the NU list the local singularity is described by:
\begin{equation}
y^2 = x (x^3 + t^2)\left[(x-1)^2+t^{n-1}\right] \, ,
\end{equation}
for $n \ge 1$. The monodromy action translates into:
\begin{equation}
\tau \rightarrow -\frac{1+\tau}{\tau} \, ,\quad \rho \rightarrow \rho + n- \frac{\beta^2}{\tau}\, ,
\quad \beta \rightarrow  -1+\frac{\beta}{\tau}\, .
\end{equation}
The resolution has the structure:
\begin{equation}
\label{resoIInIVstar}
\begin{tabular}{|ccccc|}
\hline
 & $\mathfrak{su}(2)$ &$\mathfrak{so}(7)$& $\mathfrak{su}(2)$& \\
1 & 2& 3 & 2 & 1\\
\hline
\end{tabular}\, 
\begin{tabular}{|cccc|}
\hline
  $\mathfrak{e}_6$&  &$\mathfrak{su}(3)$& \\
6&1 &3&1\\\hline
\end{tabular}^{\, \otimes n} \,
\begin{tabular}{|ccccc|}
\hline
  $\mathfrak{f}_4$&&$\mathfrak{g}_2$&$\mathfrak{sp}(1)$&\\
5 &1&3&2&2\\\hline
\end{tabular} \, .
\end{equation}
The second and third block in the above pattern appear in the resolution of the $[\mathrm{I}_{n+1}-\mathrm{IV}^*]$ model, cf.
\eqref{resoInIVstar}, associated to $k=n+9$ pointlike instantons on a $E_6$ singularity.  
It can also be checked that the resolution of $[\mathrm{II}-\mathrm{II}^*_n]$ gives the same result \eqref{resoIInIVstar}.

\subsubsection*{$[\mathrm{III}-\mathrm{II}_n^*]$ model}\label{sec:III-II_n*}

{} From the NU list we read the singularity type:
\begin{equation}
y^2 = (x^4 + t) (x^2+t^{n+1})\, .
\end{equation}
The characteristic monodromy is given by:
\begin{equation}
\tau \rightarrow -\frac{1}{\tau} \, ,\quad \rho \rightarrow \rho + n- \frac{\beta^2}{\tau}\, ,
\quad \beta \rightarrow  1+\frac{\beta}{\tau}\, .
\end{equation}
Resolving yields:
\begin{align}
\label{resoIInIIIstar}
&
\begin{tabular}{|ccccc|}
\hline
 & $\mathfrak{su}(2)$ &$\mathfrak{so}(7)$& $\mathfrak{su}(2)$& \\
1 & 2& 3 & 2 & 1\\
\hline
\end{tabular}\, 
\begin{tabular}{|cccccc|}
\hline
$\mathfrak{e}_7$& &$\mathfrak{su}(2)$&$\mathfrak{so}_7$ 
&$\mathfrak{su}(2)$& \\
7& 1& 2& 3& 2& 1
\\\hline
\end{tabular}\, \times \\ \nonumber
& \times
\begin{tabular}{|cccccc|}
\hline
$\mathfrak{e}_7$& &$\mathfrak{su}(2)$&$\mathfrak{so}_7$ 
&$\mathfrak{su}(2)$& \\
8& 1& 2& 3& 2& 1
\\\hline
\end{tabular}^{\, \oplus (n-1)}
\hspace*{-5mm} \times \hspace*{2mm}
\begin{tabular}{|cccccccccc|}
\hline
$\mathfrak{e}_7$& &$\mathfrak{su}(2)$  
  &$\mathfrak{g}_2$&  &  $\mathfrak{f}_4$ &&$\mathfrak{g}_2$&$\mathfrak{sp}(1)$&\\
8& 1& 2& 3& 1 & 5&1&3&2&2
\\\hline
\end{tabular} \, .
\end{align}
Except for the first block, this result resembles the resolution of the $[\mathrm{I}_{n+1}-\mathrm{III}^*]$ model, cf.
\eqref{resoInIIIstar}, corresponding to $k=n+10$ pointlike instantons on an $E_7$ singularity.  
Matter includes representations $\frac12(\mathbf{2},\mathbf{8},\mathbf{1}) \oplus
\frac12(\mathbf{1},\mathbf{8},\mathbf{2})$ for each $\mathfrak{su}(2) \oplus \mathfrak{so}(7) \oplus \mathfrak{su}(2)$
block. In addition, we have verified that there is an extra half-fundamental for the $\mathfrak{e}_7$ with self-intersection $-7$, as required by anomaly cancellation. The same resolution \eqref{resoIInIIIstar} is obtained for
$[\mathrm{III^*}-\mathrm{II}_n]$. 

\subsubsection*{$[\mathrm{III}-\mathrm{II}_n]$ and $[\mathrm{IV}-\mathrm{II}_n]$ models}\label{extra}

For these models we will only give the resolution for completeness. For $[\mathrm{III}-\mathrm{II}_n]$
we find
\begin{equation}
\begin{tabular}{|cccccc|}
\hline
  &  $\mathfrak{su}(2)_1$  &  $\mathfrak{su}(2)_2$&&  $\mathfrak{su}(2)_{n+1}$ &  \\
1 &      2&2&$\cdots$&2&2  \\\hline
\end{tabular} \, .
\end{equation}
Notice the similarity to the resolution of $[\mathrm{III}-\mathrm{I}_{n+1}]$ displayed in \eqref{a26}.
For $[\mathrm{IV}-\mathrm{II}_n]$ we obtain
\begin{equation}
\begin{tabular}{|ccccccc|}
\hline
  &  $\mathfrak{su}(2)$  &  $\mathfrak{su}(3)_1$&&  $\mathfrak{su}(3)_n$ & $\mathfrak{sp}(1)$ &  \\
1 &      2&2&$\cdots$&2&2&2  \\\hline
\end{tabular} \, .
\end{equation}
This result is analogous to the resolution of $[\mathrm{IV}-\mathrm{I}_{n+1}]$ shown in \eqref{a25}.

\subsection{Parabolic models of type 4}
\label{ss:para4}

In this class we find 3 models that admit a resolution, the
$[\mathrm{I}_{n-p-0}]$, already discussed in section \ref{sec:I_n-p-0}, as well as  
$[\mathrm{II}_{n-p}]$ and $[\mathrm{I}_n-\mathrm{I}_p^*]$, which are addressed below.

\subsubsection*{$[\mathrm{II}_{n-p}]$ model}\label{sec:II_n-p}

In the NU list we find two models that generalize $[\mathrm{II}_{n-0}]$, namely
the $[\mathrm{II}_{n-p}]$ degenerations. Here we consider the one classified
as type 4, with the following sextic:
\begin{equation}
y^2 = (x^2+t)(x^2 + t^{p+1}) \left[(x-1)^2+t^{n-1}\right] \, ,
\end{equation}
and monodromy:
\begin{equation}
\tau \rightarrow \tau + p \, ,\quad \rho \rightarrow \rho +n \, ,
\quad \beta \rightarrow -\beta-1 \, .
\end{equation}
The intersection $u=t=0$ in the dual F-theory model is given by the Weierstra\ss{} model:
\begin{align}
y^2 = x^3 & + \left[12(10368 t^{5+n+p}+ \cdots) u^3-12 (t^2+\cdots)u^4\right] x  \\
& + 497664(t^{6+n+p}+ \cdots)u^5 + 16 (t^3 + \cdots) u^6 +  u^7 \, . \nonumber
\end{align}
Here we have written numerical factors in the leading terms in order to stress that in this case there will be
non-generic cancellations in the discriminant $\Delta$. Computing $\Delta$ explicitly we can extract the data needed to 
perform the resolution. Proceeding as explained before, we find a chain $1231(414\cdots14)1322$ supporting 
singularities  
$\mathrm{I}_0-\mathrm{III}-\mathrm{I}^*_0-\mathrm{I}_1-(\mathrm{I}^*_1-\cdots -\mathrm{I}^*_1)-\mathrm{I}_1-
\mathrm{I}^*_0-\mathrm{IV}-\mathrm{II}$. In the central block the $-1$ curves support $\mathfrak{sp}$ algebras 
whereas the $-4$ curves support $\mathfrak{so}$ ones.
For example, for $n=p$, the full resolution takes the form:
\begin{align}
\label{resoIInp1}
&
\begin{tabular}{|cccccccccccc|}
\hline
&  $\mathfrak{su}(2)$ &$\mathfrak{so}(7)$&& $\mathfrak{so}(9)$& $\mathfrak{sp}(1)$&$\mathfrak{so}(11)$&$\mathfrak{sp}(2)$&$\cdots$&$\mathfrak{so}(2p+5)$ 
&$\mathfrak{sp}(p-1)$&$\mathfrak{so}(2p+7)$\\
1 &      2 &3&1&4 &1& 4& 1& $\cdots$&4& 1& 4 \\\hline
\end{tabular}
\times\nonumber\\
& \times 
\begin{tabular}{|ccccccccc|}
\hline
$\mathfrak{sp}(p-1)$ &$\mathfrak{so}(2p+5)$  
  &$\cdots$&  $\mathfrak{sp}(1)$&  $\mathfrak{so}(9)$ &&$\mathfrak{g}_2$&$\mathfrak{sp}(1)$&\\
 1& 4& $\cdots$& 1 & 4&1&3&2&2
\\\hline
\end{tabular} \, .
\end{align}
The matter for $\mathfrak{sp}-\mathfrak{so}$, and viceversa, is $\frac12(\bf{fund}, \bf{fund})$. 
For $\mathfrak{so}(2p+7)$ there is an additional fundamental.
In this way all gauge anomalies are canceled. The pattern is analogous to that
obtained for $k=2p+6$ instantons on a $D_{p+4}$ singularity
\cite{Intriligator:1997dh}.
When $n>p$ the resolution is instead:
\begin{align}
\label{resoIInp2}
&
\begin{tabular}{|ccccccccc|}
\hline
&  $\mathfrak{su}(2)$ &$\mathfrak{so}(7)$&& $\mathfrak{so}(9)$&$\cdots$&$\mathfrak{sp}(p-1)$&$\mathfrak{so}(2p+7)$& $\mathfrak{sp}(p)$\\
1 &      2 &3&1&4 &   $\cdots$& 1& 4& 1 \\\hline
\end{tabular}\,
\begin{tabular}{|cc|}
\hline
$\mathfrak{so}(2p+8)$& $\mathfrak{sp}(p)$\\
4& 1
\\\hline
\end{tabular}^{\, \oplus (n-p-1)} \hspace*{-5mm}\times \nonumber\\
& \times 
\begin{tabular}{|cccccccccc|}
\hline
$\mathfrak{so}(2p+7)$&$\mathfrak{sp}(p-1)$ &$\mathfrak{so}(2p+5)$  
  &$\cdots$&  $\mathfrak{sp}(1)$&  $\mathfrak{so}(9)$ &&$\mathfrak{g}_2$&$\mathfrak{sp}(1)$&\\
4& 1& 4& $\cdots$& 1 & 4&1&3&2&2
\\\hline
\end{tabular} \, .
\end{align}
This result is similar to the resolution of  $k > 2p +6$  instantons on a $D_{p+4}$ singularity \cite{Intriligator:1997dh}.
It can be checked that the matter content guarantees anomaly cancellation. For instance, for $n=4$, $p=3$, besides 
$\frac12(\bf{fund}, \bf{fund})$ for  adjacent $\mathfrak{sp}-\mathfrak{so}$ and  $\mathfrak{so}-\mathfrak{sp}$, there 
is an additional $\frac12(\bf{1}, \bf{fund})$ for $\mathfrak{so}(13)
-\mathfrak{sp}(3)$, cf.\ appendix~\ref{sec:matter-analysis}. For $n < p$ the resolution is given exchanging $p$ with $n$ in \eqref{resoIInp2}.

As we already mentioned, NU list another $[\kod{II}_{n-p}]$ model in
their parabolic type 5 class. This model has a different monodromy,
whose action on the moduli is given by $\tau \rightarrow \tau + p$, $\rho \rightarrow \rho +\tau + 2\beta + n + p$ and
$\beta \rightarrow -\beta-\tau - p$. The resolutions of the F-theory duals
are similar to \eqref{resoIInp1} and
\eqref{resoIInp2}, but there is a difference in the ``ascending''
ramps. For concreteness, we show the particular example $n=p=6$. The
resolution chain is the same as the one in  \eqref{resoIInp1} with the
same values of $n$ and $p$. However, the gauge groups on the starting
$1\,2\,3\,1$ chain are:
\begin{equation}
\begin{tabular}{ccccc}
&   $\mathfrak{su}(2)$ &$\mathfrak{so}(7)$ &$\mathfrak{su}(2)$  &\\
1 &      2 &3 & 1 & $\cdots$ \\
\end{tabular}
\end{equation}
while the $ 4 1 \cdots 4$ chain next to it supports the
following gauge algebras:
\begin{equation}
\mathfrak{so}(12) - \mathfrak{sp}(3) - \mathfrak{so}(16) -
\mathfrak{sp}(5) - \left[\mathfrak{so}(20) - \mathfrak{sp}(6)
\right]^{\oplus 3}-\mathfrak{so}(19) \, .
\end{equation}
This is glued to the same descending ramp $14\cdots$ as in \eqref{resoIInp1}. It
is interesting that we now get additional jumps in the rank of the
gauge groups, similar to what we found for the type 5 $[\kod
I_{n-p-q}]$ in section \ref{sec:I_n-p-q}. It is likely that this
corresponds to IIA brane models which involve O6$^{\pm}$ planes, along
the lines of \cite{Heckman:2016ssk}. It would be interesting to
explore this further.

\subsubsection*{$[\mathrm{I}_n-\mathrm{I}_p^*]$ model}\label{sec:I_n-I_p*}

The local equation reads:
\begin{equation}\label{eqinipstar}
y^2  = (t + x)(t^n + (x-1)^2) (t^{p+2} + x^2)\, .
 \end{equation}
The monodromy action on the moduli is:
\begin{equation}\label{monodinipstar}
\tau \rightarrow \tau + p \, ,\quad \rho \rightarrow \rho + n \, ,\quad \beta \rightarrow -\beta\, .
 \end{equation}
The data of the F-theory dual K3 can be found as in preceding examples.

We expect this model to describe $k=n+p+6$ small instantons on a  $D_{p+4}$ singularity. 
Performing the resolution we indeed find patterns matching known results for such a configuration
\cite{Aspinwall:1997ye, Intriligator:1997dh}. 
When $n=p$ and $n>p$ the resolutions are respectively of the form in eqs.~\eqref{resoIInp1} and \eqref{resoIInp2}, 
except for the replacement of the starting $123\cdots$ by 
\begin{equation}
\begin{tabular}{ccccc}
&  & $\mathfrak{sp}(1)$ &$\mathfrak{g}_2$ & \\
1 &      2 &2 & 3 & $\cdots$ \\
\end{tabular}
\end{equation}
When $n<p$ the resolution follows exchanging $p$ with $n$ in the result explained above.

\section{A classification of T-fects and 6D SCFTs}\label{sec:classification}

In the previous sections we provided several examples of heterotic geometric and
non-geometric degenerations whose dual F-theory realization admits a
smooth resolution. 
The purpose of this section is to determine   
all models from the NU list for which such a desingularization is
possible. We will show that the examples discussed so far
essentially cover all the possible situations. In this way we obtain a
catalog of six-dimensional 
theories that characterize geometric and
non-geometric ``exotic'' defects for the $E_7 \times E_8$ gauge group. 

\subsection{Criteria for the resolutions}
\label{ss:crit}

We want to discuss a more systematic approach to the resolutions or base blow-ups, respectively. The way we will proceed is strongly influenced by toric geometry of which we will make  use of in the following. For the details on toric geometry we refer the reader to the literature, e.g.\ \cite{cox2011toric,fulton1993introduction}.

In the preceding sections, we applied sequences of base blow-ups to get rid of the non-Kodaira singularities at $u=t=0$. 
At every step of this blow-up process, the map was of the following kind
\begin{equation*}
\xi_1,\, \xi_2\quad \mapsto \quad e\,\tilde\xi_1,\,e\,\tilde \xi_2\,,
\end{equation*}
where by $\xi_1$ and $\xi_2$ we denote the respective affine base coordinates at some step in the process.
For the elliptic fibration to remain Calabi-Yau the blow-up had to involve the fiber coordinates $x$ and $y$ too:
\begin{equation}
\xi_1,\, \xi_2,\,x,\,y\quad \mapsto \quad e\,\tilde\xi_1,\,e\,\tilde \xi_2,\,e^2\,\tilde x,\,e^3\,\tilde y\,.
\end{equation}
We can summarize such a blow-up in the following weight table:
\begin{equation}
\begin{array}{|c|c|c|c|c|c|c|}
\hline
     & \xi_1 & \xi_2 & x & y & e & \sum\\\hline
 E &1 & 1 & 2 & 3 & -1 & 6 \\ \hline
\end{array}\,.
\end{equation}
Note that here and in the following, we omit tildes over the new coordinates.
The hypersurface stays Calabi-Yau because after  factoring $e^6$ off, to obtain the proper transform of the Weierstra\ss{} equation, its class changes by $6E$.

We will now generalize this procedure. For this we introduce toric (blow-up) divisors in general directions:\footnote{A single toric divisor introduced that way might render the base singular, e.g.\ $n_1=2=2\,n_2$ would generate a $\mathbb Z_2$-singularity in the base at $e=\xi_2=0$. But in the collection with all the divisors we will introduce, the final base will be smooth.}
\begin{equation*}
\begin{array}{|c|c|c|c|c|c|c|}
\hline
     & \xi_1 & \xi_2 & x & y & e & \sum\\\hline
 E & n_1 & n_2 & o & p & -1 & o+p +n_1+n_2-1 \\ \hline
\end{array}\,,
\end{equation*}
with $n_i$, $o$, $p \in \mathbb N_{>0}$ and $n_1$, $n_2$ coprime. To have the same powers of $e$ in front of $y^2$ and $x^3$, we have to set $o=2k$ and $p=3k$. Because, the proper transform of the Weierstra\ss{} equation should again be of Weierstra\ss{} form. Furthermore, the hypersurface  should stay  Calabi-Yau which is true for
\begin{equation}
6\,k= 5\,k +n_1+n_2-1 \qquad\Rightarrow\qquad k=n_1+n_2-1\,.
\end{equation}
Hence, the toric divisors we are interested in are of the form
\begin{equation}
\begin{array}{|c|c|c|c|c|c|c|}
\hline
     & \xi_1 & \xi_2 & x & y & e & \sum\\\hline
 E & n_1 & n_2 & 2(n_1+n_2-1) & 3(n_1+n_2-1) & -1 &  6(n_1+n_2-1)\\ \hline
\end{array}\,.
\end{equation}
For $e^{6(n_1+n_2-1)}$ to factor off the Weierstra\ss{} equation, it is necessary that after the blow-up $f$ and $g$ have a prefactor $e$ to the power $4(n_1+n_2-1)$ and $6(n_1+n_2-1)$ or more, respectively. Since $f$ and $g$ are polynomials in $\xi_1$ and $\xi_2$, i.e.
\begin{equation}
f=\sum_i f_i \,\xi_1^{m^1_i}\xi_2^{m^2_i}\,,\qquad g=\sum_i g_i \,\xi_1^{l^1_i}\xi_2^{l^2_i}\,,
\end{equation}
this amounts to the constraint
\begin{equation}\label{eq:allowed-blow-up-directions}
(m^1_i-4)n_1+(m^2_i-4)n_2=:\tilde{\mathbf m}_i\cdot \mathbf n\ge -4 \quad \textmd{and}\quad (l^1_i-6)n_1+(l^2_i-6)n_2=:\tilde{\mathbf l}_i\cdot \mathbf n\ge -6 
\end{equation}
for all $\tilde{\mathbf m}_i$ and $\tilde{\mathbf l}_i$.

Given  $f$ and $g$, equation \eqref{eq:allowed-blow-up-directions} tells us which blow-ups $\mathbf n^j$ we have to introduce such that all the fibers over the base are of Kodaira type. The set $\{\mathbf n^j\}$ is given by all the vectors which fulfill  \eqref{eq:allowed-blow-up-directions} and have coprime entries. However, there can be cases in which the vanishing orders of  $f$ and $g$ are too high to obtain a well-defined Weierstra\ss{} fibration. This happens when there is an infinite number of allowed $\mathbf n^j$'s. Put differently, there exists an $\mathbf n$ such that 
\begin{equation}\label{eq:criterion}
\tilde{\mathbf m}_i\cdot \mathbf n\ge 0 \quad \textmd{and}\quad \tilde{\mathbf l}_i\cdot \mathbf n\ge 0
\end{equation}
for all $\tilde{\mathbf m}_i$ and $\tilde{\mathbf l}_i$. Therefore also any multiple of $\mathbf n$ would solve \eqref{eq:allowed-blow-up-directions}. The existence of such an $\mathbf n$ would also imply that $f$ and $g$ vanish to order four and six or more along the corresponding blow-up curve, i.e.\ we would have a whole curve of fibers which are beyond the 
Kodaira types. 
In this work, we are only interested in elliptic fibrations of a very restricted kind, cf.\ equation \eqref{K378}. Therefore, we can give a very simple criterion on 
$a$, $b$, $c$ and $d$ or to be more precise on their vanishing orders $\mu(a)$, $\mu(b)$, $\mu(c)$ and $\mu(d)$. 
The criterion reads:

\smallskip

{\em The solution set to \eqref{eq:allowed-blow-up-directions} is finite iff $\mu(a)<4$ or $\mu(b)<6$ or $\mu(c)<10$ or $\mu(d)<12$.}

\smallskip
 
\noindent 
By virtue of \eqref{igusa_abcd} the relevant data can be related to the vanishing orders of the Igusa-Clebsch invariants with a little subtlety for $\mu(b)$. 

Having the set of necessary blow-ups $\{ \mathbf n^j \}$ it is also simple to read off the vanishing orders of $f$ and $g$ along the exceptional curves  $\{ e_j \}$. The vanishing order of $f$ along $e_j=0$ is given by
\begin{equation}
\min_i\left( \{ \tilde{\mathbf m}_i\cdot\mathbf n^j+4\}\right)
\end{equation}
and for $g$ by
\begin{equation}
\min_i\left( \{ \tilde{\mathbf l}_i\cdot\mathbf n^j+6\}\right)
\end{equation}
The vanishing orders of the discriminant can be obtained in a similar fashion. We collect the powers of the polynomial in the discriminant  given by
\begin{equation}
\Delta=\sum_i \Delta_i \,\xi_1^{p^1_i}\xi_2^{p^2_i}\,.
\end{equation}
From the vectors in the set $\{\mathbf p_i\}$ we subtract $(12,12)^T$ to obtain $\{\tilde{\mathbf p}_i\}$. The vanishing order of the discriminant along the divisor $e_j=0$ is then:
\begin{equation}
\min_i\left( \{ \tilde{\mathbf p}_i\cdot\mathbf n^j+12\}\right)\,.
\end{equation}

\subsubsection{Two examples}

\begin{figure}[t]
\begin{center}
\includegraphics[scale=0.6]{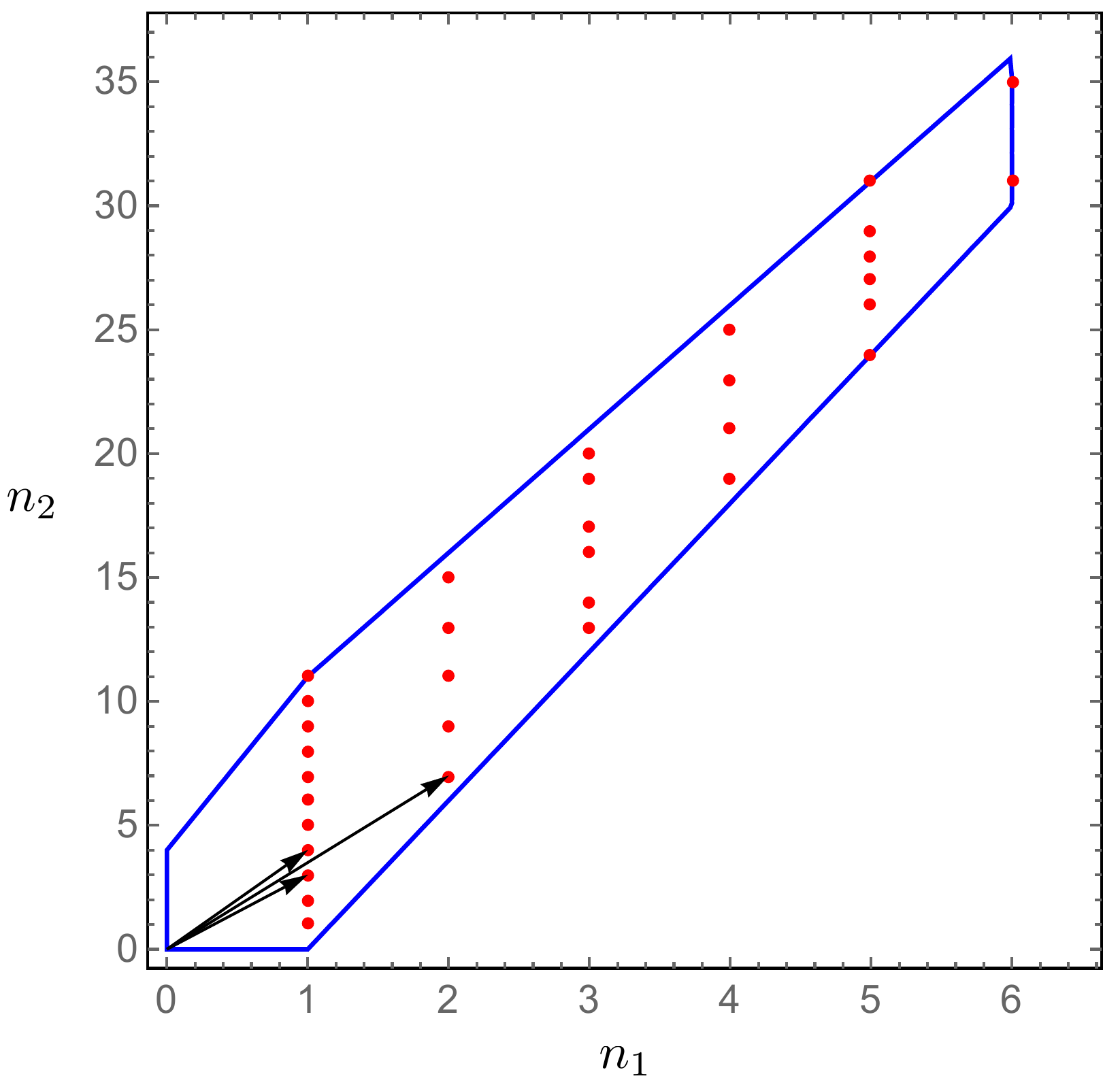}
\caption{Blow-up divisors for the resolution of the dual of the
  $\kod{[II^{\ast}-I_1]}$ model. The blue solid lines join the vertices
  given in \eqref{vertices} for $n=1$. The red dots are the points
  that satisfy the coprime conditions and are thus the blow-up
  divisors. We show the vectors corresponding to the points $(1,3)$,
  $(1,4)$ and to the divisor $(2,7)$ between them, with
  self-intersection $-1$.  }\label{Fig:toric}
\end{center}\end{figure}

To illustrate the above procedure, we will work out two examples in
detail. We start with the $[\kod{II}^{\ast}-\kod I_n]$ singularity on the heterotic side, already considered in section \ref{sss:InE8}. Mapping it to F-theory we get 
\begin{equation}
f=u^3\,t^{10+n}(\ldots)+ u^4\,t^{5+n}(\ldots)
\end{equation}
and 
\begin{equation}
g=u^5\,t^{10+n}(\ldots)+ u^6\,t^{5}(\ldots)+ u^7\,.
\end{equation}
We only gave the relevant terms because any term with a higher power in $t$ gives a weaker constraint in \eqref{eq:allowed-blow-up-directions} than the ones shown. The inequalities from $f$ are 
\begin{equation}
(6+n,-1)\cdot \mathbf n\ge -4\,,\quad (1+n,0)\cdot \mathbf n\ge -4\,,
\end{equation}
and those from $g$ are
\begin{equation}
(4+n,-1)\cdot \mathbf n\ge -6\,,\quad (-1,0)\cdot \mathbf n\ge -6\,,\quad (-6,1)\cdot \mathbf n\ge -6\,.
\end{equation}
Together with the positivity constrain for $n_1$ and $n_2$, the
solution set is given by a lattice polytope with the following
vertices:
\begin{equation}\label{vertices}
\{(0,0), \,(1,0), (6,30),\,(6,30+6n),\,(1,10+n),\,(0,4) \}\, ,
\end{equation}
However, not all lattice points in this polytope become blow-up divisors. First of all the directions $(1,0)$ and $(0,1)$ correspond to $t$ and $u$, respectively. Furthermore, because of the coprime condition, we only take the first point as a generator if we have a ray which goes through several points. As an example consider the vertices $(6,30)$ and $(6,30 +6n)$. In both cases we find six points lying on the ray going through them, but only the first points give rise to blow-up divisors.
Notice also that the points $(1,j)$, $j=1,\ldots, 10+n$ with coprime components, are contained in the polytope and actually 
correspond to the $e_j$ divisors
used in section \ref{sss:InE8}. Clearly there are additional points
associated to further blow-up divisors. The example corresponding to $n=1$ is
illustrated in figure \ref{Fig:toric}.

With the basic toric description at hand, we can show how the repeating blocks in the resolution in eq.~\eqref{resoInE8} do arise
and why there is a symmetry in the pattern of the self-intersections  of the curves. We will consider $n \ge 1$ in what follows. To begin, we find that
the $(n+1)$ divisors corresponding to $(1,5+j)$, $j=0,\ldots, n$, support type $\mathrm{II}^*$ fibers with self intersection
number $(-11)$ for $j=0,n$, and $(-12)$ for other $j$. 
 Next we should remember that toric information is invariant under $SL(d,\mathbb Z)$ transformations with $d$ the dimension of the toric variety which is $2$ in our case. 
 In particular, with
\[ \left(\begin{array}{cc} 1& 0\\ -k & 1\end{array}\right)\]
we can map all the `wedges' spanned by $(1,6+k)$ and $(1,5+k)$ to $(1,6)$ and $(1,5)$ which shows why we obtain the same self-intersection numbers between the $(-12)$- or $(-11)$-curves. Furthermore, with 
\[ \left(\begin{array}{cc} -4-n & 1\\ -6n -25 & 6\end{array}\right)\]
we can map the wedge spanned by $(0,1)$ and $(1,5+n)$ to $(1,6)$ and $(1,5)$.\footnote{Although the two parts of the polytope are not fully identical after applying this linear map to the first one, we are only interested in the first points of all the rays generated by the points in the polytope. These points lie also in the truncated piece.}
There is one further transformation,
\[ \left(\begin{array}{cc} 6 & -1\\ 31 & -5\end{array}\right)\,,\]
mapping $(0,-1)$, $(1,5)$ to $(1,5)$, $(1,6)$ which explains why most of the first self-intersections agree with those between the $(-12)$-curves.

To conclude with this example we give now the vectors corresponding to the (toric) blow-up divisors and their self-intersections between $(1,5)$ and $(1,6)$:
\begin{equation}
\begin{aligned}\{(6, 31):&\, -1,\,
 (5, 26):\, -2,\,
 (4, 21):\, -2,\,
 (3, 16):\, -3,\,
 (5, 27):\, -1,\,
 (2, 11):\, -5,\,\\
 &(5, 28):\, -1,\,
 (3, 17):\, -3,\,
 (4, 23):\, -2,\,
 (5, 29):\, -2,\,
 (6, 35):\, -1 \}\,.
 \end{aligned}
\end{equation}
Recall that in two dimensions the self-intersection number $-a_j$ of a toric divisor associated to $\mathbf n^j$ satisfies 
$a_j \mathbf n^j = \mathbf n^{j+1} + \mathbf n^{j-1}$.

As a second example we want to consider the $\kod{[III]}$ singularity
on the heterotic side, with the  monodromy $\tau \rightarrow \rho$, $\rho\rightarrow \tau + \rho - 2\beta$, $\beta \rightarrow \rho - \beta$. After mapping it to  F-theory we obtain
\begin{equation}
f=u^3\,t^{10}(\ldots)+ u^4\,t^{4}(\ldots)
\end{equation}
and 
\begin{equation}
g=u^5\,t^{12}(\ldots)+ u^6\,t^{6}(\ldots)+ u^7\,.
\end{equation}
In addition to the positivity constraint, we get only two inequalities from $f$ and $g$,
\begin{equation}
(6,-1)\cdot \mathbf n\ge -4\,,\quad (-6,1)\cdot \mathbf n\ge -6\,.
\end{equation}
These constraints are not enough to give a bounded solution set. Therefore, we will always end up with a curve of fibers which are beyond Kodaira type if we try to blow up the base to resolve the singularity at $u=t=0$. Furthermore,  in this example the vanishing orders are
$\mu(a)=4$, $\mu(b)=6$, $\mu(c)=10$ and $\mu(d)=12$, so that according to the criterion 
established before this model indeed
was not expected to have a resolution.

\subsection{A catalog of T-fects}
\label{ss:catalog}

We now briefly summarize the Namikawa-Ueno models for which we were
able to construct the dual CY resolution. The full list of NU models is reproduced in appendix \ref{app:NUlist}.
A simple way to determine whether a model admits a resolution is to apply the criterion stated in section \ref{ss:crit},
based on the vanishing orders of the coefficients $a$, $b$, $c$ and $d$ that enter in the elliptic fibration defined in
equation \eqref{K378}. The model has a resolution iff $\mu(a)<4$ or $\mu(b)<6$ or $\mu(c)<10$ or $\mu(d)<12$.

\subsubsection{Elliptic type 1}

The elliptic type 1 models of NU are characterized by monodromies that are of
finite order in the mapping class group of the genus-two surface and
contain twists around the $\gamma$ cycle (see figure \ref{Fig:genus2}). Therefore, the corresponding action of
$Sp(4,\mathbb{Z})$ elements on the Siegel upper half plane results
in mixing of the $(\tau,\rho,\beta)$ moduli and the models are thus non-geometric. Disregarding the trivial
monodromies, from a total of 18 types we
find only 4 models whose F-theory duals admit a smooth CY
resolution. We list them in table \ref{tab:elliptic1} together with
the vanishing orders of the coefficients of the dual elliptic fibration, from which we can
easily verify that the criterion discussed in the previous section is satisfied.
The explicit resolutions of these models were presented in section \ref{ss:nongeo}.
 
\begin{table}[h!]\begin{center}
\renewcommand{\arraystretch}{1.5}
\begin{tabular}{|c|c|c|c|c|}
\hline
NU model & $\mu(a)$ & $\mu(b)$ & $\mu(c)$ & $\mu(d)$  \\
\hline \hline
$\kod{[I_{0-0-0}]}$  &0& 0&0&0 \\ \hline 
$\kod{[V]}$  &2& 3&5&6 \\ \hline 
$\kod{[VII]}$ &2& 3&5&6 \\ \hline 
 $\kod{[VIII-1]} $&$\infty$& $\infty$&4&$\infty$ \\ \hline 
$\kod{[IX-1]}$ &$\infty$& $\infty$&8&$\infty$  \\\hline 
\end{tabular}
\caption{Elliptic type 1 models.}
  \label{tab:elliptic1}\end{center}\end{table}

\subsubsection{Elliptic type 2}

The NU list of type 2 models is given by all degenerations of type
$[\mathrm{K}_1-\mathrm{K}_2-m]$, with $m \geq 0$, where $\mathrm {K}_1$ and $\mathrm {K}_2$  are one of the Kodaira type singularities
for the two genus-one components of $\Sigma$, plus additional sporadic
models denoted as $[2\mathrm{K}-m]$ and
$[\mathrm{K}_1-\mathrm{K}_2-\alpha]$. None of the latter,
nor any of the models with $m\neq0$, give rise to smooth models.
Using again the notation $[\mathrm{K}_1-\mathrm{K}_2-0]\equiv
[\mathrm{K}_1-\mathrm{K}_2] $, we find a total of 20 models that
satisfy our criterion, listed in
table \ref{tab:elliptic2}. As we discussed in the previous sections,
the models of type $\kod{[I_0-K_2]}$ correspond to a configuration of
$k=\mu(d)$ pointlike instantons on the $\kod{K_2}$ singularity and the
resolutions are explicitly worked out in section \ref{ss:I0E8} and in
appendix \ref{app:ADE}. The remaining models are non-geometric since
their monodromy involves a non-trivial action on the torus volume. 
However, as we discussed in section \ref{ss:33} and \ref{ss:dual}, many of these
models lead to the same resolutions as the geometric ones.
\begin{table}[t]\begin{center}
\renewcommand{\arraystretch}{1.5}
\begin{tabular}{|c|c|c|c|c||c|c|c|c|c|}
\hline
NU model & $\mu(a)$ & $\mu(b)$ & $\mu(c)$ & $\mu(d)$&NU model & $\mu(a)$ & $\mu(b)$ & $\mu(c)$ & $\mu(d)$  \\
\hline \hline
$\kod{[I_0-I_0]}$  &0&0&0&0 &$\kod{[II-IV]}$ &3& 3&6&6 \\ \hline 
  $\kod{[I_0-II]}$ &1& 1&2&2 &$\kod{[I_0^{\ast}-II]}$ &3& 4&8&8 \\ \hline 
 $\kod{[I_0-III]} $&1& 2&3&3 &$\kod{[II-IV^{\ast}]}$ &5& 5&10&10 \\ \hline 
$\kod{[I_0-IV]}$ &2& 2&4&4  &$\kod{[II-III^{\ast}]}$ &4& 7&11&11 \\\hline 
$\kod{[I_0-I_0^{\ast}]}$ &2& 3&6&6  &$\kod{[III-III]}$ &2& 4&6&6 \\\hline 
$\kod{[I_0-IV^{\ast}]}$ &4& 4&8&8  &$\kod{[IV-III]}$ &3& 4&7&7 \\\hline 
$\kod{[I_0-III^{\ast}]}$ &3& 6&9&9  &$\kod{[I_0^{\ast}-III]}$ &3& 5&9&9 \\\hline 
$\kod{[I_0-II^{\ast}]}$ &5& 5&10&10 &$\kod{[IV^{\ast}-III]}$ &5& 6&11&11  \\\hline 
$\kod{[II-II]}$ &2& 2&4&4 &$\kod{[IV-IV]}$ &4& 4&8&8  \\\hline 
$\kod{[II-III]}$ &2&3&5&5 &$\kod{[I_0^{\ast}-IV]}$ &4& 5&10&10 \\\hline 
\end{tabular}
\caption{Elliptic type 2 models.}\label{tab:vanishing-orders_ell-type-2}
  \label{tab:elliptic2}\end{center}\end{table}

\subsubsection{Parabolic type 3}

In this class we found additional models in which the monodromy factorizes as
the product of two monodromies of Kodaira type for the two tori of
$\Sigma$, one of which is either $\mathrm{I}_n$ or $\mathrm{I}_n^{\ast}$ (the only
parabolic elements in the Kodaira list) and the other is of elliptic type. There are also 
models labeled $[\mathrm{K}_1-\mathrm{II}_{n}]$ or $[\mathrm{K}_1-\mathrm{II}_{n}^*]$
that mix all moduli but have a Kodaira type $\kod{K_1}$ monodromy for $\tau$. 

Altogether the 19 models that can be resolved are listed in table \ref{tab:parabolic3}. 
These models admit a resolution for all $n$. The
models of type $[\kod{I}_n-\kod K_2]$ or $[\kod{K}_1-\kod{I}_n]$ again correspond to
$k=\mu(d)$ pointlike instantons on the $\kod{K}_i$ singularity and their
resolution is shown in appendix \ref{app:ADE}. The resolution for 
$[\kod{II}_{n-0}]$ and other non-trivial examples are given in section \ref{ss:para3}.
In this class we also discover dual models. Concretely, starting with the fifth row in table \ref{tab:parabolic3}, the
models in the same row have the same resolution.

\begin{table}[h!]\begin{center}
\renewcommand{\arraystretch}{1.5}
\begin{tabular}{|c|c|c|c|c||c|c|c|c|c|}
\hline
NU model & $\mu(a)$ & $\mu(b)$ & $\mu(c)$ & $\mu(d)$&NU model & $\mu(a)$ & $\mu(b)$ & $\mu(c)$ & $\mu(d)$  \\
\hline \hline
$[\kod I_{n-0-0}]$  &0&0&$n$&$n$ &
 $[\kod{II}-\kod I_n] $&$1+n$& 1&$2+n$&$2+n$ \\ \hline 
$[\kod{III}-\kod I_n]$ &1& $2+n$&$3+n$&$3+n$  &
$[\kod{III}-\kod{II}_n]$ &1& $2+n$&$3+n$&$4+n$ \\ \hline 
$[\kod{IV}-\kod{I}_n]$ &$2+n$& 2&$4+n$&$4+n$  &
$[\kod{IV}-\kod{II}_n]$ &$2+n$& 2&$4+n$&$5+n$ \\ \hline 
$[\kod{II}_{n-0}]$ &2& 3&$5+n$&$6+n$ &
& & & & \\ \hline 
$[\kod I_n- \kod I_0^{\ast}]$ &2& 3&$6+n$&$6+n$  &
$[\kod{I}_0-\kod I_n^{\ast}]$ &2&3&$6+n$&$6+n$ \\ \hline
$[\kod{IV}^{\ast}-\kod I_n]$ &$4+n$& 4&$8+n$&$8+n$ &
$[\kod{II}-\kod I_n^{\ast}]$ &3&4&$8+n$&$8+n$ \\ \hline
$[\kod{III}^{\ast}-\kod I_n]$ &3& $6+n$&$9+n$&$9+n$ &
$[\kod{III}-\kod I_n^{\ast}]$ &3& 5&$9+n$&$9+n$ \\ \hline 
$[\kod{II}^{\ast}-\kod I_n]$ &$5+n$& 5&$10+n$&$10+n$ &
$[\kod{IV}-\kod I_n^{\ast}]$ &4& 5&$10+n$&$10+n$ \\ \hline 
$[\kod{IV}^{\ast}-\kod{II}_n]$ &$3+n$& 4&$7+n$&$9+n$ &
$[\kod{II}-\kod{II}_n^{\ast}]$ &$3+n$& 4&$7+n$&$9+3n$  \\ \hline 
$[\kod{III}^{\ast}-\kod{II}_n]$ &3& $5+n$&$8+n$&$11+n$  &
$[\kod{III}-\kod{II}_n^{\ast}]$ &3& $5+n$&$8+n$&$10+2n$ \\ \hline 
\end{tabular}
\caption{Parabolic type 3 models.}
  \label{tab:parabolic3}\end{center}\end{table}

\subsubsection{Parabolic type 4}

This class includes degenerations associated to parabolic Kodaira
singularities for both the genus-one components of $\Sigma$, of type
$[\kod K_1-\kod K_2-m]$ with $\kod{K_{1,2}} = \kod{I}_n,
\kod{I}_n^{\ast}$, plus additional degenerations of type
$[2\kod{K}_1-m]$, $[\kod{II}_{n-p}]$ and $[\kod{III}_n]$. We find only
3 models that admit a dual smooth resolution, listed in table
\ref{tab:parabolic4}. The explicit resolution of the
$[\kod{I}_{n-p-0}]$ model is given in section \ref{sec:I_n-p-0}, 
while the $[\kod{II}_{n-p}]$ and $[\kod{I}_n-\kod{I}_p^{\ast}]$ models are discussed
in section \ref{ss:para4}.

\begin{table}[h!]\begin{center}
\renewcommand{\arraystretch}{1.5}
\begin{tabular}{|c|c|c|c|c|}
\hline
NU model & $\mu(a)$ & $\mu(b)$ & $\mu(c)$ & $\mu(d)$\\
\hline \hline
$[\kod{I}_{n-p-0}]$  &0&0&$n+p$&$n+p$ \\ \hline 
 $[\kod{I}_n-\kod I_p^{\ast}]$  &2&3&$6+n+p$&$6+n+p$ \\ \hline
 $[\kod{II}_{n-p}]$  &2&3&$5+n+p$&$6+n+p$ \\ \hline 
\end{tabular}
\caption{Parabolic type 4 models.}
  \label{tab:parabolic4}\end{center}\end{table}

\subsubsection{Parabolic type 5}

The final class in the NU list is that of parabolic type 5 models,
which includes just 6 degenerations. Only 2 of them admit a smooth
resolution, and they are listed in table \ref{tab:parabolic5}. 
The resolution of the first (geometric) model is presented in section \ref{sec:I_n-p-q}. 
Notice that the parabolic type 5 $[\kod{II}_{n-p}]$ is different from
the one listed in table \ref{tab:parabolic4}. The
differences between the two models are discussed in section \ref{ss:para4}.

\begin{table}[h!]\begin{center}
\renewcommand{\arraystretch}{1.5}
\begin{tabular}{|c|c|c|c|c|}
\hline
NU model & $\mu(a)$ & $\mu(b)$ & $\mu(c)$ & $\mu(d)$\\
\hline \hline
$[\kod{I}_{n-p-q}]$  &0&0&$n+p+q$&$n+p+q$ \\ \hline 
$[\kod{II}_{n-p}]$  {\scriptsize$p=2k+l$, $ l=0,1$ }&2&3&$5+l+2k+n$&$6+l+2k+n$ \\ \hline
\end{tabular}
\caption{Parabolic type 5 models.}
  \label{tab:parabolic5}\end{center}\end{table}

This concludes the list of all genus-two degenerations in the NU list
that correspond to geometric and non-geometric heterotic local models,
whose F-theory duals admit a smooth resolution. Out of the 120 
entries in the NU list, we find a total of 49 
models.

\section{Final comments}\label{sec:conclusions}

In this paper we have studied compactifications to six dimensions of
the $E_8\times E_8$ heterotic string leaving an $E_8\times E_7$
subgroup unbroken. 
We have focused on configurations which are (up to degeneration
points) locally described by a $T^2$ fibration over a complex
one-dimensional base with a smooth $SU(2)$ structure bundle, patched
together using arbitrary elements of $SO^+(2,3,\mathbb{Z})$ (an order
four subgroup of the T-duality group $O(2,3,\mathbb{Z})$). This gives
rise generically to backgrounds without a global classical geometric
interpretation. At certain points in the base, the fibration (or
bundle data on it) will degenerate, and will no longer have --- in any
T-duality frame --- an interpretation in terms of the heterotic string
on a smooth $T^2$ with a smooth vector bundle on top. Our goal in this
paper has been to characterize the physics arising from such singular
points.

We have made use of the fact that for backgrounds preserving
$E_8\times E_7$ symmetry, the geometric data of the heterotic string
on $T^2$ can be encoded in the geometry of a genus-two (sextic)
Riemann surface. One can then define a six dimensional theory by
fibering this genus-two Riemann surface over a complex one-dimensional
base. For monodromies in $SO^+(2,3,\mathbb{Z})$, or equivalently
$Sp(4,\ZZ)$, one can classify the ways in which such fibration can
degenerate \cite{ogg66,Namikawa:1973yq}. Using heterotic/F-theory
duality to reinterpret these degenerations of the sextic as
degenerations of the dual F-theory K3, fibered over the
same 
base, we can read off the low energy physics at the degeneration
point.

We have encountered two noteworthy surprises in performing the
systematic analysis of the full set of degenerations of
sextics. First, we have found that many, sometimes very exotic looking,
non-geometric degenerations are described by the same low energy
physics. 
Often these are given by the long-understood configurations of
pointlike instantons sitting on ADE singularities. It would be very
interesting to understand the origin of this phenomenon in heterotic
language.

A second remarkable point is that not all of the possible
degenerations of sextics admit an F-theory dual that can be smoothed
out by a finite number of blow-ups. In these cases we cannot determine
the low energy physics using F-theory techniques. It would be very
interesting to find out which kind of SCFTs these theories might
correspond to, assuming that they give consistent backgrounds. Indeed,
except for the lack of smooth resolutions in the F-theory description,
we have found no evidence that these backgrounds are ill-defined.

\medskip

In addition to clarifying the two points just mentioned, there are
various directions for further study, of which we now highlight a
few. The most obvious one is probably to examine the case of
non-geometric compactifications of the heterotic string down to four
dimensions. Heterotic/F-theory duality will likely be an invaluable
tool in this situation too.

We note that non-geometric string backgrounds have been studied in the
past by a variety of approaches, and it is compelling to figure out
possible implications of our concrete and explicit results for these
different lines of investigation. In particular, our geometrization of
the duality group contrasts with the viewpoint advocated in doubled
formalisms such as double field theory
\cite{Hull:2006va,Hull:2007jy}\footnote{See for example
  \cite{Aldazabal:2013sca,Hohm:2013bwa} for reviews and a list of
  references.}, where by extending the spacetime coordinates one finds
extra degrees of freedom that need to be projected out. A potentially
related question is the role of the genus-two surface in the heterotic
formulation, which in a sense simultaneously encodes the physical
heterotic $T^2$ and its T-dual. Can this genus-two curve be given a
direct interpretation in the heterotic string, instead of being an
auxiliary construct parameterizing the moduli space? If so, one may
expect that there is some analog of the genus-two construction for
heterotic compactifications breaking the symmetry further than
$E_7\times E_8$. It is important 
to find this
generalization if it exists.

Along related lines, some of our non-geometric models involve elliptic
finite-order monodromies for the moduli that should admit a
description in terms of asymmetric orbifolds at some point in moduli
space \cite{Condeescu:2012sp,Condeescu:2013yma} and such ``double
elliptic'' T-folds have been used in the context of generalized
Scherk-Schwarz reductions
\cite{Dabholkar:2002sy,Hassler:2014sba,Hassler:2014mla}.

The description of non-geometric degenerations in terms of
dual F-theory models could be complemented with the explicit solutions
for our T-fects.
 The simplest example is that of the exotic brane
discussed for example in \cite{deBoer:2012ma,Hassler:2013wsa}, but 
it would be interesting to obtain local solutions
with arbitrary $Sp(4,\mathbb{Z})$ monodromy, along the lines of \cite{Lust:2015yia}. Another question is how to understand in the non-geometric
heterotic context the fact that 6d (1,0) superconformal field theories
do not possess any marginal deformations
\cite{Louis:2015mka,Cordova:2016xhm}.

Finally, we have focused on the $E_8\times E_8$ heterotic
string. Performing a similar analysis for the $SO(32)$ heterotic
string is feasible, and could potentially shed some light on some of
the open problems just mentioned. More ambitiously, let us mention
that the same genus-two technology that has played a key role in our
analysis also appears in the study of non-perturbative IIB solutions
with monodromies in a subset of the U-duality group
\cite{Martucci:2012jk,Braun:2013yla,Candelas:2014jma}. Understanding
the physics of U-duality defects in the type II context should be very
interesting. It might, for instance, be possible that
U-fold defects in IIB increase the set of SCFTs constructible from
F-theory beyond those considered in
\cite{Heckman:2013pva}.\footnote{As a simple example of the potential
  interest of the construction, notice that by T-dualizing once along
  the elliptic fiber of an appropriate Weierstra\ss{} model in IIB, one
  constructs the $E$-type $(0,2)$ SCFTs in IIA string theory as
  non-geometric T-folds.}

\vspace{1cm}
\textbf{Acknowledgments:} We are grateful to Ling Lin, Ruben Minasian,
David Morrison, Erik Plauschinn, Raffaele Savelli, Stefan Theisen,
Alessandro Tomasiello and Timo Weigand for useful discussions. We
thank Daniel Junghans for expert nomenclature counseling. This research is supported by the Munich Excellence Cluster for Fundamental Physics ``Origin and the Structure of the Universe'' and by the ERC Advanced Grant 32004~--~Strings and Gravity. 
A.F.~thanks the Alexander von Humboldt Foundation for a grant VEN/1067599 STP, as well as the Max-Planck-Institut f\"ur Physik, 
the Ludwig-Maximillians-Universit\"at, and the Max-Planck-Institut f\"ur Gravitationsphysik, for hospitality and support at various stages of this work.
I.G.-E.~and C.M.~would like to thank the Aspen Center for Physics,
where parts of this research were carried out, for a stimulating
working atmosphere. S.M. would like to thank the Mainz Institute for Theoretical Physics (MITP) for its hospitality and its partial support during the completion of this work.

\appendix

\section{Other ADE singularities}\label{app:ADE}

In this appendix we further study the local heterotic models that represent
pointlike instantons on $ADE$ singularities, together with the
resolutions of the dual F-theory models, from which we read the
corresponding non-perturbative enhancements. This procedure was
illustrated in section \ref{ss:I0E8} for the genus-two
$\kod{[I_0-II^{\ast}]}$ model, representing ten pointlike instantons
on a $E_8$ type singularity. 
In sections \ref{sec:I_n-p-0} and \ref{ss:para4} we also discussed the 
$[\mathrm{I}_{n-p-0}]$ and $[\mathrm{I}_n-\mathrm{I}_p^*]$ models, associated respectively
to pointlike instantons on $A$ and $D$ type singularities. 
Here we list the remaining cases, and compare
with the results of Aspinwall and Morrison for the $E_8\times E_8$ heterotic string \cite{Aspinwall:1997ye}.

\subsection*{$[\kod{III}^{\ast}-\kod I_n]$ model}

For $n=0$, the local genus-two model is:
\begin{equation}
y^2 = x(x^2+t^3)\left(x^2+\alpha  x+1\right)  \, ,
 \end{equation}
with a monodromy action given by:
\begin{equation}
\tau \rightarrow -\frac{1}{\tau} \, ,\quad \rho \rightarrow \rho - \frac{\beta^2}{\tau}\, ,\quad \beta \rightarrow
\frac{\beta}{\tau}
\, .
\end{equation}
The geometry of the F-theory dual model, close to $u=t=0$ is:
\begin{align}
f_{K3} & = t^{9} u^3 +t^3 u^4 \, , \quad g_{K3}  = t^{9} u^5+t^6u^6+u^7 \, ,\\
\Delta_{K3} &=  -u^9 \left(4 t^{27}+12 t^{21} u+27 t^{18} u+66 t^{15} u^2+27 t^{12} u^3+58 t^9 u^3+54 t^6 u^4+27 u^5\right)\, .
\end{align}
The resolution now requires a total of 14 blowups in the base and
gives the following chain of Kodaira curves:
\begin{equation}
\kod{[III^{\ast}]-I_0-II-IV-I_0^{\ast}-II-IV^{\ast} -II-I_0^{\ast}-II-IV^{\ast} -II-I_0^{\ast}-IV-II} \, .
\end{equation}
From the study of monodromy covers we then find the algebras:
\begin{equation}
\begin{tabular}{|cccccccccccccc|}
\hline
 &&  $\mathfrak{sp}(1)$ &$\mathfrak{g}_2$&& $\mathfrak{f}_4$&
  &$\mathfrak{g}_2$&& $\mathfrak{f}_4$ &&$\mathfrak{g}_2$&$\mathfrak{sp}(1)$&\\
1 & 2&  2 &3&1&5 &1&3&1&5&1&3&2&2\\\hline
\end{tabular}\, .
\end{equation} 
This pattern slightly differs from the result for $k=9$ pointlike instantons on a $E_7$
singularity given in \cite{Aspinwall:1997ye}. We actually find that the $-3$ curve at the middle supports
$\mathfrak{g}_2$ with a fundamental $\mathbf 7$. Each block $\mathfrak{sp}(1) \oplus \mathfrak{g}_2$
has matter content $\frac12(\mathbf{2},\mathbf{1}) \oplus
\frac12(\mathbf{2},\mathbf{7})$. 

To obtain the case of $k=9+n$, as explained in section
\ref{ss:I0E8}, we need to consider the model $[\kod{III}^{\ast}-\kod{I}_n]$. This
introduces in the resolution a chain of $n$ $\kod{III^{\ast}}$ fibers
which needs additional resolutions. This can be done by a total of $14
+ 6n$ blowups and we find:
\begin{align}
\label{resoInIIIstar}
\begin{tabular}{|cccccccccc|}
\hline
&&  $\mathfrak{sp}(1)$ &$\mathfrak{g}_2$&& $\mathfrak{f}_4$&&$\mathfrak{g}_2$&$\mathfrak{su}(2)$& \\
1 & 2&     2 &3&1&5 &   1& 3& 2& 1 \\\hline
\end{tabular}\,
&
\begin{tabular}{|cccccc|}
\hline
$\mathfrak{e}_7$& &$\mathfrak{su}(2)$&$\mathfrak{so}(7)$ 
&$\mathfrak{su}(2)$& \\
8& 1& 2& 3& 2& 1
\\\hline
\end{tabular}^{\, \oplus (n-1)} \times \nonumber\\
& \times 
\begin{tabular}{|cccccccccc|}
\hline
$\mathfrak{e}_7$& &$\mathfrak{su}(2)$  
  &$\mathfrak{g}_2$&  &  $\mathfrak{f}_4$ &&$\mathfrak{g}_2$&$\mathfrak{sp}(1)$&\\
8& 1& 2& 3& 1 & 5&1&3&2&2
\\\hline
\end{tabular} \, .
\end{align}
The gauge algebra is now in agreement with \cite{Aspinwall:1997ye}. 
There is matter only for the $\mathfrak{sp}(1) \oplus \mathfrak{g}_2$ and 
$\mathfrak{su}(2) \oplus \mathfrak{so}(7) \oplus \mathfrak{su}(2)$ clusters. Concretely,
$\frac12(\mathbf{2},\mathbf{1}) \oplus
\frac12(\mathbf{2},\mathbf{7})$ for the former
and 
$\frac12(\mathbf{2},\mathbf{8},\mathbf{1}) \oplus
\frac12(\mathbf{1},\mathbf{8},\mathbf{2})$ for the latter.

\subsection*{$[\kod{IV}^{\ast}-\kod I_n]$ model}

 The
local $[\kod{I}_0-\kod{IV}^{\ast}]$ NU model is:
\begin{equation}
y^2  = (x^3+t^4)(x^2 + \alpha x+1) \, ,
 \end{equation}
with monodromy
\begin{equation}
\tau \rightarrow -\frac{1+\tau}{\tau}\, ,\quad \rho \rightarrow \rho - \frac{\beta^2}{\tau} \, ,\quad \beta \rightarrow
\frac{\beta}{\tau}
\, .
\end{equation}
The geometry of the F-theory dual model, close to $u=t=0$ is schematically:
\begin{align}
f_{K3} & = t^{8} u^3+ t^4u^4\, , \quad g_{K3}  = t^8u^5 + t^4 u^6 + u^7\, ,\\
\Delta_{K3} &= -u^9 \left(4 t^{24}+12 t^{20} u+12 t^{16} u^2+27 t^{16} u+4 t^{12} u^3+54 t^{12} u^2+81 t^8 u^3+54 t^4 u^4+27 u^5\right)\, .\nonumber
\end{align}
We now get the following chain:
\begin{equation}
\kod{[III^{\ast}] - I_0 - II- IV- I_0^{\ast} -II- IV^{\ast}-II-I_0^{\ast} -IV-II}
\, ,
\end{equation}
from which we determine the gauge algebra:
\begin{equation}
\begin{tabular}{|cccccccccc|}
\hline
 &&  $\mathfrak{sp}(1)$ &$\mathfrak{g}_2$&& $\mathfrak{f}_4$&
  &$\mathfrak{g}_2$& $\mathfrak{sp}(1)$ &\\
1 & 2&     2 &3&1&4 &1&3&2&2\\\hline
\end{tabular} \, ,
\end{equation}
giving the theory of 8 pointlike instantons on the $E_6$
singularity. As before, matter for $\mathfrak{sp}(1) \oplus \mathfrak{g}_2$ 
consists of $\frac12(\mathbf{2},\mathbf{1}) \oplus \frac12(\mathbf{2},\mathbf{7})$.
We find an extra fundamental for the $\mathfrak{f}_4$ with self-intersection $-4$.

The model $[\kod{IV}^{\ast}-\kod{I}_n]$, $n\ge 1$,  introduces a chain of
$\kod{IV^{\ast}}$ fibers, whose resolution requires a total of
$10+4n$ blowups:
\begin{equation}
\label{resoInIVstar}
\begin{tabular}{|ccccccccc|}
\hline
 &&  $\mathfrak{sp}(1)$ &$\mathfrak{g}_2$&& $\mathfrak{f}_4$&
  &$\mathfrak{su}(3)$& \\
1 & 2&     2 &3&1&5&1&3&1\\\hline
\end{tabular}\, 
\begin{tabular}{|cccc|}
\hline
  $\mathfrak{e}_6$&  &$\mathfrak{su}(3)$& \\
6&1 &3&1\\\hline
\end{tabular}^{\, \otimes (n-1)} \,
\begin{tabular}{|ccccc|}
\hline
  $\mathfrak{f}_4$&&$\mathfrak{g}_2$&$\mathfrak{sp}(1)$&\\
5 &1&3&2&2\\\hline
\end{tabular} \, .
\end{equation}
This gives the theory of $k= 8 +n$ pointlike instantons, as originally found in \cite{Aspinwall:1997ye}.
The only matter is $\frac12(\mathbf{2},\mathbf{1}) \oplus \frac12(\mathbf{2},\mathbf{7})$ 
for each $\mathfrak{sp}(1) \oplus \mathfrak{g}_2$ block.

\subsection*{$[\kod I_0^{\ast}-\kod I_n]$ model}

 The
local model for $n=0$ is:
\begin{equation}\label{eqi0i0star}
y^2  = (x^3+\beta t^2x+t^3) (x^3+\alpha x+1)\, .
 \end{equation}
The monodromy action on the moduli leaves $\tau$ and
$\rho$ invariant, while it acts on the Wilson line as $\beta \rightarrow
-\beta$. The geometry of the F-theory dual model, close to $u=t=0$ is:
\begin{align}
f_{K3} & = t^{6} u^3 +t^2 u^4 \, , \quad g_{K3}  = t^{6} u^5+t^3u^6+u^7 \, ,\\
\Delta_{K3} &=  -u^9 \left(4 t^{18}+12 t^{14} u+27 t^{12} u+12 t^{10} u^2+54 t^9 u^2+85 t^6 u^3+54 t^3 u^4+27 u^5\right)\, .
\end{align}
We get the following resolution:
\begin{equation}
\kod{[III^{\ast}] - I_0 - II- IV-I_0^{\ast}-IV-II} \, ,
\end{equation}
and we find the algebra
\begin{equation}
\begin{tabular}{|cccccc|}
\hline
 &&  $\mathfrak{sp}(1)$ &$\mathfrak{g}_2$&$\mathfrak{sp}(1)$ &\\
1 & 2&     2 &2&2&2 \\\hline
\end{tabular} \, ,
\end{equation}
which gives the non-perturbative enhancement for $k=6$ pointlike
instantons on a $D_4$ singularity. The gauge factors and the matter representations
are the same as in the $\mathrm{III}-\mathrm{III}$ model presented in section \ref{ss:33}.

Adding $n$ more instantons, thus
considering the $[\kod I_0^{\ast}-\kod I_n]$ model, results in a
chain of additional $n$ $\mathrm{I}_0^{\ast}$ singularities, whose resolution gives:
\begin{equation}
\begin{tabular}{|ccccc|}
\hline
 &&   $\mathfrak{sp}(1)$ & $\mathfrak{g}_2$& \\
1 &2&   2 &3  & 1 \\\hline
\end{tabular} \, 
\begin{tabular}{|cc|}
\hline 
$\mathfrak{so}(8)$& \\
4&1 \\\hline
\end{tabular}^{\, \oplus (n-1)}\, 
\begin{tabular}{|ccc|}
\hline
 $\mathfrak{g}_2$& $\mathfrak{sp}(1)$&\\
3& 2 &2 \\\hline
\end{tabular} \, ,
\end{equation}
for a total of $6+2n$ blowups.
Matter is just $\frac12(\mathbf{2},\mathbf{1}) \oplus \frac12(\mathbf{2},\mathbf{7})$ 
for each $\mathfrak{sp}(1) \oplus \mathfrak{g}_2$ cluster. 

\subsection*{$[\kod{IV}-\kod I_n]$ model}

 The
local model for $\kod{[I_0-IV]}$ is:
\begin{equation}
y^2  = (x^3+t^2)(x^3+\alpha x+1)\, ,
 \end{equation}
with monodromy action
\begin{equation}
\tau\rightarrow -\frac{1}{1+\tau} \, ,\quad \rho \rightarrow \rho
-\frac{\beta^2}{1+\tau}\, ,\quad \beta\rightarrow
-\frac{\beta}{1+\tau}  \, .
\end{equation}
The geometry of the F-theory dual model, close to $u=t=0$ is:
\begin{align}
f_{K3} & = t^{4} u^3+ t^2u^4\, , \quad g_{K3}  = t^4u^5 + t^2 u^6 + u^7\, ,\\
\Delta_{K3} &= -u^9 \left(4 t^{12}+12 t^{10} u+12 t^8 u^2+27 t^8 u+4 t^6 u^3+54 t^6 u^2+81 t^4 u^3+54 t^2 u^4+27 u^5\right)\, .\nonumber
\end{align}
From the resolution we get the following chain:
\begin{equation}
\kod{[III^{\ast}] - I_0 - II-  IV-II} \, ,
\end{equation}
and gauge algebra: 
\begin{equation}
\begin{tabular}{|cccc|}
\hline
 & &  $\mathfrak{sp}(1)$ & \\
1 & 2&     2 &2 \\\hline
\end{tabular} \, .
\end{equation}
From the $[\kod{IV}-\kod{I}_n]$ model we get $n$ additional $\kod{IV}$ fibers
and we recover the theory for $k=4+n$ pointlike instantons:
\begin{equation}\label{a25}
\begin{tabular}{|cccccccc|}
\hline
 & &  $\mathfrak{sp}(1)$  &  $\mathfrak{su}(3)_1$&&  $\mathfrak{su}(3)_{n-1}$ & $\mathfrak{sp}(1)$ & \\
1 & 2&     2&2&$\cdots$&2&2 &2 \\\hline
\end{tabular} \, ,
\end{equation}
which again agrees with \cite{Aspinwall:1997ye}. Matter consists of bifundamentals for adjacent $\mathfrak{su}(3)$'s, together with
$(\mathbf{2},\mathbf{1}) \oplus (\mathbf{2}+\mathbf{1}, \mathbf{3})$ for the  $\mathfrak{sp}(1) \oplus \mathfrak{su}(3)$'s
at the corners.

\subsection*{$[\kod{III}-\kod{I}_n]$ model}

For this final example we skip the details. The resolution is found to be:
\begin{equation}\label{a26}
\begin{tabular}{|ccccccc|}
\hline
 & &  $\mathfrak{su}(2)_1$  &  $\mathfrak{su}(2)_2$&&  $\mathfrak{su}(2)_{n}$ & \\
1 & 2&     2&2&$\cdots$&2 &2 \\\hline
\end{tabular} \, ,
\end{equation}
Matter is given by bifundamentals for neighboring factors, plus two additional fundamentals
for the left- and right-most $\mathfrak{su}(2)$'s.

\section{Map for vanishing Wilson line}\label{app:mapE8E8}

In this section we consider the duality map \eqref{map78} in the limit
$\beta \rightarrow 0$. This corresponds to the splitting of the
genus-two curve into two tori, whose mapping class groups geometrize
the $SL(2,\mathbb{Z})_{\tau}\times SL(2,\mathbb{Z})_{\rho}$ subgroup of the
$O(2,2,\mathbb{Z})$ T-duality group in the absence of Wilson lines
(see Figure \ref{Fig:genus2}).

From \cite{Malmendier:2014uka}, we have that, setting $\beta = 0$:
\begin{equation}
a = -\frac{E_4(\tau)E_4(\rho)}{48} \, ,\quad
b=-\frac{E_6(\tau)E_6(\rho)}{864} \, , \quad c=0\, , \quad d =
\eta(\tau)^{24}\eta(\rho)^{24} \, .
\end{equation}
Here $E_4$ and $E_6$ are the modular forms of weights four
and six for the two
$SL(2,\mathbb{Z})$ groups and $\eta$ is the Dedekind
$\eta$-function. In fact, this form of the duality map agrees with
the formulation in \cite{McOrist:2010jw}, based on the construction
of a Shioda-Inose structure for the K3 surface
\cite{Clingher:2146c}. The data about the $\tau$ and $\rho$ moduli can
be encoded in two Weierstra\ss{} equations that describe the two
genus-one components of the split genus-two curve:
\begin{equation}
y^2 = x^3 + f_{\tau} x\,w^4 + g_{\tau}w^6 \, , \quad \tilde y^2 = \tilde x^3 +
f_{\rho} \tilde x\,\tilde w^4 +g_{\rho}\tilde w^6 \, ,
\end{equation}
by the identifications 
\begin{equation}
f= -\frac13 E_4 \, , \quad g = -\frac{2}{27}E_6 \, , \quad 
\eta^{24} = -\frac{27}{4}\frac{\Delta}{1728} \, ,
\end{equation}
where $\Delta = 4f^3 + 27 g^2$ is the discriminant of the Weierstra\ss{}
equation. We thus get that the dual K3 is described by 
\begin{equation}
y^2 = x^3 -\frac{3}{16}f_{\tau}f_{\rho} x u^4w^4 +\frac{\Delta_{\tau}\Delta_{\rho}}{16^4}u^5w^6
-\frac{27}{128}g_{\tau}g_{\rho} u^6w^6 +u^7w^6 \, .
\end{equation}
It is easy to show that this indeed satisfies the relations
\eqref{map88}, originally obtained in \cite{LopesCardoso:1996hq}.
After a rescaling
$(u,w)\rightarrow (2^{-6}u, 2^7w)$ we find
\begin{equation}\label{mapE8E8mcorist}
y^2 = x^3 -3f_{\tau}f_{\rho} x
u^4w^4+\frac{\Delta_{\tau}\Delta_{\rho}}{16}u^5 w^6
-\frac{27}{2}g_{\tau}g_{\rho} u^6w^6+u^7w^6 \, ,
\end{equation}
which is a particular case of the expression given in
\cite{McOrist:2010jw}. More generally, one can allow the coefficient
of the $u^7$ term to transform as a section of a nontrivial line
bundle, corresponding to a different distribution of the point-like
instantons between the two $E_8$ factors. The map
\eqref{mapE8E8mcorist} is modified as follows:
\begin{equation}\label{generalmcoristmap}
y^2 = x^3 -3f_{\tau}f_{\rho} x u^4w^4+ d_{\tau}d_{\rho}u^5w^6
-\frac{27}{2}g_{\tau}g_{\rho} u^6w^6+ e_{\tau} e_{\rho}u^7w^6 \, ,
\end{equation}
where $\Delta_{\tau} = 4 d_{\tau} e_{\tau}$ and $\Delta_{\rho} = 4
d_{\rho}e_{\rho}$.

As a check, one can obtain the resolution for the dual of the NU model
$[\mathrm{I}_n-\mathrm{II}^{\ast}]$ in the $\beta \rightarrow 0$ limit
by using \eqref{generalmcoristmap}. For this we can take for example
$f_{\tau} = t^4$, $g_{\tau} = t^5$, $f_{\rho} = -3$,
$g_{\rho} = 2+t^n$, and set $d_{\tau} = t^5$,
$e_{\tau} = t^5(27 + 4t^2)$, $d_{\rho} = 27 t^n$, $e_{\rho} = 4+t^n$.
With this choice we engineer $\mathrm{II}^\ast$ and $\mathrm{I}_n$
singularities in $\tau$ and $\rho$, respectively. We then find that at
$t=0$ there is a ``vertical'' $\kod{II^{\ast}}$ fiber intersecting the
two ``horizontal'' $\kod{II^{\ast}}$ fibers at $u=0$ and $v=0$, with
additional $n$ instantons coalesced at the $u=t=0$ intersection
\cite{Aspinwall:1997ye}. Resolving both intersections at $u=t=0$ and
$v=t=0$ gives precisely the chain derived in section \ref{ss:I0E8},
corresponding to $10+n$ pointlike instantons on top of the $E_8$
singularity.

\section{Igusa-Clebsch invariants}\label{app:IgusaClebschinvariants}

We collect here the expressions of the Igusa-Clebsch invariants for an
hyperelliptic curve in terms of
the coefficients of the general sextic \eqref{eq:hyperelliptic-curve}.
Similar formulas have appeared in \cite{Klemm:2015iya}.
{\allowdisplaybreaks {\scriptsize 
\begin{align}
I_2 &= 6 c_3^2-16 c_2 c_4+40 c_1 c_5-240 c_0 c_6\, ,\label{I2coeff}\\
I_4 & = 48 c_6 c_2^3+4 c_4^2 c_2^2-12 c_3 c_5 c_2^2+300 c_0 c_5^2
      c_2+4 c_1 c_4 c_5 c_2-180 c_1 c_3 c_6 c_2-504 c_0 c_4 c_6 c_2  +48 c_0 c_4^3-12 c_1 c_3 c_4^2-80 c_1^2 c_5^2 \nonumber\\
&\quad +1620 c_0^2 c_6^2+36 c_1
  c_3^2 c_5 -180 c_0 c_3 c_4 c_5+324 c_0 c_3^2 c_6  +300 c_1^2 c_4 c_6 -540 c_0
  c_1 c_5 c_6 \, ,\\
I_6 & =-36 c_5^2 c_2^4-160 c_4 c_6 c_2^4-24 c_4^3 c_2^3-96 c_0 c_6^2 c_2^3+76 c_3 c_4 c_5 c_2^3+60 c_3^2 c_6 c_2^3+616 c_1 c_5 c_6 c_2^3+8 c_3^2 c_4^2 c_2^2+26 c_1 c_3 c_5^2 c_2^2 \nonumber\\
&\quad-640 c_0 c_4 c_5^2 c_2^2-900 c_1^2 c_6^2 c_2^2-24 c_3^3 c_5 c_2^2+28 c_1 c_4^2 c_5 c_2^2+424 c_0 c_4^2 c_6 c_2^2+492 c_1 c_3 c_4 c_6 c_2^2-876 c_0 c_3 c_5 c_6 c_2^2 \nonumber\\
&\quad-160 c_0 c_4^4 c_2+76 c_1 c_3 c_4^3 c_2+1600 c_0 c_1 c_5^3 c_2+330 c_0 c_3^2 c_5^2 c_2+64 c_1^2 c_4 c_5^2 c_2+3060 c_0 c_1 c_3 c_6^2 c_2+20664 c_0^2 c_4 c_6^2 c_2 \nonumber\\
&\quad+492 c_0 c_3 c_4^2 c_5 c_2-238 c_1 c_3^2 c_4 c_5 c_2-198 c_1 c_3^3 c_6 c_2-640 c_1^2 c_4^2 c_6 c_2-18600 c_0^2 c_5^2 c_6 c_2-468 c_0 c_3^2 c_4 c_6 c_2-1860 c_1^2 c_3 c_5 c_6 c_2 \nonumber\\
&\quad+3472 c_0 c_1 c_4 c_5 c_6 c_2-36 c_1^2 c_4^4+60 c_0 c_3^2 c_4^3-320 c_1^3 c_5^3+2250 c_0^2 c_3 c_5^3-119880 c_0^3 c_6^3-24 c_1 c_3^3 c_4^2+176 c_1^2 c_3^2 c_5^2-900 c_0^2 c_4^2 c_5^2 \nonumber\\
&\quad-1860 c_0 c_1 c_3 c_4 c_5^2-10044 c_0^2 c_3^2 c_6^2+2250 c_1^3 c_3 c_6^2-18600 c_0 c_1^2 c_4 c_6^2+59940 c_0^2 c_1 c_5 c_6^2+72 c_1 c_3^4 c_5+616 c_0 c_1 c_4^3 c_5 \nonumber\\
&\quad+26 c_1^2 c_3 c_4^2 c_5-198 c_0 c_3^3 c_4 c_5+162 c_0 c_3^4 c_6-96 c_0^2 c_4^3 c_6-876 c_0 c_1 c_3 c_4^2 c_6-2240 c_0 c_1^2 c_5^2 c_6+330 c_1^2 c_3^2 c_4 c_6 \nonumber\\
&\quad+1818 c_0 c_1 c_3^2 c_5 c_6+1600 c_1^3 c_4 c_5 c_6+3060 c_0^2
  c_3 c_4 c_5 c_6 \, ,\\
I_{10} & = 3125 c_6^4 c_1^6+256 c_5^5 c_1^5-3750 c_3 c_4 c_6^3 c_1^5-2500 c_2 c_5 c_6^3 c_1^5+2000 c_3 c_5^2 c_6^2 c_1^5+2250 c_4^2 c_5 c_6^2 c_1^5-1600 c_4 c_5^3 c_6 c_1^5-128 c_3^2 c_5^4 c_1^4 \nonumber\\
&\quad-192 c_2 c_4 c_5^4 c_1^4-22500 c_0 c_2 c_6^4 c_1^4+144 c_3 c_4^2 c_5^3 c_1^4+2250 c_2 c_3^2 c_6^3 c_1^4+1500 c_0 c_4^2 c_6^3 c_1^4+2000 c_2^2 c_4 c_6^3 c_1^4+2250 c_0 c_3 c_5 c_6^3 c_1^4 \nonumber\\
&\quad-27 c_4^4 c_5^2 c_1^4-900 c_2 c_4^3 c_6^2 c_1^4+825 c_3^2 c_4^2 c_6^2 c_1^4-50 c_2^2 c_5^2 c_6^2 c_1^4-1700 c_0 c_4 c_5^2 c_6^2 c_1^4-900 c_3^3 c_5 c_6^2 c_1^4-2050 c_2 c_3 c_4 c_5 c_6^2 c_1^4 \nonumber\\
&\quad+108 c_4^5 c_6 c_1^4+320 c_0 c_5^4 c_6 c_1^4+160 c_2 c_3 c_5^3 c_6 c_1^4+1020 c_2 c_4^2 c_5^2 c_6 c_1^4+560 c_3^2 c_4 c_5^2 c_6 c_1^4-630 c_3 c_4^3 c_5 c_6 c_1^4-1600 c_0 c_2 c_5^5 c_1^3 \nonumber\\
&\quad+144 c_2^2 c_3 c_5^4 c_1^3+160 c_0 c_3 c_4 c_5^4 c_1^3+27000 c_0^2 c_3 c_6^4 c_1^3+16 c_3^4 c_5^3 c_1^3-36 c_0 c_4^3 c_5^3 c_1^3-6 c_2^2 c_4^2 c_5^3 c_1^3-80 c_2 c_3^2 c_4 c_5^3 c_1^3-1350 c_0 c_3^3 c_6^3 c_1^3 \nonumber\\
&\quad-1600 c_2^3 c_3 c_6^3 c_1^3+19800 c_0 c_2 c_3 c_4 c_6^3 c_1^3+15600 c_0 c_2^2 c_5 c_6^3 c_1^3-1800 c_0^2 c_4 c_5 c_6^3 c_1^3+18 c_2 c_3 c_4^3 c_5^2 c_1^3-4 c_3^3 c_4^2 c_5^2 c_1^3+108 c_3^5 c_6^2 c_1^3 \nonumber\\
&\quad-120 c_0 c_3 c_4^3 c_6^2 c_1^3+410 c_0^2 c_5^3 c_6^2 c_1^3+560 c_2^2 c_3 c_4^2 c_6^2 c_1^3-12330 c_0 c_2 c_3 c_5^2 c_6^2 c_1^3-630 c_2 c_3^3 c_4 c_6^2 c_1^3+1020 c_2^2 c_3^2 c_5 c_6^2 c_1^3 \nonumber\\
&\quad-13040 c_0 c_2 c_4^2 c_5 c_6^2 c_1^3+160 c_2^3 c_4 c_5 c_6^2 c_1^3+1980 c_0 c_3^2 c_4 c_5 c_6^2 c_1^3-72 c_2 c_3 c_4^4 c_6 c_1^3+16 c_3^3 c_4^3 c_6 c_1^3-36 c_2^3 c_5^3 c_6 c_1^3-208 c_0 c_3^2 c_5^3 c_6 c_1^3 \nonumber\\
&\quad+9768 c_0 c_2 c_4 c_5^3 c_6 c_1^3+24 c_2 c_3^3 c_5^2 c_6 c_1^3-682 c_0 c_3 c_4^2 c_5^2 c_6 c_1^3-746 c_2^2 c_3 c_4 c_5^2 c_6 c_1^3+144 c_0 c_4^4 c_5 c_6 c_1^3+24 c_2^2 c_4^3 c_5 c_6 c_1^3 \nonumber\\
&\quad+356 c_2 c_3^2 c_4^2 c_5 c_6 c_1^3-72 c_3^4 c_4 c_5 c_6 c_1^3+2000 c_0^2 c_3 c_5^5 c_1^2-27 c_2^4 c_5^4 c_1^2+560 c_0 c_2 c_3^2 c_5^4 c_1^2-50 c_0^2 c_4^2 c_5^4 c_1^2+1020 c_0 c_2^2 c_4 c_5^4 c_1^2 \nonumber\\
&\quad+43200 c_0^2 c_2^2 c_6^4 c_1^2-32400 c_0^3 c_4 c_6^4 c_1^2-4 c_2^2 c_3^3 c_5^3 c_1^2-746 c_0 c_2 c_3 c_4^2 c_5^3 c_1^2+24 c_0 c_3^3 c_4 c_5^3 c_1^2+18 c_2^3 c_3 c_4 c_5^3 c_1^2+256 c_2^5 c_6^3 c_1^2 \nonumber\\
&\quad-9720 c_0 c_2^2 c_3^2 c_6^3 c_1^2-6480 c_0^2 c_2 c_4^2 c_6^3 c_1^2+540 c_0^3 c_5^2 c_6^3 c_1^2-10560 c_0 c_2^3 c_4 c_6^3 c_1^2-27540 c_0^2 c_3^2 c_4 c_6^3 c_1^2-31320 c_0^2 c_2 c_3 c_5 c_6^3 c_1^2 \nonumber\\
&\quad+144 c_0 c_2 c_4^4 c_5^2 c_1^2-4 c_2^3 c_4^3 c_5^2 c_1^2-6 c_0 c_3^2 c_4^3 c_5^2 c_1^2+c_2^2 c_3^2 c_4^2 c_5^2 c_1^2-27 c_2^2 c_3^4 c_6^2 c_1^2-192 c_0^2 c_4^4 c_6^2 c_1^2+4816 c_0 c_2^2 c_4^3 c_6^2 c_1^2 \nonumber\\
&\quad-128 c_2^4 c_4^2 c_6^2 c_1^2-4536 c_0 c_2 c_3^2 c_4^2 c_6^2 c_1^2+248 c_0 c_2^3 c_5^2 c_6^2 c_1^2+15417 c_0^2 c_3^2 c_5^2 c_6^2 c_1^2+8748 c_0^2 c_2 c_4 c_5^2 c_6^2 c_1^2+162 c_0 c_3^4 c_4 c_6^2 c_1^2 \nonumber\\
&\quad+144 c_2^3 c_3^2 c_4 c_6^2 c_1^2+3942 c_0 c_2 c_3^3 c_5 c_6^2 c_1^2+16632 c_0^2 c_3 c_4^2 c_5 c_6^2 c_1^2-192 c_2^4 c_3 c_5 c_6^2 c_1^2+10152 c_0 c_2^2 c_3 c_4 c_5 c_6^2 c_1^2-576 c_0 c_2 c_4^5 c_6 c_1^2 \nonumber\\
&\quad+16 c_2^3 c_4^4 c_6 c_1^2+24 c_0 c_3^2 c_4^4 c_6 c_1^2-1700 c_0^2 c_2 c_5^4 c_6 c_1^2-4 c_2^2 c_3^2 c_4^3 c_6 c_1^2-682 c_0 c_2^2 c_3 c_5^3 c_6 c_1^2-12330 c_0^2 c_3 c_4 c_5^3 c_6 c_1^2 \nonumber\\
&\quad+248 c_0^2 c_4^3 c_5^2 c_6 c_1^2-6 c_2^3 c_3^2 c_5^2 c_6 c_1^2-5428 c_0 c_2^2 c_4^2 c_5^2 c_6 c_1^2+144 c_2^4 c_4 c_5^2 c_6 c_1^2-2412 c_0 c_2 c_3^2 c_4 c_5^2 c_6 c_1^2+3272 c_0 c_2 c_3 c_4^3 c_5 c_6 c_1^2 \nonumber\\
&\quad-108 c_0 c_3^3 c_4^2 c_5 c_6 c_1^2-80 c_2^3 c_3 c_4^2 c_5 c_6 c_1^2+18 c_2^2 c_3^3 c_4 c_5 c_6 c_1^2+2250 c_0^2 c_2^2 c_5^5 c_1-2500 c_0^3 c_4 c_5^5 c_1-900 c_0^2 c_3^3 c_5^4 c_1-630 c_0 c_2^3 c_3 c_5^4 c_1 \nonumber\\
&\quad-2050 c_0^2 c_2 c_3 c_4 c_5^4 c_1-77760 c_0^3 c_2 c_3 c_6^4 c_1+38880 c_0^4 c_5 c_6^4 c_1-72 c_0 c_2 c_3^4 c_5^3 c_1+160 c_0^2 c_2 c_4^3 c_5^3 c_1+24 c_0 c_2^3 c_4^2 c_5^3 c_1 \nonumber\\
&\quad+1020 c_0^2 c_3^2 c_4^2 c_5^3 c_1+356 c_0 c_2^2 c_3^2 c_4 c_5^3 c_1+21384 c_0^2 c_2 c_3^3 c_6^3 c_1+46656 c_0^3 c_3 c_4^2 c_6^3 c_1+6912 c_0 c_2^4 c_3 c_6^3 c_1-3456 c_0^2 c_2^2 c_3 c_4 c_6^3 c_1 \nonumber\\
&\quad-21888 c_0^2 c_2^3 c_5 c_6^3 c_1+15552 c_0^3 c_3^2 c_5 c_6^3 c_1+31968 c_0^3 c_2 c_4 c_5 c_6^3 c_1-192 c_0^2 c_3 c_4^4 c_5^2 c_1-80 c_0 c_2^2 c_3 c_4^3 c_5^2 c_1+18 c_0 c_2 c_3^3 c_4^2 c_5^2 c_1 \nonumber\\
&\quad-486 c_0 c_2 c_3^5 c_6^2 c_1-5760 c_0^2 c_2 c_3 c_4^3 c_6^2 c_1-1800 c_0^3 c_2 c_5^3 c_6^2 c_1+5832 c_0^2 c_3^3 c_4^2 c_6^2 c_1-2496 c_0 c_2^3 c_3 c_4^2 c_6^2 c_1+16632 c_0^2 c_2^2 c_3 c_5^2 c_6^2 c_1 \nonumber\\
&\quad-31320 c_0^3 c_3 c_4 c_5^2 c_6^2 c_1+2808 c_0 c_2^2 c_3^3 c_4 c_6^2 c_1-6318 c_0^2 c_3^4 c_5 c_6^2 c_1-21888 c_0^3 c_4^3 c_5 c_6^2 c_1-4464 c_0 c_2^3 c_3^2 c_5 c_6^2 c_1 \nonumber\\
&\quad+15264 c_0^2 c_2^2 c_4^2 c_5 c_6^2 c_1-640 c_0 c_2^4 c_4 c_5 c_6^2 c_1-22896 c_0^2 c_2 c_3^2 c_4 c_5 c_6^2 c_1+768 c_0^2 c_3 c_4^5 c_6 c_1+320 c_0 c_2^2 c_3 c_4^4 c_6 c_1+2250 c_0^3 c_3 c_5^4 c_6 c_1 \nonumber\\
&\quad-72 c_0 c_2 c_3^3 c_4^3 c_6 c_1+144 c_0 c_2^4 c_5^3 c_6 c_1+1980 c_0^2 c_2 c_3^2 c_5^3 c_6 c_1+15600 c_0^3 c_4^2 c_5^3 c_6 c_1-13040 c_0^2 c_2^2 c_4 c_5^3 c_6 c_1-108 c_0 c_2^2 c_3^3 c_5^2 c_6 c_1 \nonumber\\
&\quad+10152 c_0^2 c_2 c_3 c_4^2 c_5^2 c_6 c_1+3942 c_0^2 c_3^3 c_4 c_5^2 c_6 c_1+3272 c_0 c_2^3 c_3 c_4 c_5^2 c_6 c_1-640 c_0^2 c_2 c_4^4 c_5 c_6 c_1-96 c_0 c_2^3 c_4^3 c_5 c_6 c_1 \nonumber\\
&\quad-4464 c_0^2 c_3^2 c_4^3 c_5 c_6 c_1-1584 c_0 c_2^2 c_3^2 c_4^2 c_5 c_6 c_1+324 c_0 c_2 c_3^4 c_4 c_5 c_6 c_1+3125 c_0^4 c_5^6-3750 c_0^3 c_2 c_3 c_5^5-46656 c_0^5 c_6^5+108 c_0 c_2^5 c_5^4 \nonumber\\
&\quad+825 c_0^2 c_2^2 c_3^2 c_5^4+2000 c_0^3 c_2 c_4^2 c_5^4-900 c_0^2 c_2^3 c_4 c_5^4+2250 c_0^3 c_3^2 c_4 c_5^4-13824 c_0^3 c_2^3 c_6^4+34992 c_0^4 c_3^2 c_6^4+62208 c_0^4 c_2 c_4 c_6^4 \nonumber\\
&\quad+108 c_0^2 c_3^5 c_5^3+16 c_0 c_2^3 c_3^3 c_5^3-1600 c_0^3 c_3 c_4^3 c_5^3+560 c_0^2 c_2^2 c_3 c_4^2 c_5^3-630 c_0^2 c_2 c_3^3 c_4 c_5^3-72 c_0 c_2^4 c_3 c_4 c_5^3-1024 c_0 c_2^6 c_6^3 \nonumber\\
&\quad-8748 c_0^3 c_3^4 c_6^3-13824 c_0^4 c_4^3 c_6^3-8640 c_0^2 c_2^3 c_3^2 c_6^3-17280 c_0^3 c_2^2 c_4^2 c_6^3-32400 c_0^4 c_2 c_5^2 c_6^3+9216 c_0^2 c_2^4 c_4 c_6^3+3888 c_0^3 c_2 c_3^2 c_4 c_6^3 \nonumber\\
&\quad+46656 c_0^3 c_2^2 c_3 c_5 c_6^3-77760 c_0^4 c_3 c_4 c_5 c_6^3+256 c_0^3 c_4^5 c_5^2-128 c_0^2 c_2^2 c_4^4 c_5^2+16 c_0 c_2^4 c_4^3 c_5^2+144 c_0^2 c_2 c_3^2 c_4^3 c_5^2-27 c_0^2 c_3^4 c_4^2 c_5^2 \nonumber\\
&\quad-4 c_0 c_2^3 c_3^2 c_4^2 c_5^2+729 c_0^2 c_3^6 c_6^2+108 c_0 c_2^3 c_3^4 c_6^2+9216 c_0^3 c_2 c_4^4 c_6^2-4352 c_0^2 c_2^3 c_4^3 c_6^2-8640 c_0^3 c_3^2 c_4^3 c_6^2+27000 c_0^4 c_3 c_5^3 c_6^2 \nonumber\\
&\quad+512 c_0 c_2^5 c_4^2 c_6^2+8208 c_0^2 c_2^2 c_3^2 c_4^2 c_6^2-192 c_0^2 c_2^4 c_5^2 c_6^2-27540 c_0^3 c_2 c_3^2 c_5^2 c_6^2+43200 c_0^4 c_4^2 c_5^2 c_6^2-6480 c_0^3 c_2^2 c_4 c_5^2 c_6^2 \nonumber\\
&\quad-4860 c_0^2 c_2 c_3^4 c_4 c_6^2-576 c_0 c_2^4 c_3^2 c_4 c_6^2+5832 c_0^2 c_2^2 c_3^3 c_5 c_6^2-3456 c_0^3 c_2 c_3 c_4^2 c_5 c_6^2+768 c_0 c_2^5 c_3 c_5 c_6^2+21384 c_0^3 c_3^3 c_4 c_5 c_6^2 \nonumber\\
&\quad-5760 c_0^2 c_2^3 c_3 c_4 c_5 c_6^2-1024 c_0^3 c_4^6 c_6+512 c_0^2 c_2^2 c_4^5 c_6-64 c_0 c_2^4 c_4^4 c_6-576 c_0^2 c_2 c_3^2 c_4^4 c_6+1500 c_0^3 c_2^2 c_5^4 c_6-22500 c_0^4 c_4 c_5^4 c_6 \nonumber\\
&\quad+108 c_0^2 c_3^4 c_4^3 c_6+16 c_0 c_2^3 c_3^2 c_4^3 c_6-1350 c_0^3 c_3^3 c_5^3 c_6-120 c_0^2 c_2^3 c_3 c_5^3 c_6+19800 c_0^3 c_2 c_3 c_4 c_5^3 c_6+162 c_0^2 c_2 c_3^4 c_5^2 c_6 \nonumber\\
&\quad-10560 c_0^3 c_2 c_4^3 c_5^2 c_6+24 c_0 c_2^4 c_3^2 c_5^2 c_6+4816 c_0^2 c_2^3 c_4^2 c_5^2 c_6-9720 c_0^3 c_3^2 c_4^2 c_5^2 c_6-576 c_0 c_2^5 c_4 c_5^2 c_6-4536 c_0^2 c_2^2 c_3^2 c_4 c_5^2 c_6 \nonumber\\
&\quad+6912 c_0^3 c_3 c_4^4 c_5 c_6-2496 c_0^2 c_2^2 c_3 c_4^3 c_5
  c_6+2808 c_0^2 c_2 c_3^3 c_4^2 c_5 c_6+320 c_0 c_2^4 c_3 c_4^2 c_5
  c_6-486 c_0^2 c_3^5 c_4 c_5 c_6-72 c_0 c_2^3 c_3^3 c_4 c_5 c_6 \, . \label{I10coeff}
\end{align}
}
}

\section{Ogg-Namikawa-Ueno classification}\label{app:NUlist}

We list all the Ogg-Namikawa-Ueno types of degenerations of the genus-two
fibers \cite{ogg66,Namikawa:1973yq}, in the notation of \cite{Namikawa:1973yq}.  For each model we list the order of vanishing of the
Igusa-Clebsch invariants and the homological monodromy.

\paragraph*{Type 1 (elliptic)}

{\footnotesize
\begin{center}
\begin{longtable}{|M{2cm} |M{5cm}| M{1cm}| M{1cm} |M{1cm} |M{1cm}
  |M{3.5cm} N|}
\hline
Type                    &  Local model &  $\mu(I_2)$ & $\mu(I_4)$ &$\mu(I_6)$& $\mu(I_{10})$ & Monodromy &      \\
\hline

 $[I_{0-0-0}]$      &  $y^2 = x^5+\alpha  x^3+\beta  x^2+\gamma x+1$ & 0&  0   &   0   &    0 &
{\tiny $\begin{pmatrix} 1 & 0 & 0 & 0 \\
 0 & 1 & 0 & 0 \\
 0 & 0 & 1 & 0 \\
 0 & 0 & 0 & 1 \end{pmatrix}$} &  \\[25pt]\hline
 $[I^{\ast}_{0-0-0}]$&$y^2 = t^5+\gamma  t^4 x+\beta  t^3 x^2+\alpha  t^2 x^3+x^5$ &   4  &  8   &   12   &    20  &{\tiny $\begin{pmatrix} -1 & 0 & 0 & 0 \\
 0 & -1 & 0 & 0 \\
 0 & 0 & -1 & 0 \\
 0 & 0 & 0 & -1 \end{pmatrix}$} & \\[25pt]\hline
 $[II]$      &  $y^2 = t^3+\beta  t^2 x^2+\alpha  t x^4+x^6$ & 3  &  6   &   9   &    15 &{\tiny $\begin{pmatrix} 0 & 1 & 0 & 0 \\
 1 & 0 & 0 & 0 \\
 0 & 0 & 0 & 1 \\
 0 & 0 & 1 & 0 \end{pmatrix}$} &\\[25pt]\hline
 $[III]$      & $y^2 = t^2+\alpha  t x^3+x^6$ &  2  &  4   &   6   &    10 &{\tiny $\begin{pmatrix} 0 & -1 & 0 & 0 \\
 1 & -1 & 0 & 0 \\
 0 & 0 & -1 & -1 \\
 0 & 0 & 1 & 0 \end{pmatrix}$} &  \\[25pt]\hline
$[IV]$      &  $y^2 = t \left(t^2+\alpha  t x^3+x^6\right)$ & 4  &  8   &   12   &    20 &{\tiny $\begin{pmatrix} 0 & 1 & 0 & 0 \\
 -1 & 1 & 0 & 0 \\
 0 & 0 & 1 & 1 \\
 0 & 0 & -1 & 0 \end{pmatrix}$} & \\[25pt]\hline
 $[V]$      &   $y^2 = x^6 +t $ & 1  &  2   &   3   &    5   & {\tiny $\begin{pmatrix} 0 & 0 & 1 & 0 \\
 0 & 0 & 1 & 1 \\
 -1 & 1 & 0 & 0 \\
 0 & -1 & 0 & 0 \end{pmatrix}$} & \\[25pt]\hline
 $[V^{\ast}]$      &$y^2 = x^6 +t^5 $ &   5  &  10   &   15   &25&{\tiny $\begin{pmatrix} 0 & 0 & -1 & 0 \\
 0 & 0 & -1 & -1 \\
 1 & -1 & 0 & 0 \\
 0 & 1 & 0 & 0 \end{pmatrix}$} &   \\[25pt]\hline
 $[VI]$      &  $y^2 = x \left(t^2+\alpha  t x^2+x^4\right) $ & 2  &  4   &   6   &    10  &{\tiny $\begin{pmatrix} 0 & -1 & 1 & 0 \\
  1& 0 & 0 & -1 \\
 0 & 0 & 0 & -1 \\
 0 & 0 & 1 & 0 \end{pmatrix}$}  & \\[25pt]\hline
 $[VII]$      &  $y^2 = x \left(t+x^4\right) $ & 1  &  2   &   3   &    5   & {\tiny $\begin{pmatrix} 0 & 1 & 1 & 0 \\
 1 & -1 & 0 & 1 \\
 -1 & 1 & 1 & 0 \\
 0 & -1 & 0 & 0 \end{pmatrix}$} &\\[25pt]\hline
 $[VII^{\ast}]$      &$y^2 =x \left(t^5+x^4\right) $ &   5  &  10   &15& 25  &{\tiny $\begin{pmatrix} 0 & -1 & -1 & 0 \\
 -1 & 1 & 0 & -1 \\
 1 & -1 & -1 & 0 \\
 0 & 1 & 0 & 0 \end{pmatrix}$} &  \\[25pt]\hline
$\begin{aligned} &[VIII-k]\\[-5pt]  &k=1,2\end{aligned}$
                                       &$y^2 = x^5+t^{2k-1}$ &   $\infty$  &  $\infty$   &
                                                                      $\infty$
                                                                                             &
                                                                                               $8 k -4$  &{\tiny $\begin{pmatrix} 0 & 1 & 1 & 0 \\
 1 & 0 & 0 & 1 \\
 -1 & 1 & 1 & 0 \\
 0 & -1 & 0 & 0 \end{pmatrix}^{2k-1}$} &  \\[25pt]\hline
$\begin{aligned} &[VIII-k]\\[-5pt]  &k=3,4\end{aligned}$
                                       &$y^2 = x^5+t^{2k+1}$ &   $\infty$  &  $\infty$   &
                                                                      $\infty$
                                                                                             &
                                                                                               $8 k +4$  &{\tiny $\begin{pmatrix} 0 & 1 & 1 & 0 \\
 1 & 0 & 0 & 1 \\
 -1 & 1 & 1 & 0 \\
 0 & -1 & 0 & 0 \end{pmatrix}^{2k+1}$} &  \\[25pt]\hline
$\begin{aligned} &[IX-k]\\[-5pt]  &k=1,2,3,4\end{aligned}$      &$y^2 = x^5 +t^{2k} $ &    $\infty$ &  $\infty$   &   $\infty$   & $8k$  &{\tiny $\begin{pmatrix} 0 & 1 & 1 & 1 \\
 0 & 0 & 1 & 0 \\
 0 & 0 & 0 & 1 \\
 -1 & 0 & 0 & -1 \end{pmatrix}^{k}$} &  \\[25pt]
\hline
\end{longtable}
\end{center}}

\paragraph*{Type 2 (elliptic)}

{\scriptsize
\begin{center}
\begin{longtable}{|M{2.1cm}| M{5.6cm}| M{0.5cm} |M{1cm} |M{1cm}
  |M{1.2cm}| M{3cm}| N}
\hline
Type                    &  Local model &  $\mu(I_2)$ & $\mu(I_4)$ &$\mu(I_6)$& $\mu(I_{10})$ & Monodromy &      \\
\hline
 
 $ [I_0-  I_0-m]$\footnote{
Following {\tt http://www.math.u-bordeaux1.fr/{$\sim$}qliu/articles/errata-NU.pdf}, we  corrected a typo in the local equation in the NU list.
 } $(m>0)$     &  $\left(x^3+\alpha  x+1\right)
                                \left(\beta  x t^{4 m}+t^{6
                                m}+x^3\right)$ & 0&  $4m$  &    $4m$    &     $12m$  &
{\tiny $\begin{pmatrix} 1 & 0 & 0 & 0 \\
 0 & 1 & 0 & 0 \\
 0 & 0 & 1 & 0 \\
 0 & 0 & 0 & 1 \end{pmatrix}$} &  \\[25pt]\hline
 $[I_0^{\ast}-I_0^{\ast}-m]$&$\begin{aligned} \big(t^3 &+\alpha  t^2
                              (x-1)+(x-1)^3\big)\times  \\ &\big(\beta  x t^{4 m+2}+t^{6 m+3}+x^3\big)\end{aligned}$& 0& $4+4m$  &    $4+4m$    &     $12+12m$&
{\tiny $\begin{pmatrix} -1 & 0 & 0 & 0 \\
 0 & -1 & 0 & 0 \\
 0 & 0 & -1 & 0 \\
 0 & 0 & 0 & -1 \end{pmatrix}$} &  \\[25pt]\hline
 $[I_0-I_0^{\ast}-m]$      &$\left(x^3+\alpha  x+1\right) \left(\beta  x t^{4 m+2}+t^{6 m+3}+x^3\right)$& 0&  $2+4m$&$2+4m$ &     $6+12m$  &
{\tiny $\begin{pmatrix} 1 & 0 & 0 & 0 \\
 0 & -1 & 0 & 0 \\
 0 & 0 &1 & 0 \\
 0 & 0 & 0 & -1 \end{pmatrix}$} &  \\[25pt]\hline
 $[2I_0-m]$      &$\alpha  t^{2 m+4} \left(x^2-t\right)+t^{3 m+6}+\left(x^2-t\right)^3$& 3&  $10+4m$&$13+4m$ &     $27+12m$  &
{\tiny $\begin{pmatrix}
0 & 1 & 0 & 0 \\
 1 & 0 & 0 & 0 \\
 0 & 0 & 0 & 1 \\
 0 & 0 & 1 & 0\end{pmatrix}$} &  \\[25pt]\hline
 $[2I_0^{\ast}-m]$      &$\alpha  t^{2 m+3} \left(x^2-t\right)+x t^{3 m+4}+\left(x^2-t\right)^3$& 3&  $8+4m$&$11+4m$ &     $21+12m$  &
{\tiny $\begin{pmatrix} 
0 & -1 & 0 & 0 \\
 1 & 0 & 0 & 0 \\
 0 & 0 & 0 & -1 \\
 0 & 0 & 1 & 0\end{pmatrix}$} &  \\[25pt]\hline
 $[I_0-II-m]$      &  $\left(x^2+\alpha  x+1\right) \left(t^{6 m+1}+x^3\right)$ & 0&  $1+6m$&$1+6m$ &     $2+12m$  &
{\tiny $\begin{pmatrix} 1 & 0 & 1 & 0 \\
 0 & 1 & 0 & 0 \\
 -1 & 0 & 0 & 0 \\
 0 & 0 & 0 & 1 \end{pmatrix}$} &  \\[25pt]\hline
 $[I_0-II^{\ast}-m]$      &$\left(x^2+\alpha  x+1\right) \left(t^{6 m+5}+x^3\right)$& 0&  $5+6m$&$5+6m$ &     $10+12m$  &
{\tiny $\begin{pmatrix} 0 & 0 & -1 & 0 \\
 0 & 1 & 0 & 0 \\
 1 & 0 &1 & 0 \\
 0 & 0 & 0 & 1 \end{pmatrix}$} &  \\[25pt]\hline
 $[I_0-IV-m]$      &  $\left(x^2+\alpha  x+1\right) \left(t^{6 m+2}+x^3\right)$ & 0&  $2+6m$&$2+6m$ &     $4+12m$  &
{\tiny $\begin{pmatrix} 0 & 0 & 1 & 0 \\
 0 & 1 & 0 & 0 \\
 -1 & 0 & -1 & 0 \\
 0 & 0 & 0 & 1 \end{pmatrix}$} &  \\[25pt]\hline
 $[I_0-IV^{\ast}-m]$      &$\left(x^2+\alpha  x+1\right) \left(t^{6 m+4}+x^3\right)$& 0&  $4+6m$&$4+6m$ &     $8+12m$  &
{\tiny $\begin{pmatrix} -1 & 0 & -1 & 0 \\
 0 & 1 & 0 & 0 \\
 1 & 0 &0 & 0 \\
 0 & 0 & 0 & 1 \end{pmatrix}$} &  \\[25pt]\hline
\multirow{2}{2.1cm}[-0.5em]{ $[I_0^{\ast}-II-m]$ }&$t \left(x^2+\alpha  x+1\right) \left(t^{6 m+4}+x^3\right)$& $2$&  $8+6m$&$10+6m$ & $18+12m$  &
\multirow{2}{*}{{\tiny $\begin{pmatrix} 
1 & 0 & 1 & 0 \\
 0 & -1 & 0 & 0 \\
 -1 & 0 &0 & 0 \\
 0 & 0 & 0 & -1\end{pmatrix}$}} & \\[8pt] \cline{2-6}
&  $\left(t^{6 m+1}+(x-1)^3\right) \left(t^3+\alpha  t^2 x+x^3\right)$& $0$&  $3+6m$&$3+6m$ & $8+12m$  & &  \\[8pt]\hline
\multirow{2}{2.1cm}[-0.5em]{ $[I_0^{\ast}-II^{\ast}-m]$ }&$t \left(x^2+\alpha  x+1\right) \left(t^{6 m+8}+x^3\right)$& $2$&  $12+6m$&$14+6m$ & $26+12m$  &
\multirow{2}{*}{{\tiny $\begin{pmatrix} 
 0 & 0 & -1 & 0 \\
 0 & -1 & 0 & 0 \\
 1 & 0 &1 & 0 \\
 0 & 0 & 0 & -1 \end{pmatrix}$}} & \\*[8pt] \cline{2-6}
& $\left(t^{6 m+5}+(x-1)^3\right) \left(t^3+\alpha  t^2 x+x^3\right)$& $0$&  $7+6m$&$7+6m$ & $16+12m$  & &  \\[8pt]\hline
 $[I_0^{\ast}-II^{\ast}-\alpha]$ &$t \left(t^2+x^3\right) \left(x^2+\alpha  x+1\right)$& 2&  6&8 &  14 &
{\tiny $\begin{pmatrix} 0 & 0 & -1 & 0 \\
 0 & -1 & 0 & 0 \\
 1 & 0 &1 & 0 \\
 0 & 0 & 0 & -1 \end{pmatrix}$} &  \\[25pt]\hline
\multirow{2}{2.1cm}[-0.5em]{ $[I_0^{\ast}-IV-m]$ }&$t \left(x^2+\alpha  x+1\right) \left(t^{6 m+5}+x^3\right)$& $2$&  $9+6m$&$11+6m$ & $20+12m$  &
\multirow{2}{*}{{\tiny $\begin{pmatrix} 
 0 & 0 & 1 & 0 \\
 0 & -1 & 0 & 0 \\
 -1 & 0 &-1 & 0 \\
 0 & 0 & 0 & -1 \end{pmatrix}$}} & \\*[8pt] \cline{2-6}
& $\left(t^{6 m+2}+(x-1)^3\right) \left(t^3+\alpha  t^2 x+x^3\right)$& $0$&  $4+6m$&$4+6m$ & $10+12m$  & &  \\[8pt]\hline
\multirow{2}{2.1cm}[-0.5em]{ $[I_0^{\ast}-IV^{\ast}-m]$  }&$t \left(x^2+\alpha  x+1\right) \left(t^{6 m+7}+x^3\right)$& $2$&  $11+6m$&$13+6m$ & $24+12m$  &
\multirow{2}{*}{{\tiny $\begin{pmatrix} 
-1 & 0 & -1 & 0 \\
 0 & -1 & 0 & 0 \\
 1 & 0 & 0& 0 \\
 0 & 0 & 0 & -1 \end{pmatrix}$}} & \\*[8pt] \cline{2-6}
& $\left(t^{6 m+4}+(x-1)^3\right) \left(t^3+\alpha  t^2 x+x^3\right)$&
                                                                       $0$&
                                                                            $6+6m$&$6+6m$
                                                                             &
                                                                               $14+12m$
                                                                                             &
                                                                                                         &  \\[8pt]\hline 
 $[I_0^{\ast}-IV^{\ast}-\alpha]$ & $t \left(t+x^3\right) \left(x^2+\alpha  x+1\right)$ & 2&  5&7&  12&
{\tiny $\begin{pmatrix} -1 & 0 & -1 & 0 \\
 0 & -1 & 0 & 0 \\
 1 & 0 & 0& 0 \\
 0 & 0 & 0 & -1 \end{pmatrix}$} &  \\[25pt]\hline
 $[I_0-III-m]$      &  $x \left(x^2+\alpha  x+1\right) \left(t^{4 m+1}+x^2\right)$ & 0&  $1+4m$&$1+4m$ &     $3+12m$  &
{\tiny $\begin{pmatrix} 0 & 0 & 1 & 0 \\
 0 & 1 & 0 & 0 \\
 -1 & 0 & 0 & 0 \\
 0 & 0 & 0 & 1 \end{pmatrix}$} &  \\[25pt]\hline
 $[I_0-III^{\ast}-m]$      &$x \left(x^2+\alpha  x+1\right) \left(t^{4 m+3}+x^2\right)$& 0&  $3+4m$&$3+4m$ &     $9+12m$  &
{\tiny $\begin{pmatrix} 0 & 0 & -1 & 0 \\
 0 & 1 & 0 & 0 \\
 1 & 0 &0 & 0 \\
 0 & 0 & 0 & 1 \end{pmatrix}$} &  \\[25pt]\hline
\multirow{2}{2.1cm}[-0.5em]{ $[I_0^{\ast}-III-m]$  }&$t x \left(x^2+\alpha  x+1\right) \left(t^{4 m+3}+x^2\right)$& $2$&  $7+4m$&$9+4m$ & $19+12m$  &
\multirow{2}{*}{{\tiny $\begin{pmatrix} 
 0 & 0 & 1 & 0 \\
 0 & -1 & 0 & 0 \\
 -1 & 0 & 0 & 0 \\
 0 & 0 & 0 & -1\end{pmatrix}$}} & \\*[8pt] \cline{2-6}
& $(x-1)  \left(t^{4 m+1}+(x-1)^2\right)
  \left(t^3+\alpha  t^2 x+x^3\right)$& $0$&  $3+4m$&$3+4m$ & $9+12m$  & &  \\[8pt]\hline
\multirow{2}{2.1cm}[-1em]{ $[I_0^{\ast}-III^{\ast}-m]$   }&$t x \left(x^2+\alpha  x+1\right) \left(t^{4 m+5}+x^2\right)$& $2$&  $9+4m$&$11+4m$ & $25+12m$  &
\multirow{2}{*}{{\tiny $\begin{pmatrix} 
 0 & 0 & -1 & 0 \\
 0 & -1 & 0 & 0 \\
 1 & 0 & 0 & 0 \\
 0 & 0 & 0 & -1\end{pmatrix}$}} & \\*[8pt] \cline{2-6}
& $(x-1) \left(t^{4 m+3}+(x-1)^2\right) \left(t^3+\alpha  t^2 x+x^3\right)$& $0$&  $5+4m$&$5+4m$ & $15+12m$  & &  \\[8pt]\hline
 $[I_0^{\ast}-III^{\ast}-\alpha]$  &$t x \left(t+x^2\right) \left(x^2+\alpha  x+1\right)$& 2&  $5$&$7$ &     $13$  &
{\tiny $\begin{pmatrix} 0 & 0 & -1 & 0 \\
 0 & -1 & 0 & 0 \\
 1 & 0 & 0 & 0 \\
 0 & 0 & 0 & -1\end{pmatrix}$} &  \\[25pt] \hline
 $[2II-m]$  &$x t^{3 m+3}+\left(x^2-t\right)^3$& 3&  $7+6m$&$10+6m$ &     $17+12m$  &
{\tiny $\begin{pmatrix} 0 & 1 & 0 & 1 \\
 1 & 0 & 0 & 0 \\
 0 & -1 & 0 & 0 \\
 0 & 0 & 1 & 0\end{pmatrix}$} &  \\[25pt]\hline
 $[2II^{\ast}-m]$  &$x t^{3 m+5}+\left(x^2-t\right)^3$& 3&  $11+6m$&$14+6m$ &     $25+12m$  &
{\tiny $\begin{pmatrix} 0 & 0 & 0 & -1 \\
 1 & 0 & 0 & 0 \\
 0 & 1 & 0 & 1 \\
 0 & 0 & 1 & 0\end{pmatrix}$} &  \\[25pt]\hline
 $[II-II-m]$  &$\left(t+(x-1)^3\right) \left(t^{6 m+1}+x^3\right)$& 0&  $2+6m$&$2+6m$ &     $4+12m$  &
{\tiny $\begin{pmatrix} 1 & 0 & 1 & 0 \\
 0 & 1 & 0 & 1 \\
 -1 & 0 & 0 & 0 \\
 0 & -1 & 0 & 0\end{pmatrix}$} &  \\[25pt]\hline
 $[II-II^{\ast}-m]$  &$\left(t^5+(x-1)^3\right) \left(t^{6 m+1}+x^3\right) $& 0&  $6+6m$&$6+6m$ &     $12+12m$  &
{\tiny $\begin{pmatrix} 
 1 & 0 & 1 & 0 \\
 0 & 0 & 0 & -1 \\
 -1 & 0 & 0 & 0 \\
 0 & 1 & 0 & 1\end{pmatrix}$} &  \\[25pt]\hline
\multirow{2}{2.1cm}[-1em]{ $[II^{\ast}-II^{\ast}-m]$     }&$ \left(t^5+(x-1)^3\right) \left(t^{6 m+5}+x^3\right)$& $0$&  $10+6m$&$10+6m$ & $20+12m$  &
\multirow{2}{*}{{\tiny $\begin{pmatrix} 
 0 & 0 & -1 & 0 \\
 0 & 0 & 0 & -1 \\
 1 & 0 & 1 & 0 \\
 0 & 1 & 0 & 1\end{pmatrix}$}} & \\*[8pt] \cline{2-6}
&  $t \left(t^2+(x-1)^3\right) \left(t^{6 m+8}+x^3\right)$& $2$&  $14+6m$&$16+6m$ & $30+12m$  & &  \\[8pt]\hline
 $[II^{\ast}-II^{\ast}-\alpha]$  &$ t \left(t^2+x^3\right) \left(t^2+(x-1)^3\right)$& $2$&  $8$&$10$ & $18$  &
{\tiny $\begin{pmatrix} 
0 & 0 & -1 & 0 \\
 0 & 0 & 0 & -1 \\
 1 & 0 & 1 & 0 \\
 0 & 1 & 0 & 1\end{pmatrix}$} &  \\[25pt]\hline
 $[II-IV-m]$  &$\left(t^2+(x-1)^3\right) \left(t^{6 m+1}+x^3\right) $& 0&  $3+6m$&$3+6m$ & $6+12m$  &
{\tiny $\begin{pmatrix} 
0 & 0 & 1 & 0 \\
 0 & 1 & 0 & 1 \\
 -1 & 0 &-1 & 0 \\
 0 & -1 & 0 & 0\end{pmatrix}$} &  \\[25pt]\hline
 $[II-IV^{\ast}-m]$  &$\left(t^4+(x-1)^3\right) \left(t^{6 m+1}+x^3\right) $& 0&  $5+6m$&$5+6m$ & $10+12m$  &
{\tiny $\begin{pmatrix} 
1 & 0 & 1 & 0 \\
 0 & -1 & 0 & -1 \\
 -1 & 0 & 0 & 0 \\
 0 & 1 & 0 & 0\end{pmatrix}$} &  \\[25pt]\hline
 $[II^{\ast}-IV-m]$  &$\left(t^2+(x-1)^3\right) \left(t^{6 m+5}+x^3\right) $& 0&  $7+6m$&$7+6m$ & $14+12m$  &
{\tiny $\begin{pmatrix} 
0 & 0 & -1 & 0 \\
 0 & 0 & 0 & 1 \\
 1& 0 & 1 & 0 \\
 0 & -1 & 0 & -1\end{pmatrix}$} &  \\[25pt]\hline
 $[II^{\ast}-IV-\alpha]$  &$\left(t+x^3\right) \left(t^2+x^3\right) $& 2&  $5$&$7$ & $12$  &
{\tiny $\begin{pmatrix} 
0 & 0 & -1 & 0 \\
 0 & 0 & 0 & 1 \\
 1& 0 & 1 & 0 \\
 0 & -1 & 0 & -1\end{pmatrix}$} &  \\[25pt]\hline
\multirow{2}{2.1cm}[-1em]{$[II^{\ast}-IV^{\ast}-m]$     }&$ \left(t^4+(x-1)^3\right) \left(t^{6 m+5}+x^3\right)$& $0$&  $9+6m$&$9+6m$ & $18+12m$  &
\multirow{2}{*}{{\tiny $\begin{pmatrix} 
 0 & 0 & -1 & 0 \\
 0 & -1 & 0 & -1 \\
 1 & 0 & 1 & 0 \\
 0 & 1 & 0 & 0\end{pmatrix}$}} & \\*[8pt] \cline{2-6}
&$t \left(t^2+(x-1)^3\right) \left(t^{6 m+7}+x^3\right)$& $2$&
                                                                   $13+6m$&$15+6m$
                                                                             &
                                                                               $28+12m$  & &  \\[8pt]\hline
 $[II^{\ast}-IV^{\ast}-\alpha]$  &$t \left(t+x^3\right) \left(t^2+(x-1)^3\right)$& 2&  $7$&$9$ & $16$  &
{\tiny $\begin{pmatrix} 
0 & 0 & -1 & 0 \\
 0 & -1 & 0 & -1 \\
 1 & 0 & 1 & 0 \\
 0 & 1 & 0 & 0\end{pmatrix}$} &  \\[25pt]\hline
 $[2IV-m]$  &$ t^{3 m+4}+\left(x^2-t\right)^3$& $3$&  $8+6m$&$11+6m$ & $19+12m$  &
{\tiny $\begin{pmatrix} 
0 & 0 & 0 & 1 \\
 1 & 0 & 0 & 0 \\
 0 & -1 & 0 & -1 \\
 0 & 0 & 1 & 0\end{pmatrix}$} &  \\[25pt]\hline
 $[2IV^{\ast}-m]$  &$t^{3 m+5}+\left(x^2-t\right)^3 $& $3$&  $10+6m$&$13+6m$ & $23+12m$  &
{\tiny $\begin{pmatrix} 
0 & -1 & 0 & -1 \\
 1 & 0 & 0 & 0 \\
 0 & 1 & 0 & 0 \\
 0 & 0 & 1 & 0\end{pmatrix}$} &  \\[25pt]\hline
 $[IV-IV-m]$  &$ \left(t^2+(x-1)^3\right) \left(x^3+t^{6 m+2}\right)$& 0&  $4+6m$&$4+6m$ & $8+12m$  &
{\tiny $\begin{pmatrix} 
0 & 0 & 1 & 0 \\
 0 & 0 & 0 & 1 \\
 -1 & 0 & -1 & 0 \\
 0 & -1 & 0 & -1\end{pmatrix}$} &  \\[25pt]\hline
 $[IV-IV^{\ast}-m]$  &$\left(t^4+(x-1)^3\right) \left(x^3+t^{6 m+2}\right) $& 0&  $6+6m$&$6+6m$ & $12+12m$  &
{\tiny $\begin{pmatrix} 
0 & 0 & 1 & 0 \\
 0 & -1 & 0 & -1 \\
 -1 & 0 & -1 & 0 \\
 0 & 1 & 0 & 0\end{pmatrix}$} &  \\[25pt]\hline
\multirow{2}{2.1cm}[0.2em]{ $[IV^{\ast}-IV^{\ast}-m]$}&$ t \left(t+(x-1)^3\right) \left(x^3+t^{6 m+7}\right)$& $2$&  $12+6m$&$14+6m$ & $26+12m$  &
\multirow{2}{*}{{\tiny $\begin{pmatrix} 
 -1 & 0 & -1 & 0 \\
 0 & -1 & 0 & -1 \\
 1 & 0 & 0 & 0 \\
 0 & 1 & 0 & 0\end{pmatrix}$}} & \\*[8pt] \cline{2-6}
&  $\left(t^4+(x-1)^3\right) \left(t^{6 m+4}+x^3\right)$& $0$&
                                                                 $8+6m$&$8+6m$
                                                                             &
                                                                               $16+12m$  & &  \\[8pt]\hline
 $[IV^{\ast}-IV^{\ast}-\alpha]$  &$t \left(t+x^3\right) \left(t+(x-1)^3\right) $& $2$&  $6$&$8$ & $14$  &
{\tiny $\begin{pmatrix} 
-1 & 0 & -1 & 0 \\
 0 & -1 & 0 & -1 \\
 1 & 0 & 0 & 0 \\
 0 & 1 & 0 & 0\end{pmatrix}$} &  \\[25pt]\hline
 $[II-III-m]$  &$ x \left(t+(x-1)^3\right) \left(t^{4 m+1}+x^2\right)$& $0$&  $2+4m$&$2+4m$ & $5+12m$  &
{\tiny $\begin{pmatrix} 
1 & 0 & 1 & 0 \\
 0 & 0 & 0 & 1 \\
 -1 & 0 & 0 & 0 \\
 0 & -1 & 0 & 0\end{pmatrix}$} &  \\[25pt]\hline
 $[II-III^{\ast}-m]$  &$x \left(t+(x-1)^3\right) \left(t^{4
                        m+3}+x^2\right)$ &0& $4+4m$ &  $4+4m$& $11+12m$  &
{\tiny $\begin{pmatrix} 
1 & 0 & 1 & 0 \\
 0 & 0 & 0 & -1 \\
 -1 & 0 & 0 & 0 \\
 0 & 1 & 0 & 0\end{pmatrix}$} &  \\[25pt]\hline
 $[II^{\ast}-III-m]$  &$x \left(t^5+(x-1)^3\right) \left(t^{4 m+1}+x^2\right) $& $0$&  $6+4m$&$6+4m$ & $13+12m$  &
{\tiny $\begin{pmatrix} 
0 & 0 & -1 & 0 \\
 0 & 0 & 0 & 1 \\
 1 & 0 & 1 & 0 \\
 0 & -1 & 0 & 0\end{pmatrix}$} &  \\[25pt]\hline
 $[II^{\ast}-III-\alpha]$  &$\left(t+x^2\right) \left(t^2+x^3\right) $& $2$&  $5$&$7$ & $11$  &
{\tiny $\begin{pmatrix} 
0 & 0 & -1 & 0 \\
 0 & 0 & 0 & 1 \\
 1 & 0 & 1 & 0 \\
 0 & -1 & 0 & 0\end{pmatrix}$} &  \\[25pt]\hline
\multirow{2}{2.1cm}[-0.5em]{$[II^{\ast}-III^{\ast}-m]$}&$x
                                \left(t^5+(x-1)^3\right) \left(t^{4
                                m+3}+x^2\right)$& $0$&  $8+4m$&$8+4m$ & $19+12m$  &
\multirow{2}{*}{{\tiny $\begin{pmatrix} 
0 & 0 & -1 & 0 \\
 0 & 0 & 0 & -1 \\
 1 & 0 & 1 & 0 \\
 0 & 1 & 0 & 0\end{pmatrix}$}} & \\*[8pt] \cline{2-6}
& $t x \left(t^2+(x-1)^3\right) \left(t^{4 m+5}+x^2\right)$& $2$&  $11+4m$&$13+4m$ & $29+12m$  & &  \\[8pt]\hline
 $[II^{\ast}-III^{\ast}-\alpha]$  &$t x \left(t+x^2\right) \left(t^2+(x-1)^3\right) $& $2$&  $7$&$9$ & $17$  &
{\tiny $\begin{pmatrix} 
0 & 0 & -1 & 0 \\
 0 & 0 & 0 & -1 \\
 1 & 0 & 1 & 0 \\
 0 & 1 & 0 & 0\end{pmatrix}$} &  \\[25pt]\hline 
 $[IV-III-m]$  &$x \left(t^2+(x-1)^3\right) \left(t^{4 m+1}+x^2\right) $& $0$&  $3+4m$&$3+4m$ & $7+12m$  &
{\tiny $\begin{pmatrix} 
0 & 0 & 1 & 0 \\
 0 & 0 & 0 & 1 \\
 -1 & 0 & -1 & 0 \\
 0 & -1 & 0 & 0\end{pmatrix}$} &  \\[25pt]\hline
 $[IV-III^{\ast}-m]$  &$x \left(t^2+(x-1)^3\right) \left(t^{4 m+3}+x^2\right) $& $0$&  $5+4m$&$5+4m$ & $13+12m$  &
{\tiny $\begin{pmatrix} 
0 & 0 & 1 & 0 \\
 0 & 0 & 0 & -1 \\
 -1 & 0 & -1 & 0 \\
 0 & 1 & 0 & 0\end{pmatrix}$} &  \\[25pt]\hline
 $[IV-III^{\ast}-\alpha]$  &$x \left(t+x^2\right) \left(t+x^3\right) $& $2$&  $5$&$7$ & $11$  &
{\tiny $\begin{pmatrix} 
0 & 0 & 1 & 0 \\
 0 & 0 & 0 & -1 \\
 -1 & 0 & -1 & 0 \\
 0 & 1 & 0 & 0\end{pmatrix}$} &  \\[25pt]\hline
 $[IV^{\ast}-III-m]$  &$x \left(t^4+(x-1)^3\right) \left(t^{4 m+1}+x^2\right)$& $0$&  $5+4m$&$5+4m$ & $11+12m$  &
{\tiny $\begin{pmatrix} 
-1 & 0 & -1 & 0 \\
 0 & 0 & 0 & 1 \\
 1 & 0 & 0 & 0 \\
 0 & -1 & 0 & 0\end{pmatrix}$} &  \\[25pt]\hline
\multirow{2}{2.1cm}[-0.5em]{$[IV^{\ast}-III^{\ast}-m]$}&$x \left(t^4+(x-1)^3\right) \left(t^{4 m+3}+x^2\right)$& $0$&  $7+4m$&$7+4m$ & $17+12m$  &
\multirow{2}{*}{{\tiny $\begin{pmatrix} 
-1 & 0 & -1 & 0 \\
 0 & 0 & 0 & -1 \\
 1 & 0 & 0 & 0 \\
 0 & 1 & 0 & 0\end{pmatrix}$}} & \\*[8pt] \cline{2-6}
& $t x \left(t+(x-1)^3\right) \left(t^{4 m+5}+x^2\right) $& $2$&  $10+4m$&$12+4m$ & $27+12m$  & &  \\[8pt]\hline
 $[IV^{\ast}-III^{\ast}-\alpha]$  &$t x \left(t+x^2\right) \left(t+(x-1)^3\right)$& $2$&  $6$&$8$ & $15$  &
{\tiny $\begin{pmatrix} 
-1 & 0 & -1 & 0 \\
 0 & 0 & 0 & -1 \\
 1 & 0 & 0 & 0 \\
 0 & 1 & 0 & 0\end{pmatrix}$} &  \\[25pt]\hline
 $[2III-m]$  &$ \left(x^2-t\right) \left(x t^{2 m+2}+\left(x^2-t\right)^2\right)$& 3&  $7+4m$&$10+4m$ & $18+12m$  &
{\tiny $\begin{pmatrix} 
0 & 0 & 0 & 1 \\
 1 & 0 & 0 & 0 \\
 0 & -1& 0 & 0 \\
 0 & 0 & 1 & 0\end{pmatrix}$} &  \\[25pt]\hline
 $[2III^{\ast}-m]$  &$\left(x^2-t\right) \left(x t^{2 m+3}+\left(x^2-t\right)^2\right)$& $3$&  $9+4m$&$12+4m$ & $24+12m$  &
{\tiny $\begin{pmatrix} 
0 & 0 & 0 & -1\\
 1 & 0 & 0 & 0 \\
 0 & 1 & 0 & 0 \\
 0 & 0 & 1 & 0\end{pmatrix}$} &  \\[25pt]\hline
 $[III-III-m]$  &$x (x-1) \left(t+(x-1)^2\right) \left(t^{4 m+1}+x^2\right) $& $0$&  $2+4m$&$2+4m$ & $6+12m$  &
{\tiny $\begin{pmatrix} 
0 & 0 & 1 & 0 \\
 0 & 0 & 0 & 1 \\
 -1 & 0 & 0 & 0 \\
 0 & -1 & 0 & 0\end{pmatrix}$} &  \\[25pt]\hline
 $[III-III^{\ast}-m]$  &$x (x-1) \left(t^3+(x-1)^2\right) \left(t^{4 m+1}+x^2\right) $& $0$&  $4+4m$&$4+4m$ & $12+12m$  &
{\tiny $\begin{pmatrix} 
0 & 0 & 1 & 0 \\
 0 & 0 & 0 & -1 \\
 -1& 0 & 0 & 0 \\
 0 & 1 & 0 & 0\end{pmatrix}$} &  \\[25pt]\hline
\multirow{2}{2.1cm}[-0.5em]{\hspace{-0.05cm}$[III^{\ast}-III^{\ast}-m]$}&$x (x-1) \left(t^3+(x-1)^2\right) \left(t^{4 m+3}+x^2\right)$& $0$&  $6+4m$&$6+4m$ & $18+12m$  &
\multirow{2}{*}{{\tiny $\begin{pmatrix} 
0 & 0 & -1 & 0 \\
 0 & 0 & 0 & -1 \\
 1 & 0 & 0 & 0 \\
 0 & 1 & 0 & 0\end{pmatrix}$}} & \\[8pt] \cline{2-6}
&$t x (x-1) \left(t+(x-1)^2\right) \left(t^{4 m+5}+x^2\right)$& $2$&  $10+4m$&$12+4m$ & $28+12m$  & &  \\[8pt]\hline
 $[III^{\ast}-III^{\ast}-\alpha]$  &$ t x (x-1) \left(t+x^2\right) \left(t+(x-1)^2\right)$& $2$&  $6$&$8$ & $16$  &
{\tiny $\begin{pmatrix} 
0 & 0 & -1 & 0 \\
 0 & 0 & 0 & -1 \\
 1 & 0 & 0 & 0 \\
 0 & 1 & 0 & 0\end{pmatrix}$} &  \\[25pt]\hline
\end{longtable}
\end{center}}

\paragraph*{Type 3 (parabolic)}

{\scriptsize
\begin{center}
\begin{longtable}{|M{2cm}| M{5cm}| M{1.cm} |M{1.2cm} |M{1.2cm}
  |M{1.2cm}| M{3cm}| N}
\hline
Type                    &  Local model &  $\mu(I_2)$ & $\mu(I_4)$
  &$\mu(I_6)$& $\mu(I_{10})$ & Monodromy &      \\
\hline
 $[I_{n-0-0}]$ $(n>0)$  &$\left(x^3+\alpha  x+1\right) \left(t^n+(x-\beta )^2\right) $& 0&  $0$&$0$ & $n$  &
{\tiny $\begin{pmatrix} 
1 & 0 & 0 & 0 \\
 0 &1 & 0 & n \\
 0 & 0 & 1 & 0 \\
 0 & 0 & 0 & 1\end{pmatrix}$} &  \\[25pt]\hline
 $[I_n-I_0-m]$ $(n, m>0)$  &$ \left(t^n+(x-1)^2\right) \left(\alpha  x t^{4 m}+t^{6 m}+x^3\right)$& $0$&  $4m$&$4m$ & $n+12m$  &
{\tiny $\begin{pmatrix} 
1 & 0 & 0 & 0 \\
 0 & 1 & 0 & n \\
 0 & 0 & 1 & 0 \\
 0 & 0 & 0 & 1\end{pmatrix}$} &  \\[25pt]\hline
 $[I_0-I_n^{\ast}-m]$  &$\begin{aligned} &( t+x)\big(t^{n+2}+x^2\big)  \times \\ &\left(\alpha  (x-1) t^{4 m}+t^{6 m}+(x-1)^3\right) \end{aligned}$& $0$&  $2+4m$&$2+4m$ & $6+n+12m$  &
{\tiny $\begin{pmatrix} 
 1 & 0 & 0 & 0 \\
 0 & -1 & 0 & -n \\
 0 & 0 & 1 & 0 \\
 0 & 0 & 0 & -1\end{pmatrix}$} &  \\[25pt]\hline
 $[I_n-I_0^{\ast}-m]$  &$\hspace{-0.15cm}\left(t^n+(x-1)^2\right) \left(\alpha  x t^{4 m+2}+t^{6 m+3}+x^3\right) $& $0$&  $2+4m$&$2+4m$ & $6+n+12m$  &
{\tiny $\begin{pmatrix} 
-1 & 0 & 0 & 0 \\
 0 & 1 & 0 & n \\
 0 & 0 & -1 & 0 \\
 0 & 0 & 0 & 1\end{pmatrix}$} &  \\[25pt]\hline
 $[I_{n-0-0}^{\ast}]$  $(n>0)$   &$t \left(x^3+\alpha  x+1\right) \left(t^n+(x-\beta )^2\right) $& $2$&  $4$&$6$ & $10+n$  &
{\tiny $\begin{pmatrix} 
-1 & 0 & 0 & 0 \\
 0 & -1 & 0 & -n \\
 0 & 0 & -1 & 0 \\
 0 & 0 & 0 & -1\end{pmatrix}$} &  \\[25pt]\hline
 $[I_0^{\ast}-I_n^{\ast}-m]$  &$\begin{aligned}&(t+x)
   \left(t^{n+2}+x^2\right) \times \\ &\left(\alpha  (x-1) t^{4 m+2}+t^{6 m+3}+(x-1)^3\right)\end{aligned}$& $0$&  $4+4m$&$4+4m$ & $12+n+12m$  &
{\tiny $\begin{pmatrix} 
-1& 0 & 0 & 0 \\
 0 & -1 & 0 & -n \\
 0 & 0 & -1 & 0 \\
 0 & 0 & 0 & -1\end{pmatrix}$} &  \\[25pt]\hline
 $[II_{n-0}]$  &$\left(t^{n-1}+(x-1)^2\right) \left(t^2+\alpha  t x^2+x^4\right) $& $1$&  $2$&$3$ & $5+n$  &
{\tiny $\begin{pmatrix} 
-1 & 0 & 0 & 0 \\
 -1 & 1 & 0 & n \\
 0 & 0 & -1 & -1 \\
 0 & 0 & 0 & 1\end{pmatrix}$} &  \\[25pt]\hline
 $[II_{n-0}^{\ast}]$  &$ t \left(t^{n-1}+(x-1)^2\right) \left(t^2+\alpha  t x^2+x^4\right)$& $3$&  $6$&$9$ & $15+n$  &
{\tiny $\begin{pmatrix} 
1 & 0 & 0 & 0 \\
1 & -1 & 0 & -n \\
 0 & 0 & 1 & 1 \\
 0 & 0 & 0 & -1\end{pmatrix}$} &  \\[25pt]\hline 
 $[II-I_n-m]$  &$\left(t^{6 m+1}+x^3\right) \left(t^n+(x-1)^2\right) $& $0$&  $1+n+6m$&$1+6m$ & $2+n+12m$  &
{\tiny $\begin{pmatrix} 
1& 0 & 1 & 0 \\
 0 & 1 & 0 & n \\
 -1 & 0 & 0 & 0 \\
 0 & 0 & 0 & 1 \end{pmatrix}$} &  \\[25pt]\hline 
 $[II^{\ast}-I_n-m]$  &$\left(t^{6 m+5}+x^3\right) \left(t^n+(x-1)^2\right) $& $0$&  $5+n+6m$&$5+6m$ & $10+n+12m$  &
{\tiny $\begin{pmatrix} 
0 & 0 & -1 & 0 \\
 0 & 1 & 0 & n \\
 1& 0 & 1 & 0 \\
 0 & 0 & 0 & 1\end{pmatrix}$} &  \\[25pt]\hline
 $[IV-I_n-m]$  &$\left(t^{6 m+2}+x^3\right) \left(t^n+(x-1)^2\right) $& $0$&  $2+n+6m$&$2+6m$ & $4+n+12m$  &
{\tiny $\begin{pmatrix} 
 0 & 0 & 1 & 0 \\
 0 & 1 & 0 & n \\
 -1 & 0 & -1 & 0 \\
 0 & 0 & 0 & 1\end{pmatrix}$} &  \\[25pt]\hline
 $[IV^{\ast}-I_n-m]$  &$ \left(t^{6 m+4}+x^3\right) \left(t^n+(x-1)^2\right)$& $0$&  $4+n+6m$&$4+6m$ & $8+n+12m$  &
{\tiny $\begin{pmatrix} 
 -1 & 0 & -1 & 0 \\
 0 & 1 & 0 & n \\
 1 & 0 & 0 & 0 \\
 0 & 0 & 0 & 1\end{pmatrix}$} &  \\[25pt]\hline
 $[II-I_n^{\ast}-m]$  &$(t+x) \left(t^{6 m+1}+(x-1)^3\right) \left(t^{n+2}+x^2\right) $& $0$&  $3+6m$&$3+6m$ & $8+n+12m$  &
{\tiny $\begin{pmatrix} 
1 & 0 & 1 & 0 \\
 0 & -1 & 0 & -n \\
 -1 & 0 & 0 & 0 \\
 0 & 0 & 0 & -1\end{pmatrix}$} &  \\[25pt]\hline
 $[II^{\ast}-I_n^{\ast}-m]$  &$(t+x) \left(t^{6 m+5}+(x-1)^3\right) \left(t^{n+2}+x^2\right) $& 0&  $7+6m$&$7+6m$ & $16+n+12m$  &
{\tiny $\begin{pmatrix} 
0 & 0 & -1 & 0 \\
 0 & -1 & 0 &-n \\
 1 & 0 & 1 & 0 \\
 0 & 0 & 0 & -1\end{pmatrix}$} &  \\[25pt]\hline
 $[II^{\ast}-I_n^{\ast}-\alpha]$  &$ t \left(t^2+x^3\right) \left(t^n+(x-1)^2\right)$& $2$&  $6+n$&$8$ & $14+n$  &
{\tiny $\begin{pmatrix} 
0 & 0 & -1 & 0 \\
 0 & -1 & 0 &-n \\
 1 & 0 & 1 & 0 \\
 0 & 0 & 0 & -1\end{pmatrix}$} &  \\[25pt]\hline
 $[IV-I_n^{\ast}-m]$  &$(t+x) \left(t^{6 m+2}+(x-1)^3\right) \left(t^{n+2}+x^2\right)$& $0$&  $4+6m$&$4+6m$ & $10+n+12m$  &
{\tiny $\begin{pmatrix} 
0 & 0 & 1 & 0 \\
 0 & -1 & 0 & -n \\
 -1 & 0 & -1 & 0 \\
 0 & 0 & 0 & -1\end{pmatrix}$} &  \\[25pt]\hline
 $[IV^{\ast}-I_n^{\ast}-m]$  &$(t+x) \left(t^{6 m+4}+(x-1)^3\right) \left(t^{n+2}+x^2\right) $& $0$&  $6+6m$&$6+6m$ & $14+n+12m$  &
{\tiny $\begin{pmatrix} 
-1 & 0 & -1 & 0 \\
 0 & -1 & 0 & -n \\
 1 & 0 & 0 & 0 \\
 0 & 0 & 0 & -1\end{pmatrix}$} &  \\[25pt]\hline
 $[IV^{\ast}-I_n^{\ast}-\alpha]$  &$ t \left(t+x^3\right) \left(t^n+(x-1)^2\right)$& $2$&  $5+n$&$7$ & $12+n$  &
{\tiny $\begin{pmatrix} 
-1 & 0 & -1 & 0 \\
 0 & -1 & 0 & -n \\
 1 & 0 & 0 & 0 \\
 0 & 0 & 0 & -1\end{pmatrix}$} &  \\[25pt]\hline
 $[IV-II_n]$  &$x \left(t+x^3\right) \left(t^n+(x-1)^2\right) $& $1$&  $2+n$&$2$ & $4+n$  &
{\tiny $\begin{pmatrix} 
0 & 0 & 1 & 0 \\
 0 & 1 & -1 & n+1 \\
 -1 & 0 & -1 & 1 \\
 0 & 0 & 0 & 1\end{pmatrix}$} &  \\[25pt]\hline
\multirow{2}{2.1cm}[-0.5em]{$[IV^{\ast}-II_n]$}&$n=0: $ $ \left(t^3+x^2\right) \left(t^4+x^3\right) $& $6$&  $11$&$16$ & $27$  &
\multirow{2}{*}{{\tiny $\begin{pmatrix} 
-1 & 0 & -1 & -1 \\
 -1& 1 & 0 & n \\
 1 & 0 & 0 & 0 \\
 0 & 0 & 0 & 1\end{pmatrix}$}} & \\*[8pt] \cline{2-6}
&$n>0 : $ $ x \left(t^2+x^3\right) \left(t^{n-1}+(x-1)^2\right)$& $2$&  $3+n$&$4$ & $7+n$  & &  \\[8pt]\hline
 $[II-II^{\ast}_n]$ $(n\geq0)$ &$\left(t+x^3\right) \left(t^{n+1}+x^2\right) $& $2+2n$&  $3+n$&$4$ & $7+n$  &
{\tiny $\begin{pmatrix} 
1 & 0 & 1 & 1 \\
 1 & -1 & 0 & -n \\
 -1 & 0 & 0 & 0 \\
 0 & 0 & 0 & -1\end{pmatrix}$} &  \\[25pt]\hline 
 $[II^{\ast}-II^{\ast}_n]$ $(n\geq0)$  &$ t x \left(t+x^3\right) \left(t^n+(x-1)^2\right)$& $3$&  $6+n$&$8$ & $14+n$  &
{\tiny $\begin{pmatrix} 
0 & 0 & -1 & 0 \\
 0 & -1 & 1 & -n \\
 1 & 0 & 1 & -1 \\
 0 & 0 & 0 & -1\end{pmatrix}$} &  \\[25pt]\hline
 $[III-I_n-m]$  &$ x \left(t^{4 m+1}+x^2\right) \left(t^n+(x-1)^2\right)$& $0$&  $1+4m$&$1+4m$ & $3+n+12m$  &
{\tiny $\begin{pmatrix} 
0 & 0 & 1 & 0 \\
 0 & 1 & 0 & n \\
 -1 & 0 & 0 & 0 \\
 0 & 0 & 0 & 1\end{pmatrix}$} &  \\[25pt]\hline
 $[III^*-I_n-m]$  &$x \left(t^{4 m+3}+x^2\right) \left(t^n+(x-1)^2\right) $& $0$&  $3+4m$&$3+4m$ & $9+n+12m$  &
{\tiny $\begin{pmatrix} 
0 & 0 & -1 & 0 \\
 0 & 1 & 0 & n \\
 1 & 0 & 0 & 0 \\
 0 & 0 & 0 & 1\end{pmatrix}$} &  \\[25pt]\hline
 $[III-I_n^{\ast}-m]$  &$\begin{aligned}&(x-1) (t+x) \left(t^{4
       m+1}+(x-1)^2\right) \times \\ &\left(t^{n+2}+x^2\right) \end{aligned}$& 0&  $3+4m$&$3+4m$ & $9+n+12m$  &
{\tiny $\begin{pmatrix} 
0 & 0 & 1 & 0 \\
 0 & -1 & 0 & -n\\
 -1 & 0 & 0 & 0 \\
 0 & 0 & 0 & -1\end{pmatrix}$} &  \\[25pt]\hline
 $[III^{\ast}-I_n^{\ast}-m]$  &$(x-1) (t+x) \times $ $\left(t^{4 m+3}+(x-1)^2\right) \left(t^{n+2}+x^2\right) $& $0$&  $5+4m$&$5+4m$ & $15+n+12m$  &
{\tiny $\begin{pmatrix} 
0 & 0 & -1 & 0 \\
 0 & -1 & 0 & -n \\
 1 & 0 & 0 & 0 \\
 0 & 0 & 0 & -1\end{pmatrix}$} &  \\[25pt]\hline
 $[III^{\ast}-I_n^{\ast}-\alpha]$  &$t x \left(t+x^2\right) \left(t^n+(x-1)^2\right) $& $2$&  $5$&$7$ & $13+n$  &
{\tiny $\begin{pmatrix} 
0 & 0 & -1 & 0 \\
 0 & -1 & 0 & -n \\
 1 & 0 & 0 & 0 \\
 0 & 0 & 0 & -1\end{pmatrix}$} &  \\[25pt]\hline
 $[III-II_n]$  $(n\geq0)$&$\left(t+x^4\right) \left(t^n+(x-1)^2\right) $& $1$&  $1$&$2$ & $3+n$  &
{\tiny $\begin{pmatrix} 
0 & 0 & 1 & 0 \\
 0 & 1 & 1 & n+1 \\
 -1 & 0 & 0 & -1 \\
 0 & 0 & 0 & 1\end{pmatrix}$} &  \\[25pt]\hline
 \multirow{2}{2.1cm}[-0.5em]{$[III^{\ast}-II_n]$ $(n\geq0)$}&$n=0: $ $t \left(t+x^2\right) \left(t+x^4\right)$& $4$&  $7$&$11$ & $18$  &
\multirow{2}{*}{{\tiny $\begin{pmatrix} 
0 & 0 & -1 & 1 \\
 1& 1 & 0 & n \\
 1 & 0 & 0 & 0 \\
 0 & 0 & 0 & 1\end{pmatrix}$}} & \\*[8pt] \cline{2-6}
&$n>0 : $ $ \left(t^3+x^4\right) \left(t^{n-1}+(x-1)^2\right)$& $3$&  $3$&$6$ & $8+n$  & &  \\[8pt]\hline
 $[III-II_n^{\ast}]$ $(n\geq0)$  &$\left(t+x^4\right) \left(t^{n+1}+x^2\right) $& $2+n$&  $3$&$5+n$ & $8+n$  &
{\tiny $\begin{pmatrix} 
0 & 0 & 1 & -1 \\
 -1 & -1 & 0 & -n \\
 -1 & 0 & 0 & 0 \\
 0 & 0 & 0 & -1\end{pmatrix}$} &  \\[25pt]\hline
 \multirow{2}{2.1cm}[-0.5em]{$[III^{\ast}-II_n^{\ast}]$ $(n\geq0)$}&$t \left(t+x^4\right) \left(t^n+(x-1)^2\right)$& $3$&  $5$&$8$ & $13+n$  &
\multirow{2}{*}{{\tiny $\begin{pmatrix} 
0 & 0 & -1 & 0 \\
 0& -1 & -1 & -n-1 \\
 1 & 0 & 0 & 1 \\
 0 & 0 & 0 & -1\end{pmatrix}$}} & \\*[8pt] \cline{2-6}
&$ \left(t^3+x^4\right) \left(t^{n+2}+x^2\right)$& $5+n$&  $9$&$14+n$ & $23+n$  & &  \\[8pt]\hline
\end{longtable}
\end{center}}

\newpage
\paragraph*{Type 4 (parabolic)}

{\scriptsize
\begin{center}
\begin{longtable}{|M{1.6cm}| M{5cm}| M{0.8cm} |M{1.6cm} |M{1.6cm}
  |M{1.2cm}| M{2.8cm}| N}
\hline
Type                    &  Local model &  $\mu(I_2)$ & $\mu(I_4)$ &$\mu(I_6)$& $\mu(I_{10})$ & Monodromy &      \\
\hline
 $[I_{n-p-0}]$  &$ (x-1) \left(t^n+x^2\right) \left(t^p+(x-\alpha )^2\right)$& 0&  $0$&$0$ & $n+p$  &
{\tiny $\begin{pmatrix} 
 1 & 0 & p & 0 \\
 0 & 1 & 0 & n \\
 0 & 0 & 1 & 0 \\
 0 & 0 & 0 & 1\end{pmatrix}$} &  \\[25pt]\hline
 $[I_n-I_p-m]$  $(m>0)$&$ \left(t^{2 m}+x\right) \left(t^p+(x-1)^2\right) \left(t^{4 m+n}+x^2\right)$& $0$&  $4m$&$4m$ & $n+p+12m$  &
{\tiny $\begin{pmatrix} 
1 & 0 & p & 0 \\
 0 & 1 & 0 & n \\
 0 & 0 & 1 & 0 \\
 0 & 0 & 0 & 1\end{pmatrix}$} &  \\[25pt]\hline
 $[I_{n-p-0}^{\ast}]$  &$t (x-1) \left(t^n+x^2\right) \left(t^p+(x-\alpha )^2\right) $& $2$&  $4$&$6$ & $10+n+p$  &
{\tiny $\begin{pmatrix} 
 -1 & 0 & -p & 0 \\
 0 & -1 & 0 & -n \\
 0 & 0 & -1 & 0 \\
 0 & 0 & 0 & -1\end{pmatrix}$} &  \\[25pt]\hline
 $[I_n^{\ast}-I_p^{\ast}-m]$  &$\begin{aligned}&(t+(x-1)) \left(t^{2
       m+1}+x\right) \times \\ &\left(t^{p+2}+(x-1)^2\right) \left(t^{4 m+n+2}+x^2\right)\end{aligned} $& $0$&  $4+4m$&$4+4m$ & $12+n+p+12m$  &
{\tiny $\begin{pmatrix} 
 -1 & 0 & -p & 0 \\
 0 & -1 & 0 & -n \\
 0 & 0 & -1 & 0 \\
 0 & 0 & 0 & -1\end{pmatrix}$} &  \\[25pt]\hline
 $[I_n-I_p^{\ast}-m]$  &$\begin{aligned}& \left(t^{2 m+1}+x\right)
   \left(t^n+(x-1)^2\right) \times \\ &\left(t^{4 m+p+2}+x^2\right)\end{aligned}$& $0$&  $2+4m$&$2+4m$ & $6+n+p+12m$  &
{\tiny $\begin{pmatrix} 
 -1 & 0 & -p & 0 \\
 0 &1 & 0 & n \\
 0 & 0 & -1 & 0 \\
 0 & 0 & 0 & 1\end{pmatrix}$} &  \\[25pt]\hline
 \multirow{2}{2.1cm}[-0.3em]{$[2I_n-m]$ $n=2k+l$ $l=0,1$} & $\begin{aligned}&\left(t^{m+1}+\left(x^2-t\right)\right) \times \\&\left(x^l t^{k+2 m+2}+\left(x^2-t\right)^2\right)\end{aligned}$& $3$&$\text{min}(3+k+2m,6+4m)$&$\text{min}(3+k+2m,9+4m)$& $6+k+2m$  &
\multirow{2}{*}{{\tiny $\begin{pmatrix} 
0 & 1 & 0 & n \\
 1& 0 & 0 & 0 \\
0 & 0 & 0 & 1 \\
 0 & 0 & 1 & 0\end{pmatrix}$}} & \\*[8pt] \cline{2-6}
&$m=0 : $ $\left(\alpha  t+x^2\right) \left(t^{k+2} x^l+\left(x^2-t\right)^2\right)$& $3$&$\text{min}(6,3+k)$&$\text{min}(9,3+k)$ & $6+k$  & &  \\[8pt]\hline
 $[2I_n^{\ast}-m]$  $n=2k+l$ $l=0,1$ &$\begin{aligned}&\left(x
     t^{m+1}+\left(x^2-t\right)\right) \times \\ &\left(x^l t^{k+2 m+3}+\left(x^2-t\right)^2\right)\end{aligned} $& $3$&  $8+4m$&$11+4m$ & $21+l+2k+12m$  &
{\tiny $\begin{pmatrix} 
0 & -1 & 0 & -n \\
 1 & 0 & 0 & 0 \\
 0 & 0 & 0 & -1 \\
 0 & 0 & 1 & 0\end{pmatrix}$} &  \\[25pt]\hline
 $[II_{n-p}]$  &$\left(t+x^2\right) \left(t^{n-1}+(x-1)^2\right) \left(t^{p+1}+x^2\right)$& $1$&  $2$&$3$ & $5+n+p$  &
{\tiny $\begin{pmatrix} 
 -1 & 0 & -p & -1 \\
 0 & 1 & 1 & n \\
 0 & 0 & -1 & 0 \\
 0 & 0 & 0 & 1\end{pmatrix}$} &  \\[25pt]\hline 
 $[III_n]$   $n=2k+l$ $l=0,1$ &$x \left(x^l t^{k-l+6}+\left(x^2-t^3\right)^2\right)$& $6$&  $12$&$18$ & $30+l+2k$  &
{\tiny $\begin{pmatrix} 
 0 & -1 & 1 & 0 \\
 1 & 0 & n & -1 \\
 0 & 0 & 0 & -1 \\
 0 & 0 & 1 & 0\end{pmatrix}$} &  \\[25pt]\hline 
\end{longtable}
\end{center}}

\paragraph*{Type 5 (parabolic)}

{\scriptsize
\begin{center}
\begin{longtable}{|M{2cm}| M{5.1cm}| M{0.8cm} |M{0.8cm} |M{0.8cm}
  |M{1.2cm}| M{4cm}| N}
\hline
Type                    &  Local model &  $\mu(I_2)$ & $\mu(I_4)$ &$\mu(I_6)$& $\mu(I_{10})$ & Monodromy &      \\
\hline
 $[I_{n-p-q}]$  &$\left(t^n+x^2\right) \left(t^p+(x-1)^2\right)\left(t^q+(x-2)^2\right)$& 0&  $0$&$0$ & $n+p+q$  &
{\tiny $\begin{pmatrix} 
 1 & 0 & p+q & -q \\
 0 & 1 & -q & n+q \\
 0 & 0 & 1 & 0 \\
 0 & 0 & 0 & 1\end{pmatrix}$} &  \\[25pt]\hline
 $[I_{n-p-q}^{\ast}]$  &$\begin{aligned}&  t \left(t^n+x^2\right) \left(t^p+(x-1)^2\right)\times\\& \left(t^q+(x-2)^2\right)\end{aligned} $& $2$&  $4$&$6$ & $10+n+p+q$  &
{\tiny $\begin{pmatrix} 
 -1 & 0 & -p-q & q \\
 0 & -1 & q & -n-q \\
 0 & 0 & -1 & 0 \\
 0 & 0 & 0 & -1\end{pmatrix}$} &  \\[25pt]\hline
 $[II_{n-p}]$ $p=2k+l$ $l=0,1$  &$\left(t^{n-1}+(x-1)^2\right) \left(t^{k+2} x^l+\left(x^2-t\right)^2\right) $& $1$&  $2$&$3$ & $5+l+n+2k$  &
{\tiny $\begin{pmatrix} 
 -1 & 0 & -p & 0 \\
 1 & 1 & p & n \\
 0 & 0 & -1 & 1 \\
 0 & 0 & 0 & 1\end{pmatrix}$} &  \\[25pt]\hline
 $[II_{n-p}^{\ast}]$  $p=2k+l$ $l=0,1$ &$ \begin{aligned} & t \left(t^{n-1}+(x-1)^2\right)\times\\& \left(t^{k+2} x^l+\left(x^2-t\right)^2\right)\end{aligned}$& $3$&  $6$&$9$ & $15+l+n+2k$  &
{\tiny $\begin{pmatrix} 
 1 & 0 & -p & 0 \\
 -1 & -1 & p & -n \\
 0 & 0 & 1 & -1 \\
 0 & 0 & 0 & -1\end{pmatrix}$} &  \\[25pt]\hline
 $[III_n]$  $n=3k+l$ $l=0,1,2$&$t^{k+2} x^l+\left(x^3-t\right)^2$& $2$&  $4$&$6$ & $10+l+3k$  &
{\tiny $\begin{pmatrix} 
 0 & 1 & -n & n \\
 -1 & -1 & 0 & 0 \\
 0 & 0 & -1 & 1 \\
 0 & 0 & -1 & 0\end{pmatrix}$} &  \\[25pt]\hline
 $[III_n^{\ast}]$ $n=3k+l$ $l=0,1,2$ &$ t \left(x t^{k+2}+\left(x^3-t\right)^2\right)$& $4$&  $8$&$12$ & $21+3k$  &
{\tiny $\begin{pmatrix} 
 0 &-1 & n & -n \\
 1 & 1 & 0 & 0 \\
 0 & 0 & 1 & -1 \\
 0 & 0 & 1 & 0\end{pmatrix}$} &  \\[25pt]\hline
\end{longtable}
\end{center}}

\section{Matter representation analysis}\label{sec:matter-analysis}

In this appendix we have a closer look at the matter representations for the gauge groups in our resolutions. 
We proceed here along the lines of \cite{Grassi:2011hq,Morrison:2012td} and work out two cases which are of particular interest to us. 
Concretely, we will determine the matter representations for the chains of
gauge algebras $\mathfrak{su}(2)-\mathfrak{so}(7)$ \cite{Morrison:2012td} and $\mathfrak{sp}(2)-\mathfrak{so}(13)-\mathfrak{sp}(3)$, 
both examples appearing in the resolution of the $\kod{[II_{4-3}]}$ model, cf.\ section~\ref{sec:II_n-p}.

For $\kod I_0^*$ singularities we have to analyze the monodromy cover given by the following equation:
\begin{equation}
\psi^3 +\left.\frac{f}{z^2}\right|_{z=0}\psi+\left.\frac{g}{z^3}\right|_{z=0}=0 \, ,
\end{equation}
as already given in table  \ref{tab:monodromycovers}, but now denoting by
$z=0$ the (local) defining equation of the curve along which the $\kod I_0^*$ singularity appears. In our case this translates to
\begin{equation}\label{eq:covering-eqn-so7}
(\psi+ 4 e_{2})(\psi^2 - 4 e_2\psi+124416 e_4 e_2+4 e_2^2)=0\, ,
\end{equation}
where $e_2=0$ and $e_4=0$ are the defining equations for the curves of the $\kod{III}$ and $\kod I_1$ singularity, respectively, intersecting the $e_3=0$ locus along which we have the $\kod I_0^*$ singularity. Since \eqref{eq:covering-eqn-so7} factorizes into two irreducible parts, we obtain an $\mathfrak{so}(7)$ along the $(-3)$-curve $e_3=0$. The discriminant of the cubic \eqref{eq:covering-eqn-so7} is given by
\begin{equation}
\delta_{e_3}=-\left(2^{11}3^5 e_2 e_4\right)\left(36 e_2 (3456 e_4 + e_2)\right)^2=:\alpha \beta^2 \, ,
\end{equation}
where $\alpha$ is the discriminant of the quadratic factor of \eqref{eq:covering-eqn-so7}. From $\delta_{e_3}$ we can read off the matter representations of $\mathfrak{so}(7)$ because the vanishing loci of $\beta$ are related to the spinor representation and the genus of the cover, i.e.\ $\deg(\alpha)/2-1$, gives the number of vector representations of $\mathfrak{so}(7)$. Therefore, we obtain two spinor representations and no vector representation. One is located at the intersection with the $\kod{III}$ singularity and one at the point $3456 e_4 + e_2=0$. Including the $\mathfrak{su}(2)$ along $e_2=0$ in this picture, we have the representations $\frac12 (\mathbf 2, \mathbf 8_s)$ and $ (\mathbf 1, \mathbf 8_s)$ under $\mathfrak{su}(2)\oplus\mathfrak{so}(7)$, with the half-bifundamental at the intersection point of $\kod{III}$ and $\kod I_0^*$. The states are precisely those
needed for anomaly cancellation of an $\mathfrak{su}(2)$ along a $(-2)$-curve and an $\mathfrak{so}(7)$ along a $(-3)$-curve.

We now discuss the second example, i.e.\ the $\mathfrak{sp}(2)-\mathfrak{so}(13)-\mathfrak{sp}(3)$-cluster of $\kod{[II_{4-3}]}$, which to our knowledge 
has not been worked out in detail in the literature.
To figure out its matter content we follow the strategy outlined in \cite{Grassi:2011hq}.
We will start by calculating the Tate cycle \cite{Grassi:2011hq} for our example and compare it to what we obtain from the Katz-Vafa procedure \cite{Katz:1996xe} to check whether all the anomalies are canceled. Then we will add the `delocalized' matter and determine the representations actually appearing.

The monodromy covers for $\kod I_n$ and $\kod I^*_m$ with $n>2$ and $m>0$, respectively, are given by\footnote{For $\kod I^*_m$ with $m$ odd there has to be an additional factor $-\frac14$ in front of $\beta$.}
\begin{equation}
\psi^2-\beta=0\,.
\end{equation}
For the $\kod I_n$ case $\beta$ is 
\begin{equation}
\beta:=-\left.\frac92\frac gf\right|_{z=0}
\end{equation}
and for an $\kod I^*_m$ singularity
\begin{equation}
\beta:=\left\{\begin{array}{ll}\delta/\gamma^3 & \textmd{for $m$ odd}\\ -\delta/\gamma^2  & \textmd{for $m$ even}\end{array}\right.
\end{equation}
with $\delta=\left.\Delta/z^{m+6}\right|_{z=0}$ the reduced discriminant where $z$ is again the (local) defining equation of the divisor along which the singularity appears. For $\kod I^*_m$ the divisor $\gamma$ is defined to be
\begin{equation}
\gamma:=-\left.\frac92\frac {g}{z\,f}\right|_{z=0}
\end{equation}
and for $\kod I_n
$ 
\begin{equation}
\gamma:=\frac{\delta}{\beta^2}
\end{equation}
with  the reduced discriminant $\delta=\left.\Delta/z^{n}\right|_{z=0}$. 
For the three singularities $\kod I_5$, $\kod I^*_3$, $\kod I_6$ at $e_8=0$, $e_9=0$, $e_{10}=0$, respectively, we obtain the following expressions for $\beta$, $\gamma$ and $\delta$:
\begin{equation}
\begin{aligned}
&\beta_{e_8}=  6 e_9 e_7\,,      & &      \gamma_{e_8}=  - 2^{13}3^7 e_9^7 e_7^6 \,,    &  &     \delta_{e_8}=  - 2^{15}3^9 e_9^9 e_7^8 \,, \\
&\beta_{e_9}=    2^{7} 3^3 e_{10}^6 e_8^5  (e_8 - 2^{5} e_{10} )   \,,     &  &     \gamma_{e_9}=6  \,,  &   &     \delta_{e_9}=2^{10} 3^6 e_{10}^6 e_8^5  (e_8 - 2^{5} e_{10} )\,,  \\
&\beta_{e_{10}}=  6 e_9 e_{11}  \,,    &  &     \gamma_{e_{10}}= 2^{8} 3^4 e_9^7 e_{11}^7 \,,    &   &     \delta_{e_{10}}=  2^{10} 3^6 e_9^9 e_{11}^9\,.
\end{aligned}
\end{equation}
In none of the above cases $\beta$ is a perfect square. Therefore, we find  that the gauge algebras are, indeed, $\mathfrak{sp}(2)$, $\mathfrak{so}(13)$ and $\mathfrak{sp}(3)$ along the three curves, as anticipated already above.

The Tate cycle for a curve $\Sigma$ was defined in \cite{Grassi:2011hq} as
\begin{equation}
Z_{\rm{Tate},\Sigma}:=\tfrac12\left.(K_B+\Sigma)\right|_\Sigma\otimes \rho_\alpha+\tfrac12 \textmd{div}(\beta_\Sigma)\otimes \rho_{\sqrt{\beta}}+\textmd{div}(\gamma_\Sigma)\otimes \rho_\gamma
\end{equation}
with $K_B$ denoting the canonical bundle of the base. The representations $\rho_\alpha$, $\rho_{\sqrt{\beta}}$, $\rho_\gamma$ for  
our gauge algebras are
\begin{equation}
\begin{aligned}
\mathfrak{sp}(2):&\quad\rho_\alpha=  \adj+\antisym  +2\cdot\fund  \,,  & &\rho_{\sqrt{\beta}}=    \antisym +\fund \,,  & &\rho_\gamma=   \fund \,,\\
\mathfrak{so}(13):&\quad\rho_\alpha=   \adj+\vect  \,,  & &\rho_{\sqrt{\beta}}=  \vect  \,,  & &\rho_\gamma=  \tfrac14\cdot \spin+\vect \,,\\
\mathfrak{sp}(3):&\quad\rho_\alpha=   \adj+\antisym      \,,  & &\rho_{\sqrt{\beta}}=   \antisym \,,  & &\rho_\gamma=  \fund \,.
\end{aligned}
\end{equation}
To check anomaly cancellation only the `local' part
\begin{equation}
Z^{\rm{loc}}_{\rm{Tate},\Sigma}:=\tfrac12 \textmd{div}(\beta_\Sigma)\otimes \rho_{\sqrt{\beta}}+\textmd{div}(\gamma_\Sigma)\otimes \rho_\gamma
\end{equation}
of the Tate cycle is needed \cite{Grassi:2011hq}.  The explicit expressions for the three cycles $Z^{\rm{loc}}_{\rm{Tate},e_i}$ are:
\begin{equation}\label{eq:loc-tate-cycles}
\begin{aligned}
Z^{\rm{loc}}_{\rm{Tate},e_8}=&\tfrac12(e_{7,8}+e_{8,9})\otimes \antisym +\tfrac12(13\,e_{7,8}+15\,e_{8,9})\otimes\fund\,,\\
Z^{\rm{loc}}_{\rm{Tate},e_9}=&\tfrac12(5\, e_{8,9}+6\,e_{9,10}+e_{9,\zeta})\otimes \vect  \,,\\
Z^{\rm{loc}}_{\rm{Tate},e_{10}}=&\tfrac12  (e_{9,10}+e_{10,11})\otimes \antisym + (7\,e_{9,10}+7\,e_{10,11})\otimes \fund\,,
\end{aligned}
\end{equation}
where $e_{i,j}$ is a short hand for the point (or divisor on the respective curve) $e_i=e_j=0$. Note that on the projective line all points are rationally equivalent.

To determine the (virtual) local matter representations we use Katz-Vafa \cite{Katz:1996xe}. The prescription they give to obtain the local matter is: decompose the adjoint representation of the gauge group, associated to the enhanced singular point, under the covering algebra related to the curves intersecting at the enhancement point; collect all the irreducible representations besides the adjoints and singlets and reduce them further to representations of the actual algebras. Furthermore, for these quaternionic representations there is an additional overall pre-factor related to the monodromy cover, i.e.\ $\frac1k$ with $k$ the degree of the cover. Along the three curves there are five such enhancement points as we see from the reduced discriminants. These points are $e_{7,8}$, $e_{8,9}$, $e_{9,\zeta}$, $e_{9,10}$, $e_{10,11}$ with
enhancements to $\kod I^*_7$, $\kod I^*_8$, $\kod I^*_4$, $\kod I^*_9$, $\kod I^*_9$ singularities which translates to the Lie algebras $\mathfrak{so}(22)$, $\mathfrak{so}(24)$, $\mathfrak{so}(16)$, $\mathfrak{so}(26)$, $\mathfrak{so}(26)$, respectively. The decompositions of the respective adjoint representations under the covering algebras are as follows:
\begin{equation}\label{eq:local-virtual-cycle}
\begin{aligned}
SO(22)&\supset SU(5)\\
\mbf{231}&\rightarrow
\mbf 1^{\oplus 67}\oplus 
\mbf{10}\oplus\bar{\mbf{10}}\oplus
\mbf{24}\oplus
\mbf{5}^{\oplus 12} \oplus \bar{\mbf{5}}^{\oplus 12}\,,\\
SO(24)&\supset SU(5)\times SO(14)\\
\mbf{276}&\rightarrow
\mbf 1\otimes \mbf 1 \oplus
\mbf{10}\otimes \mbf 1\oplus\bar{\mbf{10}}\otimes\mbf 1 \oplus
\mbf{24}\otimes\mbf 1\oplus 
\mbf{1}\otimes\mbf {91}\oplus 
 \mbf{5}\otimes\mbf{14} \oplus  \bar{\mbf{5}}\otimes  \bar{\mbf{14}}\,,\\
SO(16)&\supset S0(14)\\
\mbf{120}&\rightarrow 
\mbf{1}\oplus\mbf{14}\oplus\bar{\mbf{14}}\oplus\mbf{91}\,,\\
SO(26)&\supset SU(6)\times SO(14)\\
\mbf{325}&\rightarrow 
\mbf 1\otimes \mbf 1 \oplus
\mbf{15}\otimes \mbf 1\oplus\bar{\mbf{15}}\otimes\mbf 1 \oplus
\mbf{35}\otimes\mbf 1\oplus 
\mbf{1}\otimes\mbf {91}\oplus 
 \mbf{6}\otimes\mbf{14} \oplus  \bar{\mbf{6}}\otimes  \bar{\mbf{14}}\,.
\end{aligned}
\end{equation}
From this we can now read off the `local part' of the virtual matter cycle which collects the all the data\footnote{We omit any indices on the representations indicating to which algebras they belong. However, the zero-cycles in front of the representations and the fact that we always used the order $\mathfrak{sp}(n)\oplus\mathfrak{so}(m)$ when two irreducible algebras are involved allow for a unique identification.}
\begin{equation}\label{eq:local-virtual-cycle}
\begin{aligned}
Z^{\rm{loc}}_{\rm{virtual}}=&
e_{7,8}\cdot\tfrac12 \left(  \fund^{\oplus 13}\oplus \antisym\right)+
e_{8,9}\cdot\tfrac12 \left( (\fund\otimes\mbf1)^{\oplus2}\oplus\antisym\otimes\mbf1\oplus(\fund\oplus\mbf1)\otimes\vect \right)+\\
&\quad+
e_{9,\zeta}\cdot\tfrac12 \vect+e_{9,10}\cdot\tfrac12 \left( \antisym\otimes\mbf1\oplus\fund\otimes(\mbf1\oplus\vect)  \right)+\\
&\quad+
e_{10,11}\cdot\tfrac12 \left( \antisym\oplus\fund^{\oplus14} \right)\,.
\end{aligned}
\end{equation}
Here we should note that the one-half in front of the representations is due to the fact that all gauge algebras involved have a di-cover. Furthermore, from \eqref{eq:local-virtual-cycle} to \eqref{eq:local-virtual-cycle} we got rid of the singlets and adjoints, reduce the representations to their actual algebras and rearranged them into quaternionic representations, e.g.\ $\mbf{6}\oplus\bar{\mbf{6}}$ is $\fund$ of $\mathfrak{sp}(3)$.

Restricting $Z^{\rm{loc}}_{\rm{virtual}}$ to the algebras $\mathfrak{sp}(2)$, $\mathfrak{so}(13)$, $\mathfrak{sp}(3)$, we must have 
\begin{equation}\label{eq:anomaly-cancelation-Z}
\left.Z^{\rm{loc}}_{\rm{virtual}}\right|_{\mathfrak{sp}(2)}\sim Z^{\rm{loc}}_{\rm{Tate},e_8}\,,\quad
\left.Z^{\rm{loc}}_{\rm{virtual}}\right|_{\mathfrak{so}(13)}\sim Z^{\rm{loc}}_{\rm{Tate},e_9} \,,\quad
\left.Z^{\rm{loc}}_{\rm{virtual}}\right|_{\mathfrak{sp}(3)}\sim Z^{\rm{loc}}_{\rm{Tate},e_{10}}\,,
\end{equation}
for anomaly cancellation. Here, the restriction has to be understood as taking only the representations which are charged under the gauge algebra to which we restrict, i.e.
\begin{equation}
\begin{aligned}
\left.Z^{\rm{loc}}_{\rm{virtual}}\right|_{\mathfrak{sp}(2)}&=e_{7,8}\cdot\tfrac12 \left(  \fund^{\oplus 13}\oplus \antisym\right)+
e_{8,9}\cdot\tfrac12 \left( \fund^{\oplus15}\oplus\antisym \right)\,,\\
\left.Z^{\rm{loc}}_{\rm{virtual}}\right|_{\mathfrak{so}(13)}&=e_{8,9}\cdot\tfrac12  \vect^{\oplus5} + e_{9,\zeta}\cdot\tfrac12 \vect+e_{9,10}\cdot\tfrac12\vect^{\oplus6}\,,\\
\left.Z^{\rm{loc}}_{\rm{virtual}}\right|_{\mathfrak{sp}(3)}&= e_{9,10}\cdot\tfrac12 \left( \antisym\oplus\fund^{\oplus14}  \right)+e_{10,11}\cdot\tfrac12 \left( \antisym\oplus\fund^{\oplus14} \right)\,.
\end{aligned}
\end{equation}
Since in \eqref{eq:anomaly-cancelation-Z} the expressions have only be (rational and) Casimir equivalent in degree 2 and 4, \eqref{eq:anomaly-cancelation-Z} is indeed fulfilled. The second condition, besides \eqref{eq:anomaly-cancelation-Z}, which has to be satisfied for anomaly cancellation is
\begin{equation}
\mu_{Z_{\rm virtual}}(\mathfrak g(e_i),\mathfrak g(e_j))=e_i\cdot e_j\,,
\end{equation}
where the right hand side is the intersection number of the two curves $e_i$ and $e_j$, $\mathfrak g(e_i)$ denotes the gauge algebras along these curves and $\mu$ is the representation multiplicity \cite{Grassi:2011hq}. In our case this is trivially met because 
we have one bi-fundamental for each intersection point.

After checking the anomalies, we can finally give the full matter content for our setup. To do so we only have to add the `delocalized matter' part
\begin{equation}
\tfrac12 \sum_i \left.(K_B+e_i)\right|_{e_i}\otimes\rho_\alpha
\end{equation}
to $Z^{\rm{loc}}_{\rm{virtual}}$ and add to the degree of the so obtained cycle the adjoint representations such that 
\begin{equation}
\rho_{\rm matter}=\adj(\mathfrak g)\oplus \deg(Z_{\rm virtuel})\,.
\end{equation}
Here $\mathfrak g$ is the full algebra, i.e.\ in our case $\mathfrak{sp}(2)\oplus\mathfrak{so}(13)\oplus\mathfrak{sp}(3)$.
For $Z_{\rm virtuel}$ we find
\begin{equation}
\begin{aligned}
Z_{\rm{virtual}}=&
e_{7,8}\cdot\tfrac12 \left(  \fund^{\oplus 11}\oplus \adj\right)+
e_{8,9}\cdot\tfrac12 \left( \adj\otimes\mbf1\oplus\fund\otimes\vect \oplus\mbf1\otimes\adj\right)+\\
&+e_{9,10}\cdot\tfrac12 \left( \adj\otimes\mbf1\oplus\fund\otimes(\mbf1\oplus\vect) \oplus\mbf1\otimes\adj \right)+
e_{10,11}\cdot\tfrac12 \left( \adj\oplus\fund^{\oplus14} \right)\,.
\end{aligned}
\end{equation}
where we used that $\left.(K_B+e_i)\right|_{e_i}=-e_{i-1,i}-e_{i,i+1}$ and that all points on $\mathbb P^1$ are rationally equivalent. Hence, we obtain 
\begin{equation}
\rho_{\rm matter}=\tfrac12(\mbf 4\otimes\mbf1\otimes\mbf1 )^{\oplus 11}\oplus \tfrac12(\mbf 4\otimes\mbf{13}\otimes \mbf 1)\oplus
\tfrac12(\mbf 1\otimes(\mbf{13}\oplus\mbf1)\otimes \mbf 6)\oplus\tfrac12(\mbf 1\otimes\mbf1\otimes\mbf6 )^{\oplus 14}
\end{equation}
for the matter content. Note that the 11 and 13 out of the 14 flavor degrees of $\tfrac12(\mbf 4\otimes\mbf1\otimes\mbf1 )$ and $\tfrac12(\mbf 1\otimes\mbf1\otimes\mbf6 )$, respectively, would become charged if we would take the full $\kod{[II_{4-3}]}$ model into account.


\bibliographystyle{utphys}
\bibliography{papers}

\end{document}